\documentclass[preprint,3p]{elsarticle}
\pdfoutput=1
\usepackage{amsmath,subeqnarray,multirow,hhline,natbib,amsmath}
\newcommand{\diff}[3][]{\dfrac{\mathrm{d}^{#1}#2}{\mathrm{d}{#3}^{#1}}}

\newcommand{\pdiff}[3][]{\dfrac{\partial^{#1} #2}{\partial {#3}^{#1}}}

\def\greekbf#1{\mbox{\mathversion{bold}$#1$}}

\begin{document}

\begin{frontmatter}

\title{Finite Element Simulation of Dynamic Wetting Flows as an Interface Formation Process}

\author{J.E. Sprittles$^{a}$\footnote{Corresponding author.} and Y.D. Shikhmurzaev$^{b}$}
\address{$^a$ Mathematical Institute, University of Oxford, Oxford, OX1 3LB, U.K.}

\address{$^b$ School of Mathematics, University of Birmingham, Edgbaston, Birmingham, B15 2TT, U.K.}

\ead{$^a$ james.sprittles@gmail.com and $^b$ yulii@for.mat.bham.ac.uk}

\begin{abstract}
A mathematically challenging model of dynamic wetting as a process
of interface formation has been, for the first time, fully
incorporated into a numerical code based on the finite element
method and applied, as a test case, to the problem of capillary rise. The
motivation for this work comes from the fact that, as discovered
experimentally more than a decade ago, the key variable in dynamic
wetting flows --- the dynamic contact angle --- depends not just
on the velocity of the three-phase contact line but on the entire
flow field/geometry.  Hence, to describe this effect, it
becomes necessary to use the mathematical model that has this dependence as its
integral part. A new physical effect, termed the `hydrodynamic resist
to dynamic wetting', is discovered where the influence of the
capillary's radius on the dynamic
contact angle, and hence on the global flow, is computed. The
capabilities of the numerical framework are then demonstrated by
comparing the results to experiments on the unsteady capillary
rise, where excellent agreement is obtained. Practical
recommendations on the spatial resolution required by the
numerical scheme for a given set of non-dimensional similarity
parameters are provided, and a comparison to asymptotic results
available in limiting cases confirms that the code is converging to
the correct solution. The appendix gives a user-friendly
step-by-step guide specifying the entire implementation and
allowing the reader to easily reproduce all presented results,
including the benchmark calculations.
\end{abstract}


\end{frontmatter}

\section{Introduction}

Reliable simulation of flows in which a liquid advances over a
solid, known as dynamic wetting flows, is the key to the
understanding of a whole host of natural phenomena and
technological processes. In the technological context, the study
of these flows has often been motivated by the need to optimize
\emph{continuous} coating processes that are routinely used to
create thin liquid films on a product \cite{weinstein04}, for
example, in the coating of optical fibres
\cite{ravinutala02,simpkins03}. However, more recently,
\emph{discrete} coating, in particular inkjet printing of
microdrops \cite{vandam04}, has matured into a viable, and often
preferable, alternative to traditional fabrication processes, e.g.
in the additive manufacturing of 3D structures or the creation of
P-OLED displays \cite{calvert01,singh10}, and it is becoming a new
driving force behind research into dynamic wetting phenomena. In
most cases, such flows can be regarded as microfluidic phenomena,
where a large surface-to-volume ratio brings in interfacial
effects on the flow that are not observed at larger scales.

Obtaining accurate information about micro and nanofluidic flows
experimentally is often difficult and usually costly so that,
consequently, a desired alternative is to have a reliable theory describing the physics that is dominant for
this class of flows and incorporate it into a flexible and robust
computational tool which can quickly map the parameter space of
interest to allow a specific process to be optimized.  Such
computational software could be validated against experiments at
scales and geometries easily accessible to accurate measurement and
then used to make predictions in processes inaccessible to
experimental analysis.

The discovery that no solution exists for the moving contact-line
problem in the framework of standard fluid mechanics
\cite{huh71,dussan79} prompted a number of remedies to be
proposed, which are summarized, for example, in \cite[ch.~3
of][]{shik07}. Of these, most are what we shall refer to as
`conventional' or `slip' models, in which the no-slip condition on
the solid surface is relaxed to allow a solution to exist, with the
Navier-slip condition \cite{navier23} being the most popular choice.
As a boundary condition on the free-surface shape at the contact
line, one has to specify the contact angle formed between the free
surface and solid. In conventional models, this angle is prescribed
as a heuristic or empirical function of the contact-line speed and
material parameters of the system, e.g.\ \cite{blake69,cox86}. Such
models provide predictions that adequately describe experiments at
relatively large scales, often around the length scale of
millimetres.  It is well established that on such scales many of the
proposed models work equally well and that finding significant
deviations between their predictions, and hence ascertaining which
best captures the key physical mechanisms of dynamic wetting, is
practically impossible \cite{popescu08,seveno09}.

A physical phenomenon that gives an opportunity to distinguish
between different models came to be known under a `technological'
name of the `hydrodynamic assist of dynamic wetting' (henceforth
`hydrodynamic assist' or simply `assist'). The essence of this
effect, first observed in high accuracy experiments on the curtain
coating process \cite{blake94,clarke06}, is that  for a given liquid
spreading over a given solid at a fixed contact-line speed, the
dynamic contact angle can still be manipulated by altering the flow
field/geometry, for example, in the case of curtain coating, by
changing the flow rate or the height from which the curtain falls.
This effect has profound technological implications as it allows the
process to be optimized by off-setting the increase of the contact
angle with increasing contact-line speed by manipulating the flow
conditions and hence postponing air entrainment \cite{blake94}.

The dependence of the dynamic contact angle on the flow field has
also been reported in the imbibition of liquid into capillaries
\cite{sobolev00,sobolev01}, in the spreading of impacted drops over
solid substrates \cite{bayer06,sikalo05} and in the coating of
fibres \cite{simpkins03}. However, in many of these flows it is yet
unclear whether hydrodynamic assist actually occurs, or whether the
free surface bends significantly beneath the spatial resolution of
the experiments, whereas for curtain coating the hope of attributing
assist to the poor spatial resolution of experiments has been
quashed by careful finite element simulations which show that the degree of free-surface bending under the reported resolution of the measurements is too small to account for the observed effect and that conventional models cannot in principle describe this important
physical effect \cite{wilson06}.

Currently, the only model known to be able to even qualitatively
describe assist \cite{blake99,lukyanov07} is the model of dynamic wetting as
an interface formation process, first introduced in \cite{shik93}
and discussed in detail in  \cite{shik07}.  This model is based on
the simple physical idea that dynamic wetting, as the very name
suggests, is the process in which a fresh liquid-solid interface (a
newly `wetted' solid surface) forms. Qualitatively, the origin of
the hydrodynamic assist is that the global flow influences the
dynamics of the relaxation of the forming liquid-solid interface
towards its equilibrium state and hence the value of this
interface's surface tension at the contact line, which, together
with the surface tension on the free surface, determines the value
of the dynamic contact angle. When there is a separation of scales
between the interface formation process and the global flow, the
`moving contact-line problem' can be considered locally and
asymptotic analysis provides a speed-angle relationship which is
seen to describe experiments just as well as formulae proposed in other
models \cite{shik07}.  However, in the situation where the scale of
the global flow and that of the interface formation are no longer
separated, the influence of the flow field on the dynamic contact
angle will occur and hence no unique speed-angle relationship will
be able to describe experiments. Then, the interface formation model
becomes the only modelling tool, and, given that the processes of
practical interest are free-surface flows under the influence of, at
least, viscosity, capillarity and inertia, it is inevitable that, to
describe such flows, one needs computer simulation, i.e.\ the
development of accurate CFD codes, which, for the effect of
hydrodynamic assist to be captured, have to incorporate the
interface formation model.

Since the interface formation model came into wider circulation,
there has been considerable interest, e.g.
\cite{attane07,bayer06,sikalo05}, in using it in its entirety as a
practical tool for describing dynamic wetting phenomena,
especially on the microscale. Review articles have also referred
to the description of assist using this model as one of the main
challenges in the field \cite{blake06}.  The major obstacle in the
development of this computational tool is the mathematical
complexity of the interface formation model, as one has to solve
numerically the Navier-Stokes equations describing the bulk flow
subject to boundary conditions which are themselves partial
differential equations along the interfaces and in their turn have
to satisfy certain boundary conditions at contact lines confining the interfaces. These
conditions determine the dynamic contact angle and hence influence
the free-surface shape, which exerts its influence back on the
bulk flow. Thus, the bulk flow, the distribution of the surface
parameters along the interfaces and the dynamic contact angle that
these distributions `negotiate' become interdependent, with the
dynamic contact angle being an \emph{outcome} of the solution as opposed
to conventional models where it is an \emph{input}. This interdependency
is, on the one hand, the physical essence of the experimentally
observed effect of hydrodynamic assist to be described but, on the
other hand, it is this very interdependency that, coupled with the
nonlinearity of the bulk equations and the flow geometry, is the
reason why the model is difficult to implement robustly into a
numerical code.

Some numerical progress has been made for the computationally less
complex steady Stokes flows \cite{lukyanov07}, but what is lacking
is a step-by-step guide to the implementation of this model in the
general case, with unsteady effects in the bulk and interfaces as well as nonlinearity of the bulk equations fully implemented.  This which would pave the way for incorporating the
interface formation model into existing codes as well as developing
new ones. Therefore, the first objective of the present paper is to
address these issues and provide a digestible guide to the model's
implementation into CFD codes. Then, after giving benchmark
computational results to verify this implementation, we consider a
problem of imbibition into a capillary, compare the outcome with
experiments and predict essentially new physical effects.

A starting point in the development of the aforementioned CFD code
is to first develop an accurate computational approach for the
simulation of dynamic wetting flows using the mathematically less
complex conventional models and this was achieved in
\cite{sprittles11c}.  It was shown that many of the previous
numerical results obtained for dynamic wetting processes are
unreliable as they contain uncontrolled errors caused by not
resolving all the scales in the conventional model, most notably the
dynamics of slip and the curvature of the free surface near the
contact line.  Benchmark calculations in \cite{sprittles11c} for a range of mesh
resolutions resolved previous
misunderstandings about how to impose the dynamic contact angle and
showed that implementing it using the usual finite element ideology,
as opposed to `strong' implementation of the contact angle, works
most efficiently: it allows errors to be seen, and hence controlled,
as the computed contact angle varies from its imposed value, instead
of them being masked elsewhere in the code. Furthermore, in
\cite{sprittles11b}, we have shown that numerical artifacts which
occur at obtuse contact angles are present in both commercial
software and in academic codes where, misleadingly, they have
previously been interpreted as physical effects.  By comparing
computational results to analytic near field asymptotic ones, we
have shown that the previously obtained spikes in the distributions
of pressure observed near the contact angle are completely spurious,
and, to rectify this issue a special method, based on removing the
`hidden' eigensolutions in the problem prior to computation, has
been developed \cite{sprittles11b}. In \cite{sprittles_chem}, we
showed that our code is capable of simulating unsteady high
deformation flows by comparing to benchmark calculations published
in the literature and performed by various research groups. In
contrast, in \cite{hysing07} it has been shown that when commercial
software is used to simulate relatively simple benchmark free
surface flows, the converged solution is not the correct one.

In this series of articles
\cite{sprittles11c,sprittles11b,sprittles_chem},  our approach has
been to carefully develop a robust CFD algorithm for the simulation
of dynamic wetting flows. Thus far, this has been achieved for the
conventional models, where we validated our results by performing
mesh independence tests, comparing with asymptotic solutions in
limiting cases and, where no analytic progress was possible, to
previously obtained benchmark solutions published in the literature.
In this article, we continue this series of papers  and, for the
first time, incorporate the interface formation model into our code
in full.  Notably, it will be shown that a code implementing
the mathematically complex interface formation equations can easily recover the much
simpler conventional models by setting
certain parameters to zero.  This allows the same code to be used to
compare the predictions of dynamic wetting models for a range of
flows and hence to determine, by a comparison with experiment, where
the physics of interface formation manifests itself in a significant
way and where mathematically simpler conventional models are
adequate. Here, we focus on dynamic wetting; the extension to other
flows of interest where interface formation or disappearance also
occurs is a straightforward procedure computationally, and it will
be dealt with in forthcoming articles.

The layout of the present article is as follows. First, in
\S\ref{equations}, without limiting ourself to a particular flow
configuration, the equations describing the dynamic wetting process
are briefly recapitulated.  Then, in \S\ref{FEM}, the finite element
equations are derived for the dynamic wetting flow as an interface
formation process.  Local element matrices, and additional details
about the finite element procedure are provided in the Appendix,
which complements a `user-friendly' step-by step guide to finite
element simulation given for this class of flows in
\cite{sprittles11c} and allows one to reproduce the benchmark
simulations in \S\ref{results}. These simulations are performed for
the dynamic wetting flow through a capillary both in the case where
the meniscus motion is steady and for the unsteady imbibition of a
liquid into an initially dry capillary. Computations are checked for
convergence both as the mesh is refined and towards asymptotic
results in limiting cases. Having established the accuracy of our
approach, in \S\ref{results2} new physical effects are discovered by
considering the influence of capillary geometry on the dynamic
wetting process.  Finally, the computational tool's ability to
easily describe experimental data is shown in \S\ref{results3} and a number of
advantages over current approaches, particularly in the initial
stages of a meniscus' motion into a capillary tube, are highlighted.
Concluding remarks and areas for future research are discussed in
\S\ref{conclusions}.

\section{Modelling dynamic wetting flows as an interface formation process}\label{equations}

Consider the spreading of a Newtonian liquid, with constant density $\rho$ and viscosity $\mu$, over a chemically homogeneous smooth solid surface. The liquid is surrounded by a gas which, for simplicity, is assumed to be inviscid and dynamically passive, of a constant pressure $p_g$.  Let the flow be characterized by scales for length $L$, velocity $U$, time $L/U$, pressure $\mu U/L$ and external body force $F_{0}$. In the dimensionless form, the continuity and momentum balance equations are then given by
\begin{equation}\label{ns}
\nabla\cdot\mathbf{u} = 0,\qquad Re~\left[\pdiff{\mathbf{u}}{t} + \mathbf{u}\cdot\nabla\mathbf{u}\right] = \nabla\cdot\mathbf{P} + St~\mathbf{F},
\end{equation}
where
\begin{equation}\label{stens}
\mathbf{P} = -p\mathbf{I}+\left[\nabla\mathbf{u}+\left(\nabla\mathbf{u}\right)^{T}\right],
\end{equation}
is the stress tensor, $t$ is time, $\mathbf{u}$ and $p$ are the
liquid's velocity and pressure, $\mathbf{F}$ is the external force
density and $\mathbf{I}$ is the metric tensor of the coordinate
system.   The non-dimensional parameters are the Reynolds number
$Re=\rho U L/\mu$ and the Stokes number $St=\rho F_{0}L^{2}/(\mu
U)$.

Boundary conditions to the bulk equations are required at the
liquid-gas free surface $\mathbf{x}=\mathbf{x}_1(s_1,s_2,t)$, whose
position is to be found as part of the solution, and at the
liquid-solid interface $\mathbf{x}=\mathbf{x}_2(s_1,s_2,t)$, whose
position is known, and at other bounding surfaces which will be
specified by the problem of interest; here, $(s_1,s_2)$ are the
coordinates that parameterize the surfaces. Boundary conditions
along the free surface, the liquid-solid interface and the contact
line at which they intersect are provided by the interface formation
model \cite{shik07}, as follows.

\subsection{The interface formation model}

To represent the boundary conditions on an interface with a normal $\mathbf{n}$ in an invariant form, it is convenient to introduce a (symmetric) tensor $\mathbf{I}-\mathbf{n}\mathbf{n}$, which is essentially a metric tensor on the surface.  If an arbitrary vector $\mathbf{a}$ is decomposed into a scalar normal component $a_n=\mathbf{a}\cdot\mathbf{n}$ and a vector tangential part $\mathbf{a}_{||}$, so that $\mathbf{a}=\mathbf{a}_{||} + a_n\mathbf{n}$, we can see that, because $\mathbf{n}\cdot\left(\mathbf{I}-\mathbf{n}\mathbf{n}\right)=\mathbf{0}$, the tensor $\left(\mathbf{I}-\mathbf{n}\mathbf{n}\right)$  extracts the component of a vector which is tangential to the surface, i.e.\ $\mathbf{a}\cdot\left(\mathbf{I}-\mathbf{n}\mathbf{n}\right)=\mathbf{a}_{||}$. Hereafter, $\mathbf{n}$ is the unit normal to a surface pointing \emph{into} the liquid, and subscripts $1$ and $2$ refer to the free surface and solid surface, respectively.

The equations of interface formation on a liquid-gas free surface, which act as boundary conditions for the bulk equations (\ref{ns}), are given by
\begin{equation}\label{P_454}
\left(\pdiff{\mathbf{x}_{1}}{t}-\mathbf{v}^s_{1}\right)\cdot\mathbf{n}_1 = 0,
\end{equation}
\begin{equation}\label{P_ifm_stress}
Ca~\mathbf{n}_1\cdot\left[\nabla\mathbf{u}+\left(\nabla\mathbf{u}\right)^T\right]\cdot
\left(\mathbf{I}-\mathbf{n}_1\mathbf{n}_1\right) +\nabla\sigma_1 =\mathbf{0},
\end{equation}
\begin{equation}\label{P_ifm_stressa}
Ca\left\{p_g-p+ \mathbf{n}_1\cdot\left[\nabla\mathbf{u}+\left(\nabla\mathbf{u}\right)^T\right]\cdot
\mathbf{n}_1\right\}=\sigma_1\nabla\cdot\mathbf{n}_1,
\end{equation}
\begin{equation}\label{P_459}
\mathbf{v}^{s}_{1||}-\mathbf{u}_{||}=\frac{1+4\bar{\alpha}\bar{\beta}}{4\bar{\beta}}\nabla\sigma_1,
\end{equation}
\begin{equation}\label{P_457}
\left(\mathbf{u}- \mathbf{v}^s_{1}\right)\cdot\mathbf{n}_1 = Q\left(\rho^{s}_1-\rho^{s}_{1e}\right),
\end{equation}
\begin{equation}\label{P_458}
\epsilon\left[\pdiff{\rho^{s}_{1}}{t} + \nabla\cdot\left(\rho^{s}_{1}\mathbf{v}^{s}_{1}\right)\right] = - \left(\rho^{s}_{1}-\rho^{s}_{1e}\right),
\end{equation}
\begin{equation}\label{P_460} \sigma_{1}=\lambda(1-\rho^{s}_{1}),
\end{equation}
whilst at liquid-solid interfaces formed on a solid which moves with velocity $\mathbf{U}$, the equations of interface formation have the form
\begin{equation}\label{P_462}
\left(\mathbf{v}^s_2-\mathbf{U}\right)\cdot\mathbf{n}_2=0,
\end{equation}
\begin{equation}\label{P_463}
Ca~\mathbf{n}_2\cdot\mathbf{P}\cdot(\mathbf{I}-\mathbf{n}_2\mathbf{n}_2) + \hbox{$\frac{1}{2}$}\nabla
\sigma_{2}=\bar{\beta}\left(\mathbf{u}_{||}-\mathbf{U}_{||}\right),
\end{equation}
\begin{equation}\label{P_466}
\mathbf{v}^{s}_{2||}-\hbox{$\frac{1}{2}$}\left(\mathbf{u}_{||}+\mathbf{U}_{||}\right)=\bar{\alpha}\nabla\sigma_{2},
\end{equation}
\begin{equation}\label{P_465}
\left(\mathbf{u}-\mathbf{v}^s_2\right)\cdot\mathbf{n}_2 = Q\left(\rho^{s}_{2}-\rho^{s}_{2e}\right),
\end{equation}
\begin{equation}\label{P_464}
\epsilon\left[\pdiff{\rho^{s}_{2}}{t} + \nabla\cdot\left(\rho^{s}_{2} \mathbf{v}^{s}_{2}\right)\right] =
-\left(\rho^{s}_{2}-\rho^{s}_{2e}\right),
\end{equation}
\begin{equation}\label{P_468} \sigma_{2}=\lambda(1-\rho^{s}_{2}).
\end{equation}

\begin{figure}
\centering
\includegraphics[scale=0.60]{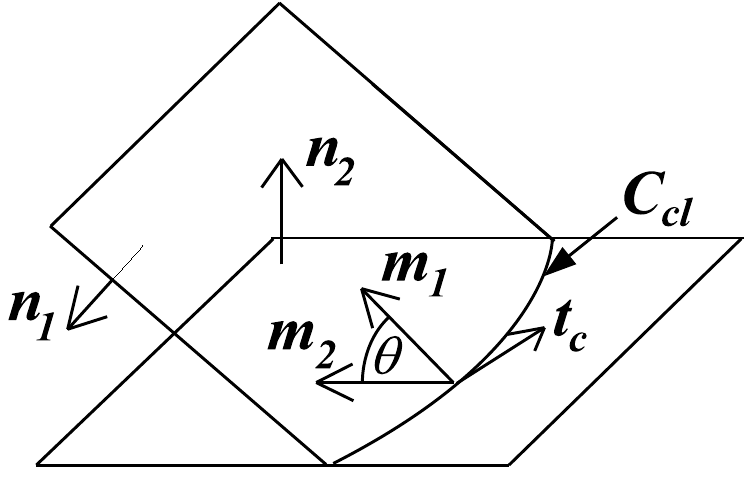}
\caption{Illustration showing the vectors associated with the liquid-gas free surface $A_1$ and the liquid-solid surface $A_2$ in the vicinity of the contact line $C_{cl}$.}
\label{F:angles}
\end{figure}

The differential term $\sigma_1\nabla\cdot\mathbf{n}_1$, where
$\sigma_1$ is the (dimensionless) surface tension on the free
surface, describing the capillary pressure in the normal-stress
equation (\ref{P_ifm_stressa}) indicates that this equation requires
its own boundary condition where the free surface terminates, i.e.\
at the contact line.  There $\mathbf{n}_2$ is known and
$\mathbf{n}_1$ must be specified by setting the dynamic contact
angle $\theta_d$ at which the free surface meets the solid surface:
\begin{equation}\label{angle}
\mathbf{n}_{1}\cdot\mathbf{n}_2 = -\cos\theta_d.
\end{equation}
This angle is determined from a force balance at the contact line, given by Young's equation \cite{young05}:
\begin{equation}\label{P_473}
\sigma_{2}+\sigma_{1}\cos\theta_d=\sigma_{S},
\end{equation}
where $\sigma_2$ and $\sigma_S$ are the surface tensions of the liquid-solid interface and solid-gas interface, respectively, and the latter is henceforth assumed to be negligible $\sigma_S\approx0$. Equations (\ref{P_458}) and (\ref{P_464}) also require a boundary condition at the contact line where the two interfaces meet, and here we have the continuity of surface mass flux
\begin{equation}\label{P_470a}
\rho^s_1\left(\mathbf{v}^s_{1||} - \mathbf{U}_{c}\right)\cdot\mathbf{m}_1 + \rho^s_2\left(\mathbf{v}^s_{2||}  - \mathbf{U}_{c}\right)\cdot\mathbf{m}_2=0
\end{equation}
where $\mathbf{U}_c$ is the (dimensionless) velocity of the
advancing contact line and $\mathbf{m}_{i}$ is the inward unit vector
tangential to surface $i$ and normal to the contact line (see
Figure~\ref{F:angles}).

The interface formation model is described in detail in the
monograph \cite{shik07} and a series of preceding papers
\cite[e.g.][]{shik93,shik97} so that here only the main ideas are
briefly recapitulated.   The surface variables are in the `surface
phase', i.e.\ physically in a microscopic layer of liquid adjacent
to the surface which is subject to intermolecular forces from two
bulk phases. In the continuum limit, this microscopic layer
becomes a mathematical surface of zero thickness with $\rho^{s}$
denoting its surface density (mass per unit area) and
$\mathbf{v}^{s}$ the velocity with which it is transported.  The
following non-dimensional parameters have been introduced
$\bar{\alpha}=\alpha\sigma/(U L)$, $\bar{\beta}=(\beta U
L)/\sigma$, $\epsilon=U\tau/L$, $\lambda =
\gamma\rho^s_{(0)}/\sigma$ and $Q =\rho^s_{(0)}/(\rho U \tau)$
which are based on phenomenological material constants $\alpha$,
$\beta$, $\gamma$, $\tau$ and $\rho_{(0)}^{s}$; in the simplest
variant of the theory, the latter are assumed to be constant and take the same
value on all interfaces; $\sigma$ is the characteristic surface
tension for which it is convenient to take the equilibrium surface
tension on the liquid-gas interface.

Estimates for the material constants have been obtained by comparing
the theory to experiments in dynamic wetting, e.g.\ in
\cite{blake02}, but could just as easily have been taken from an
entirely different process in which interface formation is key to
describing the dynamics of the flow
\cite{shik05a,shik05b,shik05c,sprittles07}. In other words, once
obtained for a particular liquid in one set of experiments, the
material constants determined can be used to describe all phenomena
involving the same fluid in which interface formation dynamics
`kicks in'.

The surface tension $\sigma_i$ is considered as a dynamic quantity
related to the surface density $\rho^s_i$ via the equations of state
in the `surface phase' ($\ref{P_460}$) and  ($\ref{P_468}$), which
are taken here in their simplest linear form.  Gradients in surface
tension influence the flow, firstly, via the stress boundary
conditions (\ref{P_ifm_stress}), (\ref{P_ifm_stressa}) and
(\ref{P_463}), i.e. via the Marangoni effect, and, secondly, in the
Darcy-type equations\footnote{The analogy with the Darcy equation is
that the tangential surface velocity $\mathbf{v}^s_{||}$ is the
average velocity of the interfacial layer and its deviation from
that generated by the outer flow is proportional to the gradient of
surface tension, which is the negative gradient of surface pressure
as $p^s=-\sigma$.} (\ref{P_459}) and (\ref{P_466}) by forcing the
surface velocity to deviate from that generated in the surface phase
by the outer flow. Mass exchange between the bulk and surface
phases, caused by the possible deviation of the surface density from
its equilibrium value $\rho^{s}_{e}$,  is accounted for in the
boundary conditions for the normal component of bulk velocity
(\ref{P_457}) and (\ref{P_465}), and in the corresponding surface
mass balance equations (\ref{P_458}) and (\ref{P_464}).

One would expect a generalized set of boundary conditions to have
the classical conditions as their limiting case. For the interface
formation model this limiting case follows from the double limit
$\epsilon \rightarrow 0,~\bar{\beta}/Ca \rightarrow \infty$.  When
the limit $\epsilon \rightarrow 0$ is applied to
(\ref{P_454})--(\ref{P_460}), the liquid-gas interface equations are
reduced to their classical form.  Notably, if we apply $\epsilon
\rightarrow 0$ to the liquid-solid equations
(\ref{P_462})--(\ref{P_468}), the conventional `slip' model is
obtained, that is the Navier-slip condition combined with
impermeability. In this limit, the surfaces are in equilibrium so
Young's equation  (\ref{P_473}) gives that the dynamic contact
angle is fixed as a constant at its equilibrium value
$\theta_d=\theta_e$. If we wish to go further with the conventional
model approach and describe the dynamic contact angle as some
function of the contact-line speed, then Young's equation must be
discarded in favour of this function. Therefore, implementing the
interface formation model into a numerical code allows one to test
all conventional models of wetting proposed in the framework of
continuum mechanics. By applying the limit $\bar{\beta}/Ca ~(= \beta
L/\mu) \rightarrow \infty$  we recover the classical equations of
no-slip and impermeability on the solid surface for which the moving
contact-line problem is known to have no solution
\cite{huh71,dussan79}.

Despite the model's complexity, in limiting cases, analytic progress
on it can be made to obtain explicit relationships for the surface
variables and, with some further assumptions, even a formula
relating the contact-line speed to the dynamic contact angle.  Such
formulae are a useful test of our numerical solutions, and we will
briefly recapitulate their outcome.

\subsection{Asymptotic formulae in a limiting case}\label{asymptotics}

When the contact-line motion can be analyzed as a local problem,
as opposed to cases where the interface formation and the bulk
flow scales are not separated so that manipulating the global
flow influences the relaxation process along the interfaces,
asymptotic progress is possible.  A full derivation of the results
we use may be found in \cite{shik07} and references therein; here
we shall just outline the main assumptions and results.

If in the steady propagation of a liquid-gas free surface over a
solid substrate in the Stokes regime ($Re\ll1$), the
characteristic length scale of the interface formation process
$l=U\tau$ is small compared to the bulk length scale $L$, we have
that our non-dimensional parameter $\epsilon\ll1$. If in the limit
$\epsilon\rightarrow 0$ we also assume that the capillary number
$Ca\rightarrow 0$, then, to leading order in $Ca$, the normal-stress boundary condition (\ref{P_ifm_stressa}) gives that the
free surface near the contact line is planar, so that the problem
may be considered locally in a wedge-shaped domain. Then, we can
identify the following three asymptotic regions:
\begin{enumerate}
\item[(a)]
The outer region, where, in a reference frame moving with the
contact line, one has a flow in a wedge in the classical
formulation, with a zero tangential-stress and a no-slip boundary,
described in \cite{moffatt64};

\item[(b)]
The intermediate region with the characteristic (dimensionless) length scale
$\epsilon$, where the surface-tension-relaxation process takes
place and where, due to smallness of $Ca$, to leading order the
influence of the bulk flow on this process can be neglected;

\item[(c)]
The inner region, with the characteristic length scale
$\epsilon Ca$, through which the surface densities and
velocities, to leading order, stay constant.

\end{enumerate}

On the free surface, at leading order in the intermediate and inner regions, one has
\begin{equation}\label{val_surface_distns}
\rho^s_1 = \rho^s_{1e},\qquad v^s_{1||} = u_f(\theta_d) ,
\end{equation}
where $u_f(\theta_d)$ is the (dimensionless) radial velocity
of the bulk flow in the far field on the liquid-gas interface
given by \cite{moffatt64}:
\begin{equation}\label{val_moffatt_fs}
u_f(\theta_d) = \frac{\sin\theta_d-\theta_d\cos\theta_d}{\sin\theta_{d}\cos\theta_{d} - \theta_{d}},
\end{equation}
so that the surface mass flux into the contact line is
$-\rho^s_{1e}u_f(\theta_d)$. Then, since the surface variables are
constant through the inner region, the boundary conditions at the
contact line (\ref{P_473}), (\ref{P_470a}) can be applied to the
distributions of the surface variables in the intermediate region.
As a result, we have two first-order ODEs to solve for the
distributions of $\rho^s_2$ and $v^s_{2||}$ along the liquid-solid interface
\begin{equation}\label{two_odes}
\diff{\rho^s_2}{\bar{s}} = 4V^2(1-v^s_{2||}),\qquad \diff{(\rho^s_2 v^s_{2||})}{\bar{s}} = -(\rho^s_2-\rho^s_{2e}) \qquad (\bar{s}>0)
\end{equation}
where $\bar{s} = s/\epsilon$ is the intermediate region's variable
and
$V^2=\bar{\beta}\epsilon/((1+4\bar{\alpha}\bar{\beta})\lambda)$ is
the non-dimensional contact-line speed, subject to (a) boundary
condition (\ref{P_470a}), now taking the form $\rho^s_2 v^s_2 =
-\rho^s_{1e}u_f(\theta_d)$ at $\bar{s}=0$, (b) the matching
condition $\rho^s_2\to\rho^s_{2e}$ as $\bar{s}\to\infty$, and, as
boundary condition (a) implicitly depends on the parameter
$\theta_d$, (c) Young's equation (\ref{P_473}), now taking the
form $\cos\theta_d=\lambda(\rho^s_2-1)$. The equilibrium contact
angle $\theta_e$ is obviously related to $\rho^s_{2e}$ by
$\cos\theta_e=\lambda(\rho^s_{2e}-1)$, which can be used to
replace $\rho^s_{2e}$ with $\theta_e$.

The above problem is easily solved numerically; however, by taking
an additional assumption $\lambda \gg1$, one can obtain an analytic
relation between the contact angle and the non-dimensional speed
of the contact line $V$:
\begin{equation}\label{val_cangle}
\cos\theta_e - \cos\theta_d = \frac{2V\left[\cos\theta_e +
\left(1-\rho^s_{1e}\right)^{-1}\left(1+\rho^s_{1e}u_f(\theta_d)\right)\right]}{V+\left[V^2+1+\cos\theta_e\left(1-\rho^s_{1e}\right)\right]^{1/2}},
\end{equation}
where
\begin{equation}\nonumber
k=2V(\rho^s_{2e})^{-1}\left[\left(V^{2}+\rho^s_{2e}\right)^{1/2}-V\right],\qquad
C=\frac{2V\left(\rho^s_{2e}+\rho^s_{1e}u_f(\theta_d)\right)}{\left(V^{2}+\rho^s_{2e}\right)^{1/2}+V}.
\end{equation}

The presence of $u_f(\theta_d)$ in (\ref{val_cangle}) highlights a
connection between the flow in the outer asymptotic region and the
value of the dynamic contact angle. When the contact line is
`insulated' from the global flow by the low-Reynolds-number
region, the flow near the contact line is completely determined by
the contact-line speed, and hence the mass flux into the contact
line that `feeds' the liquid-solid interface can be found from the
local solution. This is the case considered in
\cite{shik93,shik97a}, where the theory shows excellent agreement
of (\ref{val_cangle}) with experiments. If the outer flow is to
influence the mass flux into the contact line \cite{lukyanov07},
this will affect the value of the contact angle.

The most important length scale $L_{if}$ associated with the
interface formation process is the characteristic distance over
which the solid surface returns to equilibrium and the asymptotic
result indicates that this is given, in non-dimensional units, by
$kL_{if}/\epsilon=1$ in (\ref{val_surface_distns}) so that $L_{if}
=\epsilon\rho^s_{2e}/\left[2V(\sqrt{V^2+\rho^s_{2e}}-V)\right]$.

The expressions given above will be used in \S\ref{results} below to
compare our computations to in the situations where the underlying
assumptions of the asymptotics are satisfied.  Now, we shall
consider the development of this computational algorithm for the
general case without making any simplifying assumptions.

\section{Finite element procedure}\label{FEM}

A finite element framework for the simulation of dynamic wetting flows using the conventional models of dynamic wetting was described in \cite{sprittles11c}.  To handle the evolution of the free surface this framework uses an arbitrary Lagrangian Eulerian (ALE) scheme based on the method of spines, a computational approach which has been successfully applied to a range of coating flows over the past thirty years, e.g. in \cite{kistler83,heil04,wilson06}. In \cite{sprittles_chem}, the framework was extended for the simulation of time-dependent free surface flows, with the code providing accurate solutions for the benchmark test case \cite{basaran92,meradji01} of a freely oscillating liquid drop. This confirmed that the implementation of the viscous, inertial and capillarity effects is accurate, even when the mesh undergoes $O(1)$ deformations.

What follows is the implementation of the interface formation equations into our framework. For a more detailed description of the basic components of the framework and a user-friendly step-by-step guide to implementing dynamic wetting flows into the finite element method, the reader is referred to \cite{sprittles11c} and in particular to the Appendix which makes it possible for the interested user to reproduce the results presented there. The Appendix in the present article provides additional details of the implementation  and, in this sense, complements the Appendix in \cite{sprittles11c}, allowing one to reproduce the results of \S\ref{results} and \S\ref{results2}.

\subsubsection{Problem formulation in the ALE scheme}\label{preliminaries}

Consider how the equations of \S\ref{equations}, written in Eulerian coordinates $\mathbf{x}$, are formulated for an ALE system where the flow domain $\greekbf{\chi}=\greekbf{\chi}\left(\mathbf{x},t\right)$ deforms in time.  This deformation must be accounted for in the temporal derivatives of variables whose position in the Eulerian system is evolutionary, in particular, in the Navier-Stokes equations where the material derivative $D/Dt$ transforms as
\begin{equation}\label{mat_deriv}
\frac{D\mathbf{u}}{Dt}= \left.\pdiff{\mathbf{u}}{t}\right|_{\mathbf{x}}+\mathbf{u}\cdot\nabla\mathbf{u} = \left.\pdiff{\mathbf{u}}{t}\right|_{\greekbf{\chi}}+\left(\mathbf{u}-\mathbf{c}\right)\cdot\nabla\mathbf{u}.
\end{equation}
Here,
$\mathbf{c}\left(\greekbf{\chi},t\right)=\left.\pdiff{\mathbf{x}}{t}\right|_{\greekbf{\chi}}$
is the velocity of the ALE coordinates with respect to the fixed
reference frame.  It can be seen that, as should be expected, for
$\mathbf{c} = \mathbf{u}$ we have a Lagrangian scheme whereas for
$\mathbf{c} = \mathbf{0}$ the Eulerian system is recovered.

Before considering temporal derivatives occurring in the interface
formation equations, it is convenient to introduce the surface
gradient $\nabla^s$, which is the projection of the usual gradient
operator $\nabla$ onto the surface $\nabla^s =
\left(\mathbf{I}-\mathbf{n}\mathbf{n}\right)\cdot\nabla$. An
arbitrary surface vector $\mathbf{a}^s$ is written in terms of
components normal and tangential to the surface as $\mathbf{a}^s =
\mathbf{a}^s_{||}+a^s_n\mathbf{n}$, where $a^s_n =
\mathbf{a}^s\cdot\mathbf{n}$, so that for its divergence one has
$\nabla\cdot\mathbf{a}^s = \nabla^s\cdot\mathbf{a}^s_{||}+a^s_n
\nabla^s\cdot\mathbf{n}$.  In particular, as described in
\cite{pozrikidis04}, points on the surfaces move with the normal
surface velocity $v^s_n$, i.e.\ according to the kinematic
equation (\ref{P_454}), and an arbitrary tangential component
$\mathbf{c}^s_{||}$ which depends on the choice of mesh design.
Then, on a given surface
\begin{equation}\nonumber
\mathbf{c}^s = \mathbf{c}^s_{||} + v^s_n\mathbf{n},\qquad  \mathbf{c}^s_{||} = \pdiff{\mathbf{x}}{t}\cdot\left(\mathbf{I}-\mathbf{n}\mathbf{n}\right).
\end{equation}
Therefore, in the ALE framework the left-hand side of the surface mass balance equations (\ref{P_458}) and (\ref{P_464}), become
\begin{equation}\nonumber
\left.\pdiff{\rho^s_\gamma}{t}\right|_{\mathbf{x}_\gamma}+\nabla\cdot\left(\rho^s_\gamma\mathbf{v}^s_\gamma\right) = \left.\pdiff{\rho^s_\gamma}{t}\right|_{\greekbf{\chi}^s_\gamma}-\mathbf{c}^s_\gamma\cdot\nabla\rho^s_\gamma+\nabla\cdot\left(\rho^s_\gamma\mathbf{v}^s_\gamma\right),
\end{equation}
where $\gamma=1,2$ refers to the liquid-gas and liquid-solid
interface, respectively, and
$\greekbf{\chi}^s_\gamma=\greekbf{\chi}^s_\gamma\left(\mathbf{x},t\right)$
are the corresponding coordinates.  Then, (\ref{P_458}) and
(\ref{P_464}) take the form
\begin{equation}\label{P_458lhs}
 \epsilon\left[\pdiff{\rho^s_\gamma}{t}-\mathbf{c}^s_{\gamma ||}
 \cdot\nabla^s\rho^s_\gamma+\nabla^s\cdot
 \left(\rho^s_\gamma\mathbf{v}^s_{\gamma||}\right)
 +\rho^s_\gamma v^s_{\gamma n}\nabla^s\cdot\mathbf{n}_\gamma\right]
 +\rho^s_\gamma-\rho^s_{\gamma e} = 0,\quad (\gamma=1,2),
\end{equation}
where we have used that $\mathbf{n}_\gamma\cdot\nabla^s\rho^s_\gamma=0$.  Equations (\ref{P_458lhs}) can be rearranged to obtain
\begin{equation}\nonumber
\epsilon\left\{  \pdiff{\rho^s_\gamma}{t}+\rho^s_\gamma\nabla^s\cdot\mathbf{c}^s_{\gamma ||}+\nabla^s\cdot\left[\rho^s_\gamma\left(\mathbf{v}^s_{\gamma||}-\mathbf{c}^s_{\gamma||}\right)\right] +\rho^s_\gamma v^s_{\gamma n}\nabla^s\cdot\mathbf{n}_\gamma \right\} + \rho^s_\gamma-\rho^s_{\gamma e} = 0,\quad (\gamma=1,2).
\end{equation}

In the limiting case, where the surface moves only normal to
itself, so that $\mathbf{c}^s_{\gamma ||}=\mathbf{0}$, the usual
Eulerian equations are recovered:
$$
 \epsilon\left[\pdiff{\rho^s_\gamma}{t}+\nabla^s\cdot
 \left(\rho^s_\gamma\mathbf{v}^s_{\gamma||}\right)
 +\rho^s_\gamma v^s_{\gamma n}\nabla^s\cdot\mathbf{n}_\gamma\right]
 +\rho^s_\gamma-\rho^s_{\gamma e} =
$$
\begin{equation}\nonumber
 \epsilon\left[\pdiff{\rho^s_\gamma}{t}+\nabla\cdot
 \left(\rho^s_\gamma\mathbf{v}^s_\gamma\right)\right]
 +\rho^s_\gamma-\rho^s_{\gamma e}= 0,\quad (\gamma=1,2),
\end{equation}
whilst if the surface moves in a Lagrangian way, $\mathbf{c}^s_\gamma=\mathbf{v}^s_\gamma$, then the term under the divergence becomes identically zero, and we have
\begin{align}\nonumber
\epsilon\left(\pdiff{\rho^s_\gamma}{t}+\rho^s_\gamma\nabla^s\cdot\mathbf{v}^s_{\gamma||} + \rho^s_\gamma v^s_{\gamma n}\nabla^s\cdot\mathbf{n}_\gamma\right)+\rho^s_\gamma-\rho^s_{\gamma e}= \\ \epsilon\left(\pdiff{\rho^s_\gamma}{t}+\rho^s_\gamma\nabla\cdot\mathbf{v}^s_\gamma\right) +\rho^s_\gamma-\rho^s_{\gamma e} = 0,\quad (\gamma=1,2).
\end{align}
Here, as should be expected, there is no term representing the convection of surface density by the surface velocity, i.e.\ a term of the form $\mathbf{v}^s\cdot\nabla\rho^s$ does not appear.

Having reformulated the equations for the ALE system and introduced surface operators, we can now derive the appropriate FEM residuals.

\subsubsection{Forming the finite element residuals}

The defining feature of the FEM is that the computational domain $V$ is tessellated into a finite number of non-overlapping elements, each containing a fixed number of nodes at which the functions' values are determined.  Between these nodes the functions are approximated using interpolating functions whose functional dependence on position is chosen. In what follows, $N_p$ is the total number of nodes in $V$ at which the pressure is determined, $N_u$ is the number of nodes at which the velocity components are to be found, $N_1$ is the number of nodes on the free surface $A_1$, $N_2$ the number of nodes on the solid surface $A_2$, and $N_c$ the number of nodes along the contact line.

The procedure of generating the finite element equations is well known and a detailed explanation of how this is achieved for dynamic wetting flows described by the conventional model is given in \cite{sprittles11c}, so that here we just give the main details. Functions are approximated as a linear combination of interpolating functions each weighted by the corresponding nodal value. In particular, we use mixed interpolation so that the Ladyzhenskaya-Babu\u{s}ka-Brezzi \cite{babuska72} condition is satisfied\footnote{Equal-order methods may also be used with stabilization, e.g.\ \cite{hughes86,codina00}, but these issues lie beyond the scope of this paper.} with linear basis functions $\psi_{\textrm{j}}$ to represent pressure and quadratic ones $\phi_{\textrm{j}}$ for velocity:
\begin{equation}\nonumber
p              = \sum^{N_p}_{\textrm{j}=1} p_{\textrm{j}}\psi_{\textrm{j}}(\mathbf{x}),       \qquad
\mathbf{u}     = \sum^{N_u}_{\textrm{j}=1}\mathbf{u}_{\textrm{j}}\phi_{\textrm{j}}(\mathbf{x}), \qquad (\mathbf{x}\in V).
\end{equation}

In the Galerkin finite element method, the basis functions which are used to discretize the functions of the problem are also used as weighting functions to create the weak form of the problem, see \cite[\S 3~of][]{sprittles11c} for specific details.  Note that here we use Roman letters for the indices ($\textrm{i},\textrm{j}$, etc) to refer to the nodal values and approximating functions spanning the whole domain (globally); in the Appendix, where all the numerical details are given, these indices will be used alongside italicized ones ($i,j$, etc), which will refer to local, element-based, values and functions.

Surface variables are also approximated quadratically with basis functions on surface $A_\gamma~(\gamma=1,2)$ denoted by $\phi_{\gamma,{\textrm{j}}}(s_1,s_2)$. So, all of the interface formation variables, represented by an arbitrary surface variable $a^s_\gamma$, as well as the shape of the free surface $\mathbf{x}_1$, are approximated as
\begin{equation}\nonumber
a^{s}_\gamma = \sum^{N_\gamma}_{\textrm{j}=1}a^s_{\gamma,{\textrm{j}}} \phi_{\gamma,{\textrm{j}}}(s_1,s_2), \qquad
\mathbf{x}_1 = \sum^{N_1}_{\textrm{j}=1}\mathbf{x}_{1,{\textrm{j}}} \phi_{1,{\textrm{j}}}(s_1,s_2).
\end{equation}

To determine the free surface shape, i.e.\ the nodal values $\mathbf{x}_{1,{\textrm{j}}}$, a function $h = h(s_1,s_2)$ is found  as part of the solution at free-surface every node, so that $h_j = h_j(s_1,s_2)$ for $\textrm{j}=1,...,N_1$, which points in a direction linearly independent from both $s_1$ and $s_2$ at each node, i.e.\ `out' of the free surface.  For example, in the simplest case of a Cartesian coordinate system one could have the free surface at  $(x,y,z)=(s_1,s_2,h(s_1,s_2))$, so that $h$ is the height above the $(x,y)$-plane, or, in a two-dimensional example, one may have a polar coordinate system with the free surface given by $(\theta,r)=(s_1,h(s_1))$, in which case $h$ is the distance of the free surface from the origin for every angle $\theta$.

The basis functions used to approximated the variables are now used to derive the weak form of the problem, i.e. the finite element equations. From (\ref{ns}), the continuity of mass (incompressibility of the fluid) residuals $R^{C}_{\textrm{i}}$ are
\begin{equation}\label{incom}
R^{C}_{\textrm{i}} = \int_{V} \psi_{\textrm{i}}\nabla\cdot\mathbf{u}~dV\qquad (\textrm{i}=1,...,N_p).
\end{equation}

After projecting the momentum equations (\ref{ns}) onto the unit
basis vectors of the coordinate system
$\mathbf{e}_{\alpha}~(\alpha=1,2,3)$ and using (\ref{mat_deriv}),
our momentum residuals $R^{M,\alpha}_{\textrm{i}}$ take the form
\begin{equation}\label{ns1}
R^{M,\alpha}_{\textrm{i}} = \int_{V} \phi_{\textrm{i}}\mathbf{e}_{\alpha} \cdot \left[Re~\left(\pdiff{\mathbf{u}}{t} + \left(\mathbf{u}-\mathbf{c}\right)\cdot\nabla\mathbf{u}\right) - \nabla\cdot\mathbf{P} -
St~\mathbf{F}\right]~dV \qquad (\textrm{i}=1,...,N_u).
\end{equation}
Integrating by parts and using the divergence theorem, as shown in
\cite{sprittles11c}, one can rewrite (\ref{ns1}) in terms of
volume and surface contributions:
\begin{align}\nonumber
R^{M,\alpha}_{\textrm{i}} &= \left(R^{M,\alpha}_{\textrm{i}}\right)_V + \left(R^{M,\alpha}_{\textrm{i}}\right)_A  \qquad (\textrm{i}=1,...,N_u),\\ \nonumber
\left(R^{M,\alpha}_{\textrm{i}}\right)_V &= \int_{V}\left\{\phi_{\textrm{i}}\mathbf{e}_\alpha\cdot\left[Re~\left(\pdiff{\mathbf{u}}{t} + \left(\mathbf{u}-\mathbf{c}\right)\cdot\nabla\mathbf{u}\right) - St~\mathbf{F}\right]+
\nabla(\phi_{\textrm{i}}\mathbf{e}_\alpha):\mathbf{P}\right\}~dV, \\ \label{vol3}
\left(R^{M,\alpha}_{\textrm{i}}\right)_A &= \int_{A} \phi_{\textrm{i}}\,\mathbf{e}_\alpha\cdot\mathbf{P}\cdot\mathbf{n}~dA.
\end{align}
In (\ref{vol3}), only when node $\textrm{i}$ is on the surface $A$ will $\phi_{\textrm{i}}$ be non-zero, i.e.\ it is nodes on the surface of $V$ which contribute to the momentum residuals via (\ref{vol3}). This term allows us to incorporate stress boundary conditions naturally, i.e.\ by adding them as contributions to the momentum equations at the surfaces, which is a well known advantage of the finite element method over other numerical approaches where special boundary approximations have to be constructed.

To incorporate our free-surface stress boundary conditions into
(\ref{vol3}), equations (\ref{P_ifm_stress}) and
(\ref{P_ifm_stressa}) are rewritten into the computationally
favourable form
\begin{equation}\nonumber
Ca~\left(\mathbf{n}_1 p_g + \mathbf{n}_1\cdot\mathbf{P}\right)
+\nabla^s\cdot\left[\sigma_{1}\left(\mathbf{I}-\mathbf{n}_1\mathbf{n}_1\right)\right]=\mathbf{0},
\end{equation}
where $\sigma_{1}\left(\mathbf{I}-\mathbf{n}_1\mathbf{n}_1\right)$ is the surface stress, playing the same role on the surface as $\mathbf{P}$ does in the bulk. Then (\ref{vol3}) can be rewritten on the free surface as
\begin{equation}\nonumber
\int_{A_1} \phi_{\textrm{1,i}}\mathbf{e}_\alpha\cdot\mathbf{P}\cdot\mathbf{n}_1~dA_1 = -\frac{1}{Ca}\int_{{A}_1}
\phi_{1,\textrm{i}} \mathbf{e}_\alpha\cdot\left\{ \nabla^s\cdot\left[\sigma_{1}\left(\mathbf{I}-\mathbf{n}_1\mathbf{n}_1\right)\right] + Ca p_g \mathbf{n}_1\right\}~dA_1.
\end{equation}
It was initially suggested by Ruschak \cite{ruschak80}, and
generalized for three-dimensional problems in
\cite{ho91,cairncross00}, that, by using the surface divergence
theorem, one could lower the highest derivatives and thus both
reduce the constraints on the differentiability of the
interpolating functions which are used to approximate the free
surface shape and give a natural way to impose boundary conditions
on the shape of the surface where it meets other boundaries. This
is achieved by, first, using the chain rule:
\begin{equation}\nonumber
\int_{A_1} \phi_{\textrm{1,i}}\mathbf{e}_\alpha\cdot\mathbf{P}\cdot\mathbf{n}_1~dA_1 = -\frac{1}{Ca}\int_{{A}_1}\left\{
\nabla^s\cdot\left[\phi_{1,\textrm{i}}\sigma_1\mathbf{e}_\alpha\cdot\left(\mathbf{I}-\mathbf{n}_1\mathbf{n}_1\right)\right]-
\sigma_1\nabla^s\cdot\left(\phi_{1,\textrm{i}}\mathbf{e}_\alpha\right) + Ca p_g \phi_{1,\textrm{i}}(\mathbf{e}_\alpha\cdot\mathbf{n}_1) \right\}~dA_1,
\end{equation}
and then using the surface divergence theorem
\cite[p.~224]{aris62}, which for a surface vector $\mathbf{a}^s$
with no normal component, $\mathbf{a}^s=\mathbf{a}^s_{||}$, is
given by
\begin{equation}\label{s_div}
\int_{A}\nabla^s\cdot\mathbf{a}^s_{||}~dA = -\int_{C}\mathbf{a}^s_{||}\cdot\mathbf{m}~dC,
\end{equation}
where the unit vector $\mathbf{m}$ is the inwardly facing normal to the contour $C$ that confines $A$ (Figure~\ref{F:angles}), with $\mathbf{a}^s_{||}=\phi_{1,{\textrm{i}}}\sigma_1\mathbf{e}_\alpha\cdot\left(\mathbf{I}-\mathbf{n}_1\mathbf{n}_1\right)$, so that
\begin{equation}\nonumber
\int_{A_1} \phi_{\textrm{1,i}}\mathbf{e}_\alpha\cdot\mathbf{P}\cdot\mathbf{n}_1~dA_1 =
\frac{1}{Ca}\int_{{A}_1}\left[\sigma_1\nabla^s\cdot\left(\phi_{1,\textrm{i}}\mathbf{e}_\alpha\right)-Ca p_g \phi_{1,\textrm{i}}(\mathbf{e}_\alpha\cdot\mathbf{n}_1)\right]~dA_1 +
\frac{1}{Ca}\int_{C_1}\phi_{1,\textrm{i}}\sigma_1\mathbf{e}_\alpha\cdot\mathbf{m}_{1}~dC_1.
\end{equation}
Thus, on the free surface, the term (\ref{vol3}) in the momentum residual is now replaced by a different surface contribution and a line contribution
\begin{equation}\label{fs5}
\left(R^{M,\alpha}_{\textrm{i}}\right)_{A_{1}} =
\frac{1}{Ca}\int_{{A}_1}\left[\sigma_1\nabla^s\cdot\left(\phi_{1,\textrm{i}}\mathbf{e}_\alpha\right)-Ca p_g \phi_{1,\textrm{i}}(\mathbf{e}_\alpha\cdot\mathbf{n}_1)\right]~dA_1, \qquad
\left(R^{M,\alpha}_{\textrm{i}}\right)_{C_{1}} = \frac{1}{Ca}\int_{C_1}\phi_{1,\textrm{i}}\sigma_1\mathbf{e}_\alpha\cdot\mathbf{m}_{1}~dC_1.
\end{equation}

The same procedure of integrating by parts and using the
divergence theorem has been used on both the surface stress term
$\nabla^s\cdot\left[\sigma_{1}\left(\mathbf{I}-\mathbf{n}_1\mathbf{n}_1\right)\right]$
and the bulk stress term $\nabla\cdot\mathbf{P}$.  In both cases,
this has created contributions from the boundary of that term's
domain, i.e.\ the confining contour and surface, respectively.
Consequently, the momentum residual now contains a cascade of
scales
\begin{equation}\label{casc}
 R^{M,\alpha}_{\textrm{i}}=\left(R^{M,\alpha}_{\textrm{i}}\right)_{V}+\left(R^{M,\alpha}_{\textrm{i}}\right)_{A}+\left(R^{M,\alpha}_{\textrm{i}}\right)_{C},
\end{equation}
which represent, respectively, the volume, surface and line contributions. In particular, part of the contour $C_1$ which bounds the free surface is the contact line $C_{cl}$ where the free surface meets the solid. Other boundaries to the free surface further away from the contact line, for example an axis or plane of symmetry, are treated in the same way but are not to be formulated until specific problems are considered in \S\ref{results}.

At the contact line, it is useful to rearrange the term in the integrand of the contour integral in (\ref{fs5}) by representing the vector $\mathbf{m}_1$ in terms of a linear combination of its components parallel to $\mathbf{n}_2$ and $\mathbf{m}_2$ (Figure~\ref{F:angles}):
\begin{equation}\nonumber
\mathbf{m}_1  =
\left(\mathbf{m}_{1}\cdot\mathbf{m}_2\right)\mathbf{m}_2 +
\left(\mathbf{m}_{1}\cdot\mathbf{n}_2\right)\mathbf{n}_2.
\end{equation}
This identity can be used to make the contribution to (\ref{casc})
coming from the contact line dependent on the known vector
$\mathbf{n}_2$, the vector $\mathbf{m}_2$, varying along the
contact line but independent of the free-surface shape, and
$\theta_d$, defined by (\ref{angle}) and determined by Young's
equation (\ref{P_473}), so that
\begin{equation}\nonumber
 \left(R^{M,\alpha}_{\textrm{i}}\right)_{C_{cl}} =
\frac{1}{Ca}\int_{C_{cl}}\sigma_1\phi_{1,\textrm{i}}\mathbf{e}_\alpha\cdot\left(\mathbf{m}_2\cos\theta_d +
\mathbf{n}_2\sin\theta_d\right)~dC_{cl} \qquad
(\textrm{i}=1,...,N_c).
\end{equation}
If the contact line's tangent vector is $\mathbf{t}_c$, then
$\mathbf{m}_2=\pm\mathbf{t}_c\times\mathbf{n}_2$, with the sign
chosen to ensure the inward facing vector $\mathbf{m}_2$ is
selected. Thus, equation (\ref{angle}), which defines the contact
angle, can be applied in a \emph{natural} way, that is without
needing to drop another equation in order to make room for an
equation that would fix the shape of the free surface at the
contact line. The contact angle itself is determined from Young's
equation (\ref{P_473}), which, when put in residual form as an
integral over the contact line contour, gives
\begin{equation}\nonumber
R^{Y}_{\textrm{i}} =\int_{C_{cl}}\phi_{\textrm{1,i}}\left(\sigma_1\cos\theta_d + \sigma_2\right)~dC_{cl} \qquad (\textrm{i}=1,...,N_c).
\end{equation}

At the liquid-solid interface the approach developed in \cite{sprittles11c} is used. Instead of dropping the momentum equation normal to the solid to impose a Dirichlet condition on the normal velocity (\ref{P_465}), we use it to determine the normal stress acting on the liquid-solid interface which allows the contact line, where boundary conditions of different type meet, to be treated in a manner consistent with standard FEM ideology.  Specifically, we introduce a new unknown $\Lambda$ \footnote{Labelled $\lambda$ in \cite{sprittles11c}.} on the liquid-solid interface which is defined by the equation
\begin{equation}\label{lambda}
\Lambda=\mathbf{n}_2\cdot\mathbf{P}\cdot\mathbf{n}_2.
\end{equation}
It should be pointed out that this particular implementation simplifies the finite element procedure independently of the dynamic wetting model chosen and is particularly useful when considering a surface non-aligned with a coordinate axis. In this case, the procedure of rotating the momentum equations to align with the coordinate axes \cite{engelman82} is cumbersome whereas our approach is independent of both the free surface and the solid's shape, that is the curved nature of a surface is as easy to handle as, say, a planar surface aligned with coordinate axes.

On the liquid-solid interface the contribution to the momentum equations from the stress on the surface, which contains contributions from both the generalized Navier condition (\ref{P_463}) for the tangential stress and (\ref{lambda}) for the normal stress gives
\begin{equation}\label{ns555}
\left(R^{M,\alpha}_{\textrm{i}}\right)_{A_{2}} =\int_{A_2}
\phi_{2,\textrm{i}}\left[\Lambda~(\mathbf{e}_\alpha\cdot\mathbf{n}_2) +
\frac{\bar{\beta}}{Ca}~\mathbf{e}_\alpha\cdot\left(\mathbf{u}_{||}-\mathbf{U}_{||}\right) - \frac{1}{2Ca}\mathbf{e}_\alpha\cdot\nabla^s\sigma_2   \right]~dA_2 \qquad (\textrm{i}=1,...,N_2).
\end{equation}
where $\phi_{2,\textrm{i}}$ is an interpolating function for the liquid-solid interface corresponding to the $\textrm{i}$--th node.

In addition to the boundary conditions involving stress on each interface, we have an additional equation involving the velocity normal to each interface.  On the liquid-gas free surface this is the kinematic condition (\ref{P_454}) whose residuals $R^{K}_{\textrm{i}}$ are given by
\begin{equation}\label{w_kin}
R^{K}_{\textrm{i}} = \int_{A_1}\phi_{1,\textrm{i}}\left[\left(\pdiff{\mathbf{x}_1}{t}-\mathbf{v}^s_{1}\right)\cdot\mathbf{n}_1\right]~dA_1\qquad (\textrm{i}=1,...,N_1),
\end{equation}
whilst on the liquid-solid interface we have a condition of impermeability of the solid (\ref{P_462}) with residuals $R^{I}_{\textrm{i}}$ given by
\begin{equation}\label{w_imp}
R^{I}_{\textrm{i}} = \int_{A_2}\phi_{2,\textrm{i}}\left(\mathbf{v}^s_2-\mathbf{U}\right)\cdot\mathbf{n}_2~dA_2\qquad (\textrm{i}=1,...,N_2).
\end{equation}

In keeping with the framework presented in \cite{sprittles11c},
all momentum equations are applied at both the liquid-gas free
surface and at the liquid-solid interface and hence, once the two
boundaries meet at the contact line, the contact line conditions
can be implemented naturally, without dropping any of the
equations there. A crude, but useful, interpretation is to think
of the momentum equations as determining the bulk velocities, the
kinematic condition on the free surface as determining the
unknowns that specify its \emph{position}, and the impermeability
condition, which is the geometric constraint of the prescribed
shape, as determining the `extra' unknown $\Lambda$, i.e.\ the
\emph{normal stress}.

Thus far, the equations are assumed to determine the bulk velocity, the shape of the free surface and the normal
stress on the liquid-solid interface ($\Lambda$).  In addition to
these equations we have (\ref{P_459})--(\ref{P_460}) on the free
surface and (\ref{P_466})--(\ref{P_468}) on the liquid-solid
interface which determine the surface velocity $\mathbf{v}^s$,
surface density $\rho^s$ and surface tension $\sigma$. In
particular, we can think of the Darcy-type equations (\ref{P_459})
and (\ref{P_466}) as determining the surface velocity tangential
to the surface $\mathbf{v}^s_{||}$, which in a fully
three-dimensional flow will have two components in the
$\mathbf{e}_{s_1}$ and $\mathbf{e}_{s_2}$ direction, where
$\mathbf{e}_{s_\nu}$ is a basis vector in the direction of
increasing $s_\nu,~\nu=1,2$.   The residuals
$R^{v^s_{1||},\nu}_{\textrm{i}}$ and
$R^{v^s_{2||},\nu}_{\textrm{i}}$ are
\begin{equation}\label{w_vt1}
R^{v^s_{1||},\nu}_{\textrm{i}} = \int_{A_1}\phi_{1,\textrm{i}}\mathbf{e}_{s_\nu} \cdot\left(\mathbf{v}^{s}_{1||}-\mathbf{u}_{||}-\frac{1+4\bar{\alpha}\bar{\beta}}{4\bar{\beta}}\nabla^s\sigma_1\right)~dA_1\qquad (\textrm{i}=1,...,N_1),
\end{equation}
\begin{equation}\label{w_vt2}
R^{v^s_{2||},\nu}_{\textrm{i}}  = \int_{A_2}\phi_{2,\textrm{i}}\mathbf{e}_{s_\nu}  \cdot\left(\mathbf{v}^{s}_{2||}-\hbox{$\frac{1}{2}$}\left(\mathbf{u}_{||}+\mathbf{U}_{||}\right)-\bar{\alpha}\nabla^s\sigma_{2}\right)~dA_2\qquad (\textrm{i}=1,...,N_2).
\end{equation}

Equations (\ref{P_457}) and (\ref{P_465}) can be thought of as determining the normal component of the surface velocity on the surface $A_\gamma$ ($\gamma=1,2$) and the residuals $R^{v^s_{\gamma n}}_{\textrm{i}}$ from these equations take the same form on each interface:
\begin{equation}\label{w_vn}
R^{v^s_{\gamma n}}_{\textrm{i}} = \int_{A_\gamma}\phi_{\gamma,\textrm{i}}\left[\left(\mathbf{u}-\mathbf{v}^s_\gamma\right)\cdot\mathbf{n}_\gamma-Q\left(\rho^s_\gamma-\rho^s_{\gamma e}\right)\right]~dA_\gamma\qquad (\textrm{i}=1,...,N_\gamma;\quad\gamma=1,2).
\end{equation}

The corresponding residuals $R^{\rho^s_{\gamma}}_{\textrm{i}}$ from equations (\ref{P_458}) and (\ref{P_464}) to determine the evolution of the surface density are
\begin{align}\nonumber
R^{\rho^s_\gamma}_{\textrm{i}} =& \int_{A_\gamma}\phi_{\gamma,\textrm{i}}\left\{\epsilon\left(\pdiff{\rho^{s}_{\gamma}}{t} + \nabla^s\cdot\left[\rho^s_{\gamma}\left(\mathbf{v}^{s}_{\gamma ||} - \mathbf{c}^s_{\gamma||}\right)\right]\right. \right.  \\ \label{w_r5}
&\left.\left. +\rho^s_{\gamma}\nabla^s\cdot \mathbf{c}^s_{\gamma ||} + \rho^s_\gamma (\mathbf{v}^s_{\gamma}\cdot\mathbf{n}_\gamma)\nabla^s\cdot \mathbf{n}_\gamma\right) +\rho^{s}_{\gamma}-\rho^{s}_{\gamma e}\right\}~dA_\gamma\qquad (\textrm{i}=1,...,N_\gamma;\quad\gamma=1,2).
\end{align}
Using the standard FEM ideology, which will give us a method for applying boundary conditions on the surface flux $\rho^s\mathbf{v}^s_{||}$ where the surface meets a boundary, we integrate the divergence term in (\ref{w_r5}) by parts to obtain
\begin{align}\notag
R^{\rho^s_\gamma}_{\textrm{i}} = & \int_{A_\gamma}\left\{ - \epsilon\rho^{s}_{\gamma}\left(\mathbf{v}^{s}_{\gamma||}-\mathbf{c}^s_{\gamma||}\right)\cdot\nabla^s\phi_{\gamma,\textrm{i}}+
\phi_{\gamma,\textrm{i}}\left[\epsilon\left(\pdiff{\rho^{s}_{\gamma}}{t} + \rho^s_\gamma\nabla^s\cdot \mathbf{c}^s_{\gamma||}+ \rho^s_\gamma(\mathbf{v}^s_{\gamma}\cdot\mathbf{n}_\gamma)\nabla^s\cdot \mathbf{n}_\gamma \right)
+\rho^{s}_{\gamma}-\rho^{s}_{\gamma e}\right]\right\} ~dA_\gamma + \\
\label{w_r} & \int_{A_\gamma}\epsilon\nabla^s\cdot\left[\phi_{\gamma,\textrm{i}}\rho^s_\gamma \left(\mathbf{v}^{s}_{\gamma||}-\mathbf{c}^s_{\gamma||}\right)\right] ~dA_\gamma \qquad (\textrm{i}=1,...,N_\gamma;\quad\gamma=1,2).
\end{align}
Then, using the surface divergence theorem (\ref{s_div}) with $\mathbf{a}^s_{||}=\phi_{1,{\textrm{i}}}\rho^s\left(\mathbf{v}^s_{||}-\mathbf{c}^s_{||}\right)$ we have a contribution from both the surface and the contour bounding that surface, so that
\begin{align}\label{w_rr1}
R^{\rho^s_\gamma}_{\textrm{i}} &= \left(R^{\rho^s_\gamma}_{\textrm{i}}\right)_{A_\gamma} + \left(R^{\rho^s_\gamma}_{\textrm{i}}\right)_{C_\gamma}, \\ \label{w_rr2}
\left(R^{\rho^s_\gamma}_{\textrm{i}}\right)_{A_\gamma} &= -\int_{A_\gamma} \phi_{\gamma,\textrm{i}}\left\{\epsilon\left[\pdiff{\rho^{s}_{\gamma}}{t} + \rho^s_\gamma\nabla^s\cdot \mathbf{c}^s_{\gamma||}+ \rho^s_\gamma v^s_{\gamma n}\nabla^s\cdot \mathbf{n}_\gamma + \rho^s_\gamma\left(\mathbf{v}^s_{\gamma}\cdot\mathbf{n}_\gamma\right)\nabla^s\cdot \mathbf{n}_\gamma\right]
+\rho^{s}_{\gamma}-\rho^{s}_{\gamma e}\right\} ~dA_\gamma, \\ \label{w_rr3}
\left(R^{\rho^s_\gamma}_{\textrm{i}}\right)_{C_\gamma} &=-\int_{C_\gamma}\epsilon\phi_{\gamma,\textrm{i}}\rho^s_\gamma\left(\mathbf{v}^s_{\gamma||}-\mathbf{c}^s_{\gamma||}\right)\cdot\mathbf{m}_\gamma~dC_\gamma \qquad (\textrm{i}=1,...,N_\gamma;\quad\gamma=1,2).
\end{align}
Consequently, we are able to apply any boundary conditions on
surface mass flux by replacing the contour contribution
(\ref{w_rr3}) with the appropriate value.  In particular, at the
contact line contour $C_{cl}$ we have the surface mass flux
continuity condition (\ref{P_470a}). This condition could
potentially be applied by using it to replace either the contour
contribution to the free surface or the liquid-solid interface
equations, and to determine which way is correct we study the
structure of the interface formation equations. Specifically, an
asymptotic approach to the dynamic wetting problem for small $Ca$
and $\epsilon$ recapitulated in \S\ref{asymptotics} shows that it
is the flux of surface mass into the contact line from the free
surface, which depends on the global flow via the velocity in the
far field on the free surface $u_f(\theta_d)$, that determines the
relaxation process on the solid surface. Therefore, in a numerical
implementation of this problem one should allow the free surface
equations to determine the flux into the contact line and then use
the surface mass continuity condition (\ref{P_470a}) on the
liquid-solid side of the contact line to take this flux as the
surface mass supply into the liquid-solid interface.

Therefore, on the free surface we have
\begin{equation}\label{w_r}
\left(R^{\rho^s_1}_{\textrm{i}}\right)_{C_{cl}} =-\int_{C_{cl}}\epsilon\phi_{1,\textrm{i}}\rho^s_1\left(\mathbf{v}^s_{1||}-\mathbf{c}^s_{1||}\right)\cdot\mathbf{m}_1~dC_{cl},
\end{equation}
whilst on the liquid-solid interface,  using (\ref{P_470a}) to rewrite the flux $\rho^s_2\left(\mathbf{v}^s_{2||}-\mathbf{c}^s_{2||}\right)\cdot\mathbf{m}_2$ into this interface in terms of the flux into the contact line from the free surface, we have
\begin{equation}\label{w_r}
\left(R^{\rho^s_2}_{\textrm{i}}\right)_{C_{cl}} = \int_{C_{cl}}\epsilon\phi_{2,\textrm{i}}\rho^s_1\left(\mathbf{v}^s_{1||}-\mathbf{c}^s_{1||}\right)\cdot\mathbf{m}_1~dC_{cl}.
\end{equation}
Then, the flux into the liquid-solid interface is given in terms of the flux that goes into the contact line from the free surface.

The final residuals from the surface equations, $R^{\sigma_\gamma}_{\textrm{i}}$, are obtained from the surface equation of state (\ref{P_460}) and (\ref{P_468}), which express the surface tension as an algebraic function of the surface density:
\begin{equation}\label{w_imp}
R^{\sigma_\gamma}_{\textrm{i}} = \int_{A_\gamma}\phi_{\gamma,\textrm{i}}\left(\sigma_\gamma - \lambda\left(1-\rho^s_\gamma\right)\right)~dA_\gamma\qquad (\textrm{i}=1,...,N_\gamma;\quad \gamma=1,2).
\end{equation}

In this section, we have derived the finite element residuals required to solve dynamic wetting flows using the interface formation model. In particular we have the following coupling of unknowns
\begin{equation}
\left(\mathbf{u}_{\textrm{i}}\cdot\mathbf{e}_\alpha, p_{\textrm{i}}, h_{\textrm{i}}, \Lambda_{\textrm{i}}, \mathbf{v}^s_{\gamma ||,\textrm{i}}\cdot\mathbf{e}_\nu, v^s_{\gamma n,\textrm{i}},\rho^s_{\gamma,\textrm{i}}, \sigma_{\gamma,\textrm{i}}, \theta_d \right) \qquad \alpha=1,2,3\quad \gamma=1,2 \quad \nu=1,2,
\end{equation}
each of which has its corresponding residual
\begin{equation}
\left(R^{M,\alpha}_{\textrm{i}}, R^{C}_{\textrm{i}},R^{K}_{\textrm{i}}, R^{I}_{\textrm{i}}, R^{v^s_{\gamma ||,\nu}}_{\textrm{i}},R^{v^s_{\gamma n}}_{\textrm{i}},R^{\rho^s_\gamma}_{\textrm{i}}, R^{\sigma_\gamma}_{\textrm{i}}, R^{Y}_\textrm{i} \right) \qquad \alpha=1,2,3\quad \gamma=1,2 \quad \nu=1,2.
\end{equation}
The subscripts $\textrm{i}$ above will have different limits that are the same in both the variable and its corresponding residual, i.e.\ the same number of equations and unknowns has been assured.

\section{Validation of the code and some benchmark calculations for Problem A}\label{results}

\begin{figure}
\centering
\includegraphics[scale=0.60]{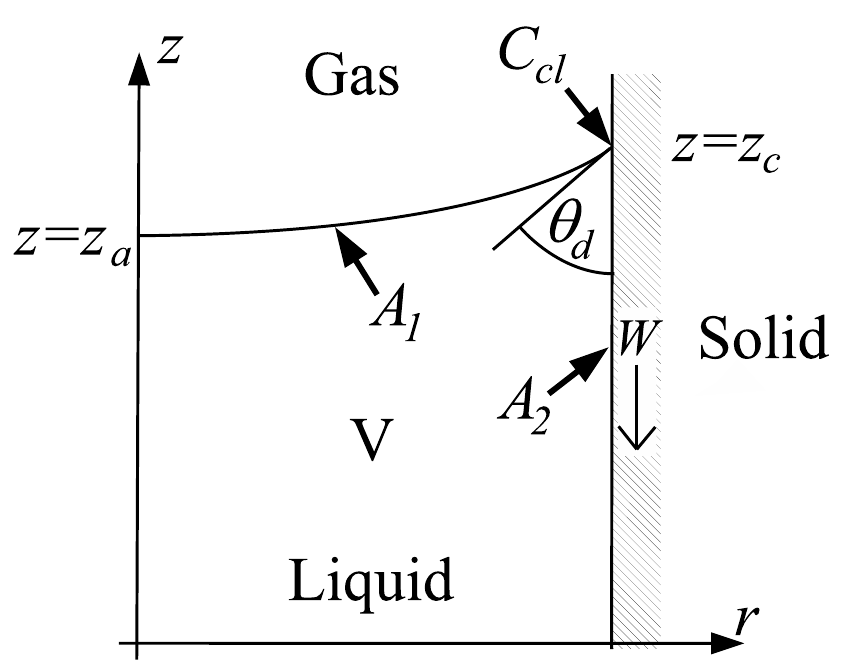}
\caption{Illustration showing the computational domain for flow through a capillary with the bulk domain $V$, liquid-gas free surface $A_1$, liquid-solid surface $A_2$ and contact line $C_{cl}$ all indicated.}
\label{F:domain}
\end{figure}

Computations are performed for cases which are axisymmetric or
two-dimensional in simple geometries so that calculations may be
easily reproduced, thus giving benchmark results for future
investigators.  In what follows, we consider a meniscus rising
against gravity through a cylindrical capillary of radius R. The
computational domain is a region in the $(r,z)$-plane, and the
free surface is parameterized by arclength $s$. Motion is assumed
to be axisymmetric about the $z$-axis which runs vertically through the centre of the capillary with the radial axis
perpendicular and located at the base of the capillary
(Figure~\ref{F:domain}). First, we consider the steady propagation
of a meniscus through a capillary (hereafter Problem A) to check
convergence of our code as the mesh is refined and to compare our
computations to the asymptotic results summarized in
\S\ref{asymptotics}.   Then, we will examine Problem A from the
view point of analyzing the influence of the radius $R$ of the
capillary on the interface formation dynamics. Finally, the
unsteady imbibition of a liquid into a capillary will be
considered (hereafter Problem B) and the results compared to
published experimental data for this type of flow.

In the $(r,z)$-plane where the solution is computed, in addition
to the equations formulated in \S\ref{equations}, on the axis of
symmetry for both Problem A and B we have the symmetry conditions
in the form of impermeability and zero tangential stress
\begin{equation}\nonumber\mathbf{u}\cdot\mathbf{n}=0, \qquad \mathbf{n}\cdot\mathbf{P}\cdot\mathbf{t}=0,\qquad (r=0,~0<z<z_a), \end{equation}
where $z_a$ is the a-priori unknown apex height. Additionally, at
the apex we have the conditions of (a) smoothness of the
free-surface shape and (b) the absence of a surface mass source or
sink:
$$
 \mathbf{n}\cdot\mathbf{e}_r=0,
 \qquad
 \rho^s_1 \mathbf{v}^s_{1||}\cdot\mathbf{e}_r=0,
 \qquad
 (r=0, ~z=z_a).
$$

Before doing the calculations we need to consider estimates for
the model's parameters for the two problems to be studied, leaving
free the radius of the capillary as a parameter, whose influence
will be examined in \S\ref{results2}.

\subsection{Typical parameter regime}

To obtain estimates for our parameters, consider the flow of a
water-glycerol mixture through a capillary of radius $R$ at speed
$U=0.01$~m~s${}^{-1}$. At 60\% glycerol this gives fluid
properties of $\rho\approx 10^{3}$~kg~m${}^{-3}$,
$\mu\approx10^{-2}$~kg~m${}^{-1}$~s${}^{-1}$ and $\sigma \approx
7\times 10^{-2}$~N~m${}^{-1}$ \cite{blake02}. On the molecular
scale, the interface is a layer of finite thickness $\ell$ and, as
discussed in \cite{blake02}, the generalized Navier equation and
Darcy-type equation are analogous to what one would have for the
averaged quantities for flow in a channel of width $\ell$. Using
this analogy, one would have $\beta\sim\mu/\ell$ and $\alpha\sim
\ell/\mu$, so that, taking the coefficients of proportionality as
unity as a first approximation, and estimating
$\rho^s_{1e}\approx0.6$ (so that $\lambda=2.5$), one has from
dynamic wetting experiments \cite{blake02} that the relaxation
time scales as $\tau= \tau_{\mu} \mu$ with $\tau_{\mu}$ extracted from
the data as $\tau_{\mu} \approx
7\times10^{-6}$~kg${}^{-1}$~m~s${}^2$. The equilibrium contact
angle is taken as $\theta_e=30^{\circ}$, so that $\rho^s_{2e} =
1.35$. Using these estimates, we have the following
parameters of what we will regard as the base state.  The
parameters independent of the length scale are
\begin{equation}\notag
\quad Ca = 10^{-2},\quad Q=0.1,\quad \rho^s_{1e} = 0.6,\quad \rho^s_{2e}=1.35,\quad \lambda = 2.5,\quad \theta_e=30^\circ,
\end{equation}
and those dependent on the length scale are
\begin{equation}\notag
Re = 10^{-2}(\mu \hbox{m})^{-1}R, \quad St= 10^{-5}(\mu \hbox{m})^{-2} R^{2},\quad \bar{\alpha} = 10^{-1}(\mu \hbox{m}) R^{-1},\quad \bar{\beta} = 10(\mu \hbox{m})^{-1} R,\quad \epsilon = 10^{-2}(\mu \hbox{m})R^{-1}.
\end{equation}

As capillary sizes of interest can range from the millimetre scale
$R=10^3~\mu$m, relevant, for example, for applications in
microgravity \cite{stange03}, right down to a few tens of
nanometres $R=10^{-2}~\mu$m \cite{sobolev01}, the flow can be
dominated by different physical factors. In particular, in large
capillaries bulk inertia becomes important as $Re>1$ and the
surface tension relaxation effect is localized ($\epsilon \ll 1$),
being important only close to the contact line, whilst for
nanoscale flows inertia becomes negligible ($Re\ll1$) and the
surface tension relaxation length becomes comparable to the
capillary's width $\epsilon\approx 1$.

Now, we will present benchmark calculations for the steady
propagation of a meniscus through a capillary (Problem A) and then
consider the time-dependent rise of a meniscus into a capillary
(Problem B), which will be compared to experiments.

\subsubsection{Problem A: steady propagation of a meniscus through a capillary}

For the steady propagation of a meniscus, the `base' of the domain
($z=0$) is sufficiently far (quantified below) from the meniscus
so that the base can be considered as a `far field'. In the far
field, the flow is fully developed and the surface variables take
their equilibrium values.
%
%
%

The velocity distribution across the capillary base is a Poiseuille profile adapted to allow for slip at the liquid-solid interface that follows from the Navier-slip condition and an additional flux of mass out of the domain via the liquid-solid interface which occurs when $Q$ is non-zero:
\begin{equation}\nonumber u=0,\qquad w= -1 + \frac{1+\bar{q}}{1/2 + (2+\bar{q})/(\bar{\beta}/Ca)}\left(1+2/(\bar{\beta}/Ca)-r^2\right) ,\qquad \bar{q}=2\epsilon Q \rho^s_{2e}. \end{equation}
The derivation of this condition is given in the Appendix.
Notably, if $\bar{q} =0$ and $\bar{\beta}/Ca\rightarrow\infty$, then $w
= 1-2r^2$ which is the usual Poiseuille profile satisfying
$dw/dr=0$ at $r=0$, $w=-1$ at $r=1$ and $\int_{r=0}^{1} w~r~dr
=0$.

Taking the radius of the capillary as $R=100~\mu$m, the parameters
dependent on the capillary width become $Re = 1$, $St= 10^{-1}$,
$\bar{\alpha} = 10^{-3}$, $\bar{\beta} = 10^3$ and $\epsilon =
10^{-4}$. In this case, $\epsilon=10^{-4}$ is so small that the
distance of the far field is determined by the need for the bulk
flow to settle as opposed to for the surface phase to relax to its
equilibrium state. It is known that putting the contact line a
distance of ten radii away from the `far field' is more than
enough for this to be satisfied \cite{bazhlekov96,sprittles11c}.
%
%

\begin{figure}
\centering
\includegraphics[scale=0.45]{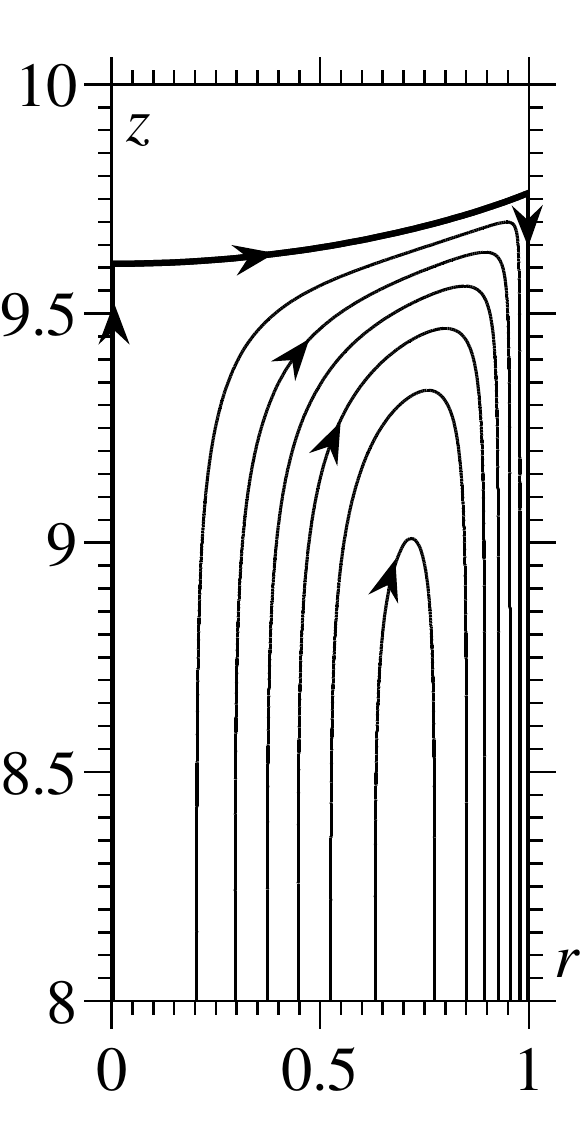}\hspace{2.5 cm}
\includegraphics[scale=0.45]{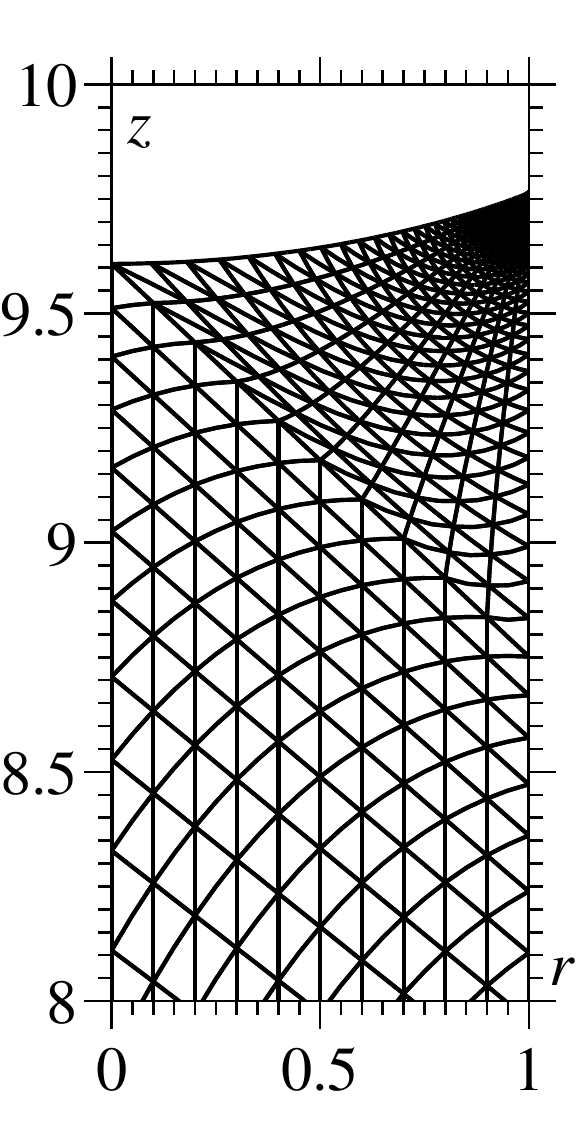}
\caption{Left: streamlines for the base state simulation in increments of $0.02$. Right: corresponding finite element mesh in this region.}
\label{F:stream_big}
\end{figure}

\begin{figure}
\centering
\includegraphics[scale=0.3]{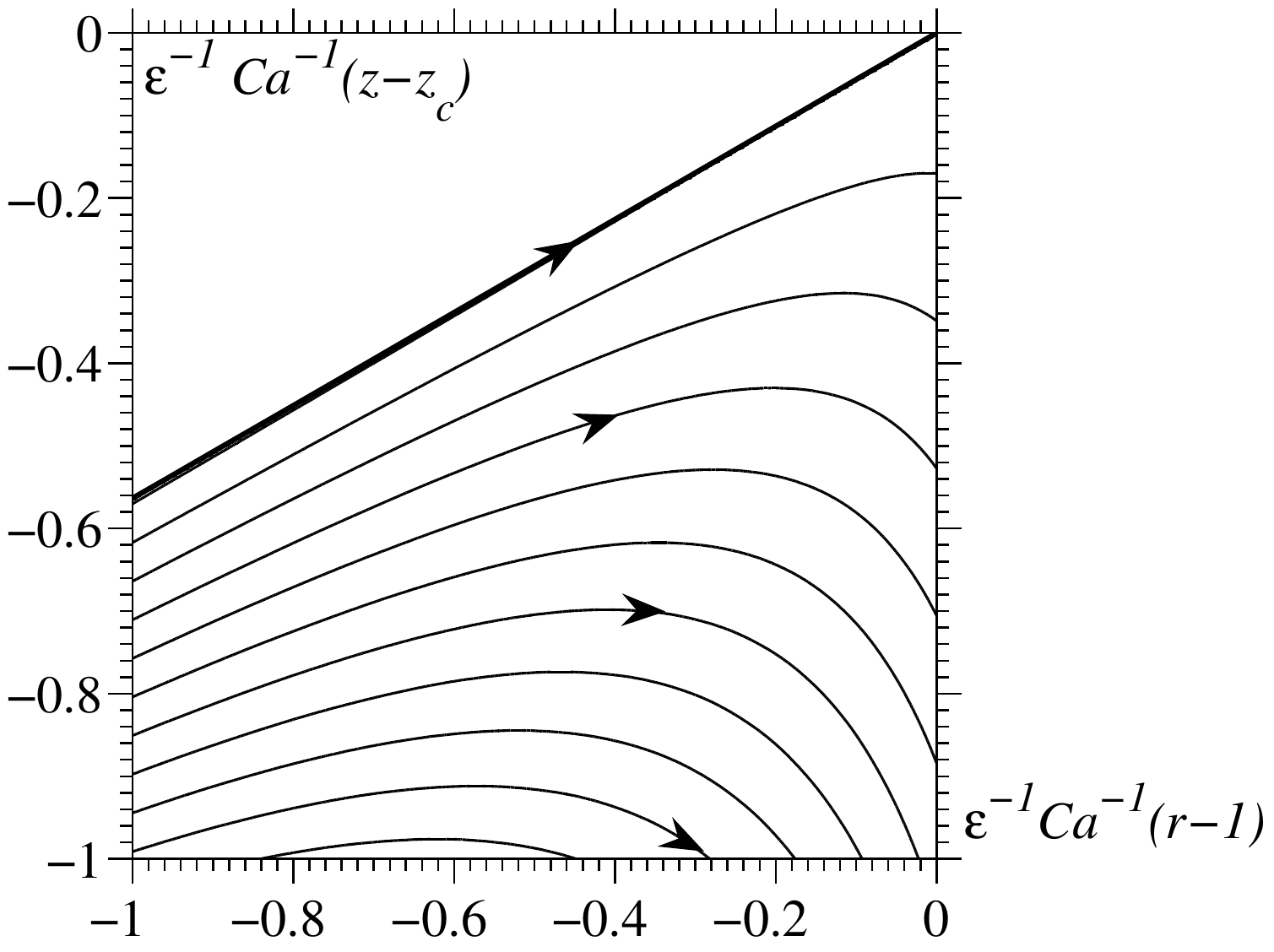}
\includegraphics[scale=0.3]{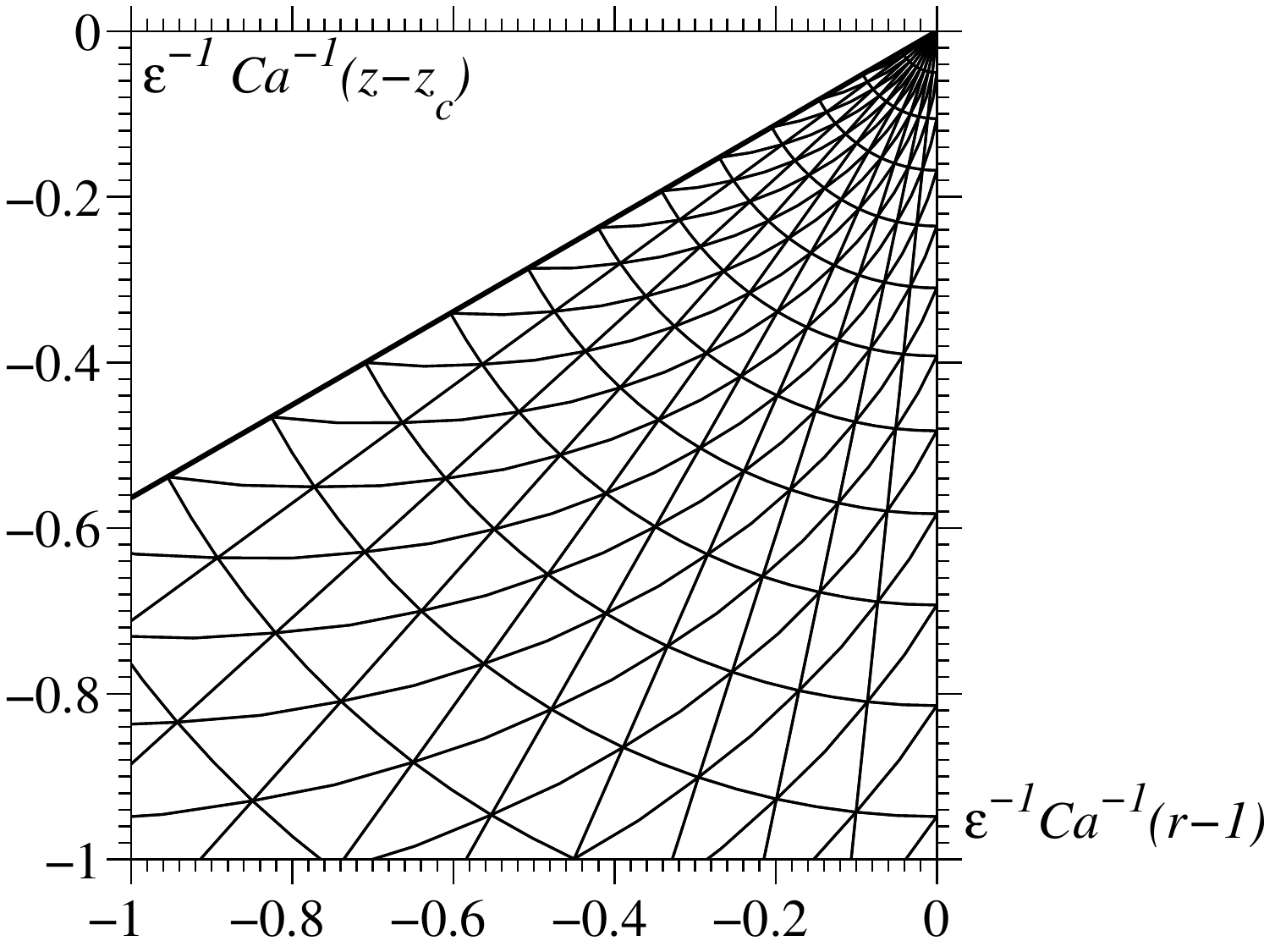}
\caption{Left: a magnified view of the flow near the contact line for the base state simulation showing the adsorption of the fluid into the liquid-solid interface. The length scale is $\epsilon Ca =10^{-6}$ and streamlines are in increments of $2.5\times10^{-3}\epsilon Ca$. Right: corresponding finite element mesh in this region.}
\label{F:stream_small}
\end{figure}

The streamlines for this flow are shown in
Figure~\ref{F:stream_big}. It can be seen that the motion of the
solid with respect to the meniscus drags fluid away from the
contact line and that this in turn leads to a flux of fluid into
the contact-line zone from the region near the free surface. In Figure ~\ref{F:stream_small}, one can see that, as predicted by experiments, the motion of the liquid in the vicinity of the contact-line is `rolling', with the liquid-solid interface formed by adsorbing fluid from the bulk. The
Poiseuille profile, indicated by parallel streamlines in Figure~\ref{F:stream_big}, is
established relatively quickly, so that the truncated far field is
certainly far enough away from the contact-line region to not
influence the results.  The mesh is graded, with small elements near the contact
line and larger ones further away, so that the essential physics
of wetting, occurring on a smaller length scale than the bulk
flow, is fully resolved whilst the problem remains computationally
tractable.   The details of the construction of this mesh are
given in \cite{sprittles11c}, where it is shown how the bipolar
coordinate system can be utilized to provide circular spines near
the contact line and straight ones further away to match with the
domain's shape.

\subsubsection{Mesh independence tests}\label{independcetests}

The mesh is constructed such that the number of elements in the body of the capillary and those near the meniscus can be separated.  This allows us to study the convergence of the surface equations to a solution without being concerned by the bulk dynamics.  The  number of nodes along the surfaces can be chosen as well as how fast the element size is increased as one moves away from the contact line. Here, we fix the increase in element size as 10$\%$, and check that the solution converges as the mesh is refined, i.e. as we increase the resolution near the contact line.

In \cite{sprittles11c}, it was shown that the dynamic contact
angle $\theta_d$ imposed into the weak formulation can differ from
the computed one $\theta_c$, i.e. the one obtained from the
computed free-surface shape, if the mesh is under resolved near
the contact line.  Rather than imposing the angle in the strong
form, and moving the errors to other less observable parts of the
numerical scheme, as has often been the case in previous works, it
was demonstrated that the difference $|\theta_d - \theta_c|$
provides a simple error-indicator for the scheme.  When
considering the conventional model, it was shown\footnote{Note
that in this previous publication the non-dimensionalization is
slightly different with $\bar{\beta}=\beta L/\mu$ whereas here it
is $\bar{\beta}/Ca = \beta L/\mu$} that to resolve both the
dynamics of slip and the free-surface shape near the contact line,
so that for the computed angle one has, say,
$|\theta_d-\theta_c|<0.1^\circ$, the smallest element size $l_{min}$
must satisfy $l_{min}\le \bar{\beta}^{-1}\min(5\times 10^{-2},
Ca)$. Then, for the flow considered, this imposes $l_{min}\le
10^{-3}\min(5\times 10^{-2},10^{-2}) = 10^{-5}$. In addition, we
now have length scales associated with the physics of interface
formation; in particular, the asymptotics for $Ca,\epsilon \ll 1$
shows that, in the cases considered, the inner most region is on
the scale $O(\epsilon Ca) = 10^{-6}$.

To study convergence, it is convenient to consider the following two angles:
\begin{itemize}
  \item[$\theta_d$:]
  The dynamic contact angle featuring in the problem
   statement; this angle is imposed into the code through the
   weak formulation, but in the mathematical problem it is a
   variable whose converged value is to be found.
  \item[$\theta_c$:] The angle which the free surface computed for a given spatial resolution of the mesh makes with the solid.
\end{itemize}

\begin{figure}
\centering
\includegraphics[scale=0.30]{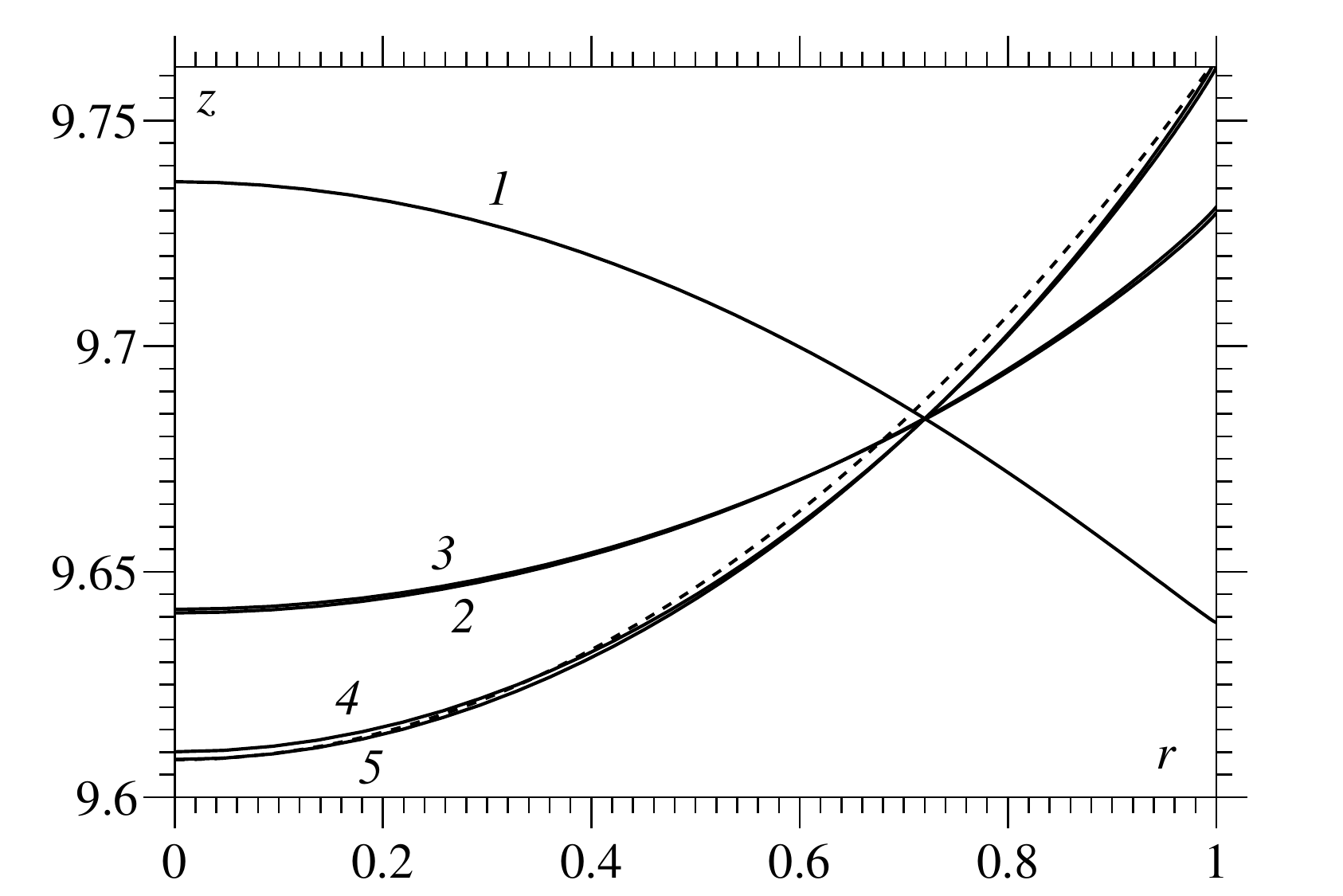}
\caption{Convergence of the free-surface shape with curves corresponding to smallest element sizes of 1:~$6\times10^{-3}$, 2:~$3\times10^{-4}$, 3:~$2\times10^{-5}$, 4:~$9\times10^{-7}$, 5:~$5\times10^{-8}$. The dashed line shows the spherical cap approximation which meets the solid at contact angle $\theta_{app} = 72.4^\circ$.}
\label{F:fs_shape}
\end{figure}

Consider 13 meshes with smallest elements ranging over five orders of magnitude from $l_{min}=5\times10^{-3}$ down to $l_{min}=5\times10^{-8}$. Coarser meshes were unable to provide a solution as the computed angle $\theta_c$ increased past $180^\circ$, despite the imposed angle $\theta_d$ being less than $60^\circ$, in the same way as observed in \cite{sprittles11c}.  The change in the free-surface shape during the mesh refinement procedure is shown in Figure~\ref{F:fs_shape}, where one can see just how bad the approximation is on our most coarse mesh, represented by curve 1. It can be seen that convergence appears to take place in two stages, so that decreasing the smallest element from $3\times 10^{-4}$ (curve 2) to $2\times 10^{-5}$ (curve 3) has little influence on the free-surface shape, but, for the reason explained below, these shapes are yet far from the converged free-surface shape (curve 5) which is obtained only when the mesh elements decrease much further.

\begin{figure}
\centering
\includegraphics[scale=0.30]{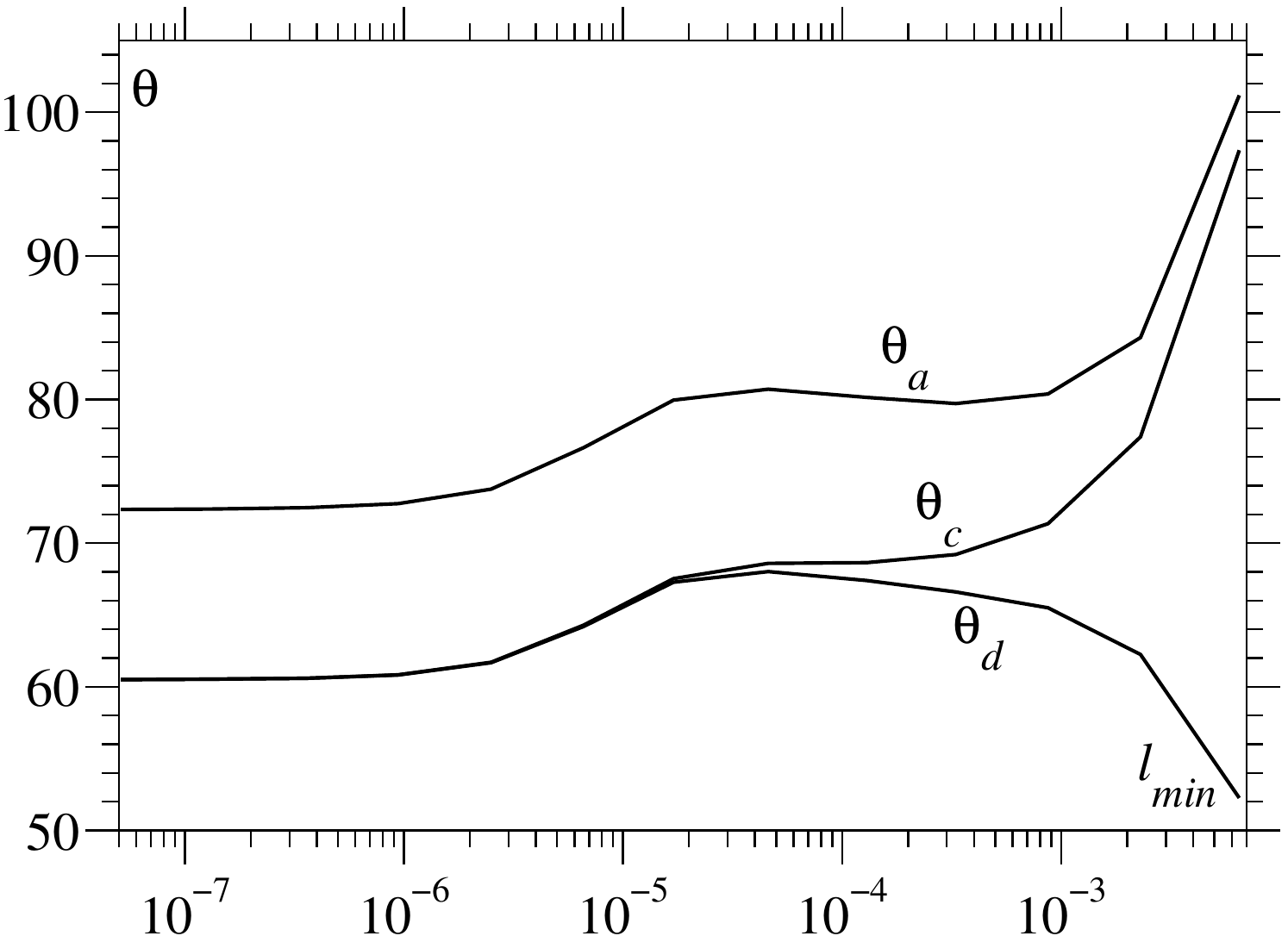}
\includegraphics[scale=0.30]{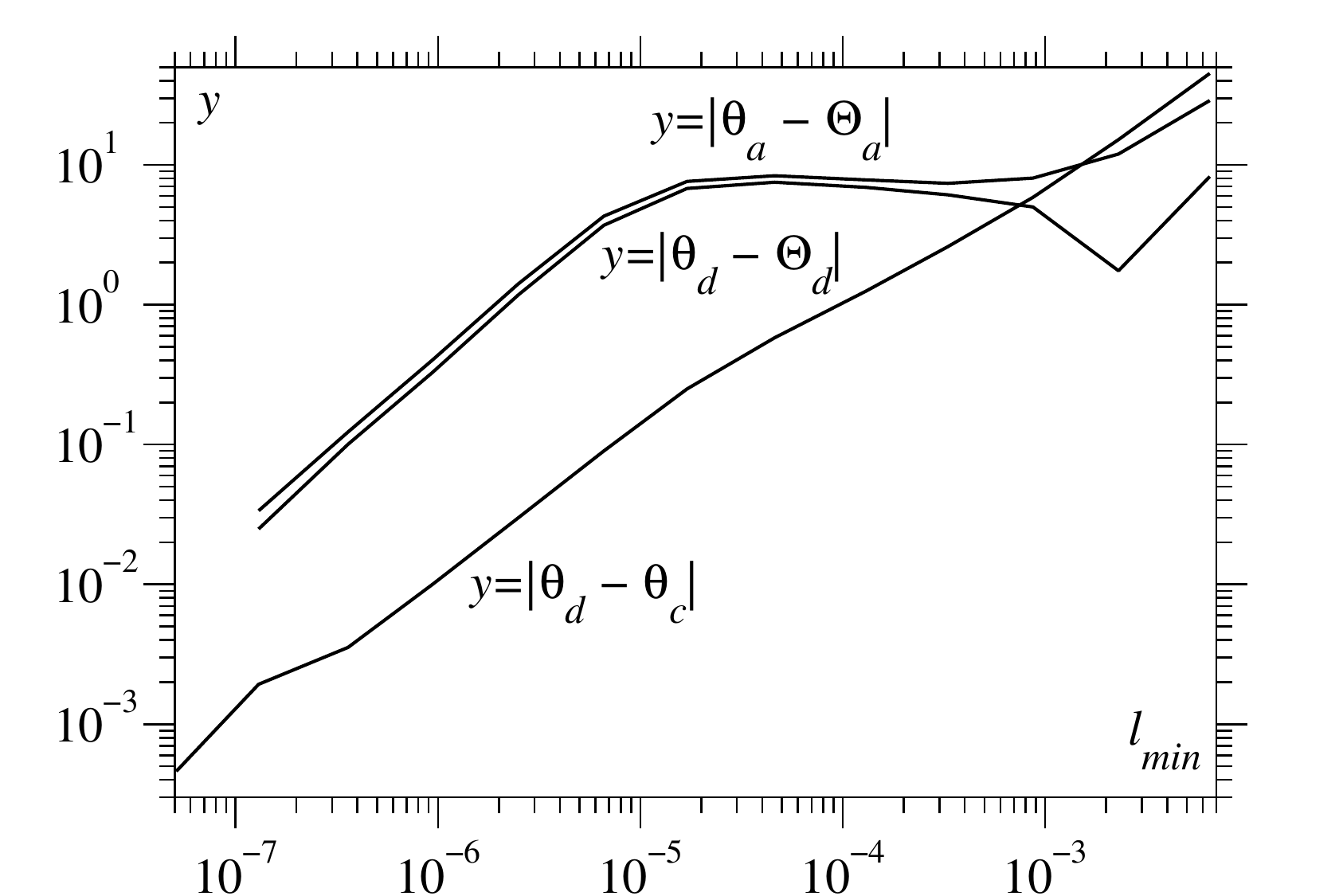}
\caption{Convergence of the dynamic, computed and apparent angles as the mesh, with smallest element size $l_{min}$, is refined for the base state simulation.  The converged values of $\theta_d$ and $\theta_a$ are $\Theta_d = 60.5^\circ$ and $\Theta_a=72.3^\circ$, respectively.}
\label{F:convergence}
\end{figure}

The two-stage convergence can be explained by examining the
aforementioned angles as the mesh is refined, as shown in
Figure~\ref{F:convergence}. One can see that at around
$l_{min}=10^{-5}$ the computed angle $\theta_c$ becomes
indistinguishable from the dynamic angle $\theta_d$ that goes into
the weak formulation, but at this stage $\theta_d$, which is a
part of the solution, is yet to converge to its final value.  This
corresponds to curves $2$ and $3$ in Figure~\ref{F:fs_shape}. More
specifically, we see that $|\theta_d - \theta_c|<0.1^\circ$ at
$l_{min}=7\times10^{-6}$, close to $10^{-6}$ predicted from
\cite{sprittles11c}, i.e.\ the mesh is already sufficiently
resolved for the computed angle $\theta_c$ to be close to the
angle $\theta_d$ that features in the mathematics of the problem
and goes into the weak formulation.  However, in our problem, the
angle $\theta_d$ is itself a variable and the
deviation of $\theta_d$ from its final value $\Theta_d=60.5^\circ$ is still
large $|\theta_d-\Theta_d|=3.8^\circ$.  The convergence towards a
final solution is then dominated by the requirement to resolve the
interface formation dynamics. We have
$|\theta_d-\Theta_d|<0.1^\circ$ at $l_{min}=4\times10^{-7}$ which
again falls close to our estimate and shows the need to resolve
the length scale $\epsilon Ca = 10^{-6}$.  This suggests a new
estimate which builds upon that given in \cite{sprittles11c}, that
for errors in the angle of the order of $0.1^\circ$, we require
\begin{equation}
l_{min}\le \min(5\times 10^{-2}\bar{\beta}^{-1}, Ca\bar{\beta}^{-1},10^{-1} \epsilon Ca ),
\end{equation}
which, respectively, ensure that the code resolves the free surface near the contact line (particularly at high $Ca$), the length scale on which slip occurs and the dynamics of interface formation.

A frequently used way of simplifying the problem of modelling the
flow through a capillary and of easily interpreting theoretical and
experimental results for this flow is to approximate the
free-surface shape as a spherical cap \cite{washburn21,ngan82}. Then, the free-surface
shape is fully specified by the difference between the contact-line height and apex height $h = z_c - z_a$. Usually, the
spherical-cap approximation is characterized by the so-called
`apparent' contact angle $\theta_a$, which is the angle that a
spherical cap fitted through the apex and the contact line makes
with the solid:
\begin{equation}
\theta_a = \pi - \arccos\left(-\frac{2h}{1+h^2}\right).
\end{equation}
A spherical cap is the free-surface shape obtained in the limit of $Ca\rightarrow 0$ and by computing $\theta_a$ we will also have a measure of the deviation from this shape due to viscous bending of the free surface.

In Figure~\ref{F:fs_shape} the spherical cap approximation for the
converged solution is plotted (dashed line).  Deviations from the
actual free-surface shape (curve 5) can be observed and the
apparent angle is considerably larger, by $11.8^\circ$, than the
dynamic contact angle, highlighting the finite capillary number
regime considered, which will increase at higher capillary
numbers. Figure~\ref{F:convergence} shows that the convergence of
the value of $\theta_a$ and its deviation from the final value,
i.e. the one corresponding to the converged free-surface shape,
closely follow  the trend of $\theta_c$. This demonstrates that
the shape of the free surface near the contact line is what
governs the global error to the free-surface shape, as expected
for small capillary numbers. In other words, the dynamics which
governs $\theta_d$ must be resolved not only to determine the flow
local to the contact line, but also to accurately predict the
\emph{entire} free-surface shape and hence the global flow.

\subsubsection{Convergence to the asymptotic solution as $\epsilon\to0$, $Ca\to0$}

Having established that our computed solution converges under mesh
refinement, we now check that this solution converges to the
asymptotic results outlined in \S\ref{asymptotics} as
$\epsilon,Ca\rightarrow0$.  As in the the asymptotics, the parameter
$V^2 = \epsilon\bar{\beta}/(\lambda(1+4\bar{\alpha}\bar{\beta}))$ is
regarded finite in the asymptotic limit considered. To check
convergence to the asymptotics, the parameters used in
\S\ref{independcetests} will remain unchanged except that we now
vary $\epsilon$ and $Ca$ and, for a start, fix $V^2=0.1$ so that
$\bar{\beta}=\bar{\alpha}^{-1}=1.25\epsilon^{-1}$.

Three data sets are considered, with $(Ca,\epsilon) =
1:(5\times10^{-2},5\times
10^{-4}),~2:(10^{-3},2.5\times10^{-2}),~3:(10^{-3},5\times10^{-4})$,
so that we can ascertain the influence of decreasing either $\epsilon$ or $Ca$ by a factor of $50$.  The simplest `integral' measure of
convergence is the dynamic contact angle whose asymptotic value for
this set of parameters is $102.1^\circ$. The obtained solutions for
$\theta_d$ for $1-3$ are $104.1^\circ$, $107.7^\circ$ and
$102^\circ$ so that the deviations from the asymptotic result are
$2^\circ$, $5.6^\circ$ and $0.1^\circ$, respectively.  Therefore, we
have the convergence of the numerically obtained dynamic contact
angle to the asymptotic one, with an indication how this convergence
is controlled by $\epsilon$ and $Ca$.

\begin{figure}
\centering
\includegraphics[scale=0.27]{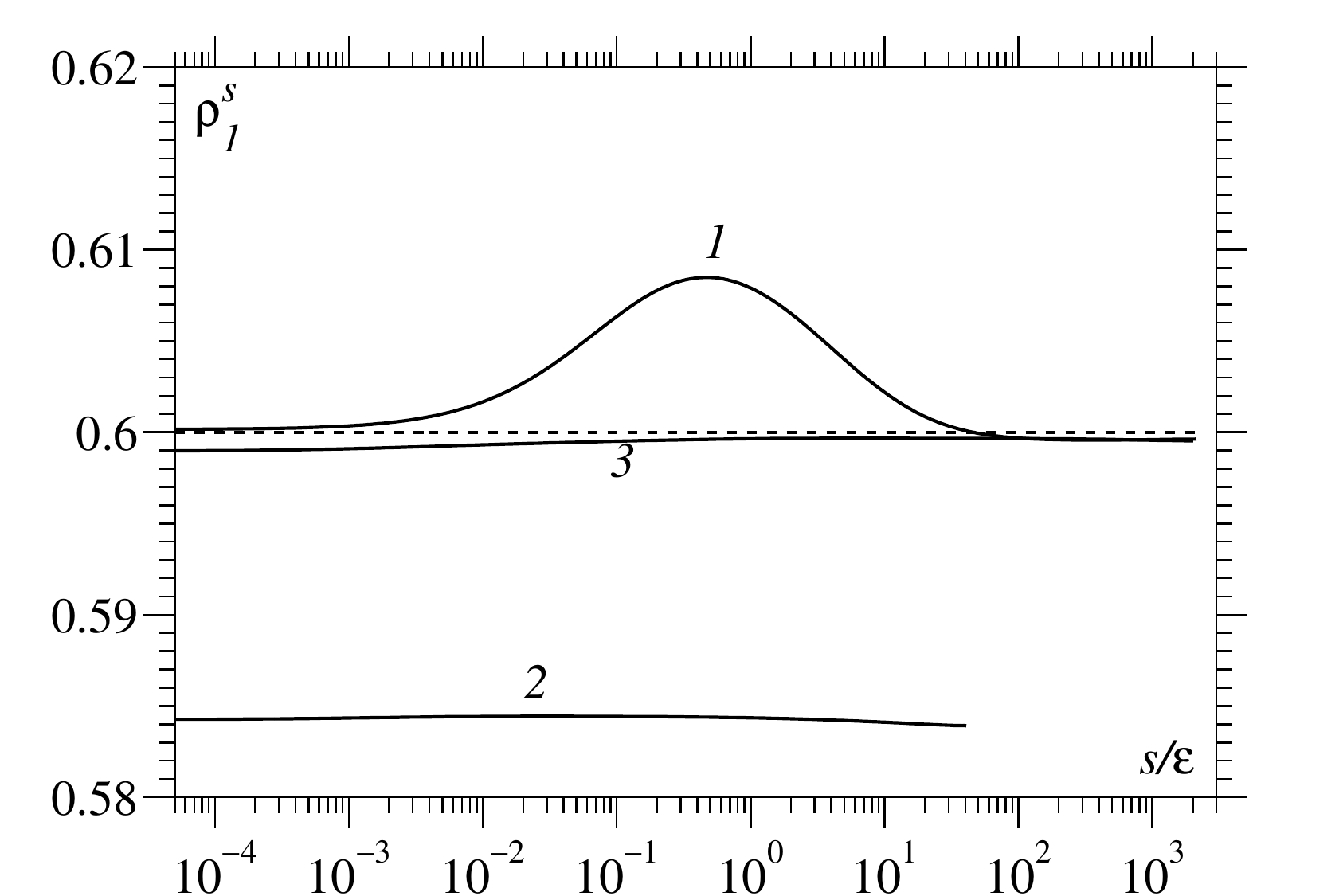}
\includegraphics[scale=0.27]{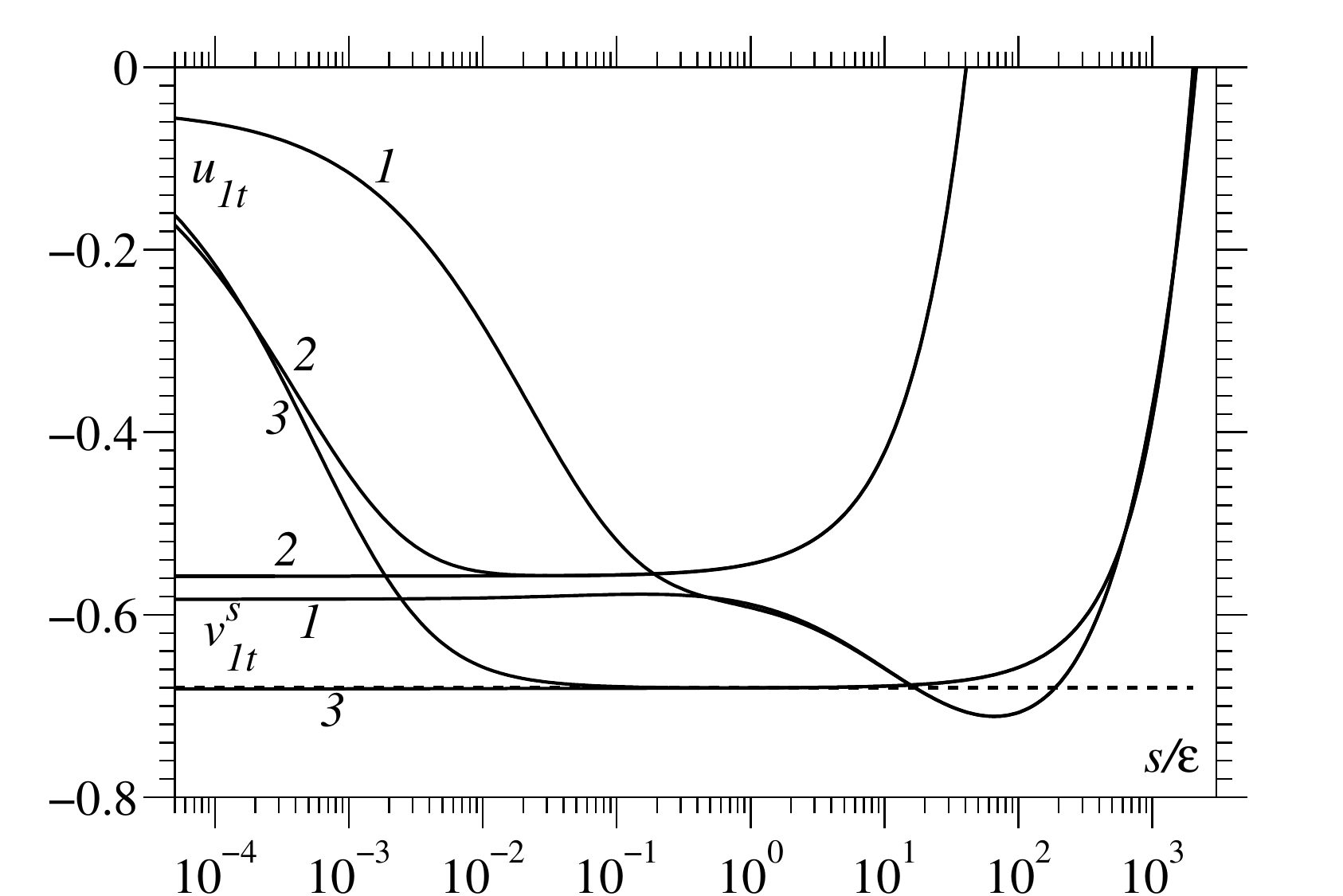}
\caption{Distribution of the surface variables along the free surface against the scaled distance from the contact line $s/\epsilon$.  Curves are for $(Ca,\epsilon) = 1:(5\times10^{-2},5\times 10^{-4}),~2:(10^{-3},2.5\times10^{-2}),~3:(10^{-3},5\times10^{-4})$.  Left: $\rho^s_1$ compared to its asymptotic (equilibrium) value (dashed line). Right: surface velocity tangential to the free surface and bulk velocity tangential to the surface compared to far-field free-surface velocity $u_f=-0.68$ (dashed line).}
\label{F:benchfs}
\end{figure}

To examine the issue of convergence in more detail, consider the
distributions of the surface parameters and the bulk velocity along
the interfaces. Deviations of the surface density on the free
surface $\rho^s_1$ from its equilibrium value $\rho^s_{1e}$ and the
distributions of the tangential components of the surface velocity
$v^s_{1t}$ and the bulk velocity tangential to the free surface
$u_{1t}$ are shown in Figure~\ref{F:benchfs}.  As one can see, the
magnitude of the variation in $\rho^s_1$ along the free surface is
governed by $Ca$, but it is $\epsilon$ that determines how far the
`level' of $\rho^s_1$ is from its equilibrium value. As
$\epsilon\to0$, $Ca\to0$, we have that
$\rho^s_1\rightarrow\rho^s_{1e}$.

In agreement with the asymptotic results, the surface velocity
$v^s_{1t}$ coincides with $u_{1t}$ in the intermediate region and
the two velocities begin to deviate from each other at around
$s/\epsilon\approx Ca$, i.e.\ in the inner region, where, as expected, $v^s_{1t}$
remains constant whereas $u_{1t}$ decreases as the contact line is
approached. This pattern is clearly seen for data set $3$, where
both the surface and bulk velocity attain the value of $u_f = -0.68$
in the outer region, $s/\epsilon \gg 1$, then, as the intermediate
asymptotic region is entered, $v^s_{1t}$ and $u_{1t}$ remain at this
level through this region $Ca\ll s/\epsilon\ll 1$, and as the inner
region is reached, the surface velocity $v^s_{1t}$ remains constant
up to the contact line, whereas $u_{1t}$ varies. Since $\rho^s_1$ is
also constant through the inner region, it is the surface mass flux
of the intermediate region that goes into the contact line and hence
determine the mass flux that feeds the liquid-solid interface.
Notably, as one could already see in Figure~\ref{F:stream_small}, for $Q\neq 0$ the
contact line is not a stagnation point for the bulk flow, so that
$u_{1t}\neq 0$ at the contact line.

\begin{figure}
\centering
\includegraphics[scale=0.27]{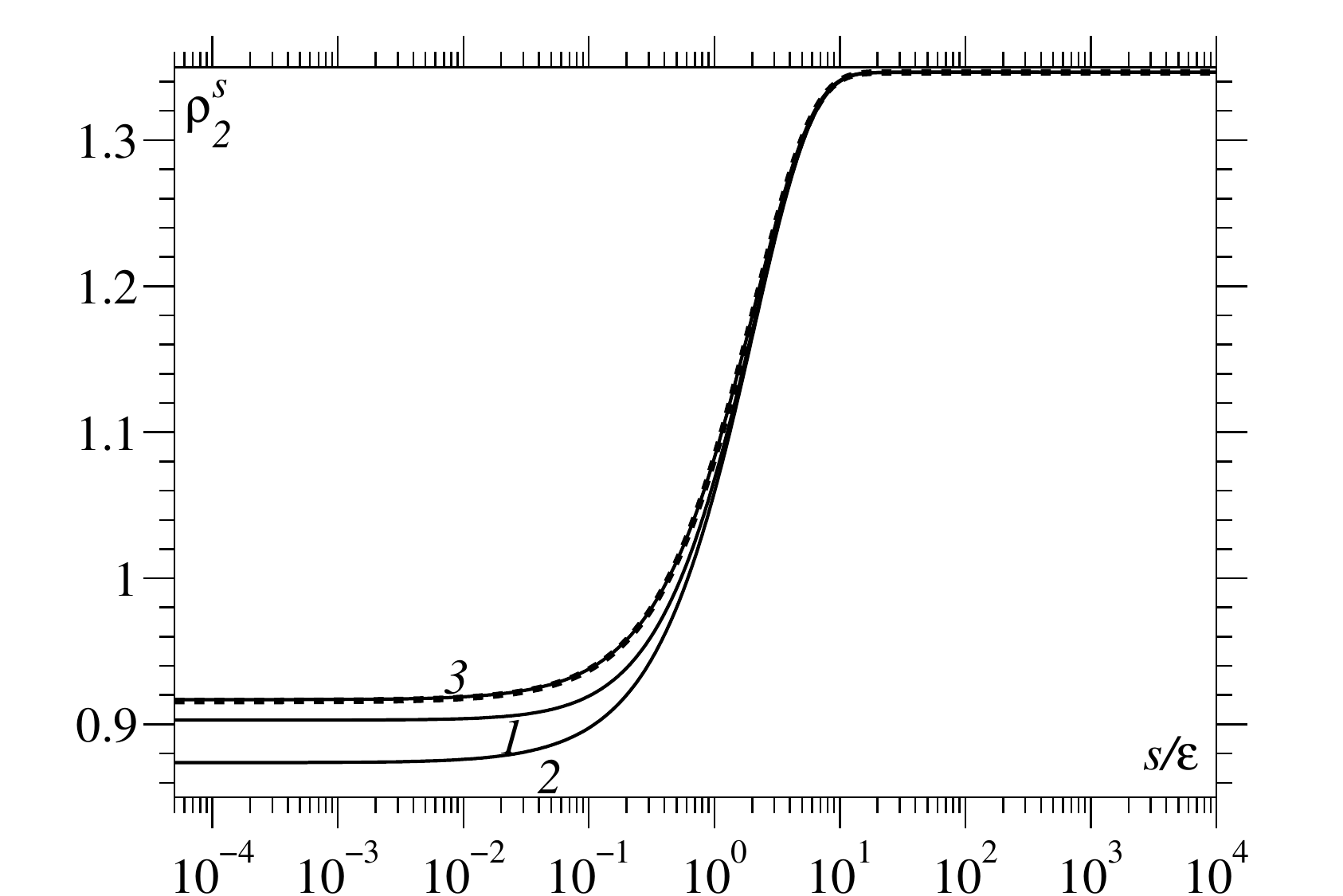}
\includegraphics[scale=0.27]{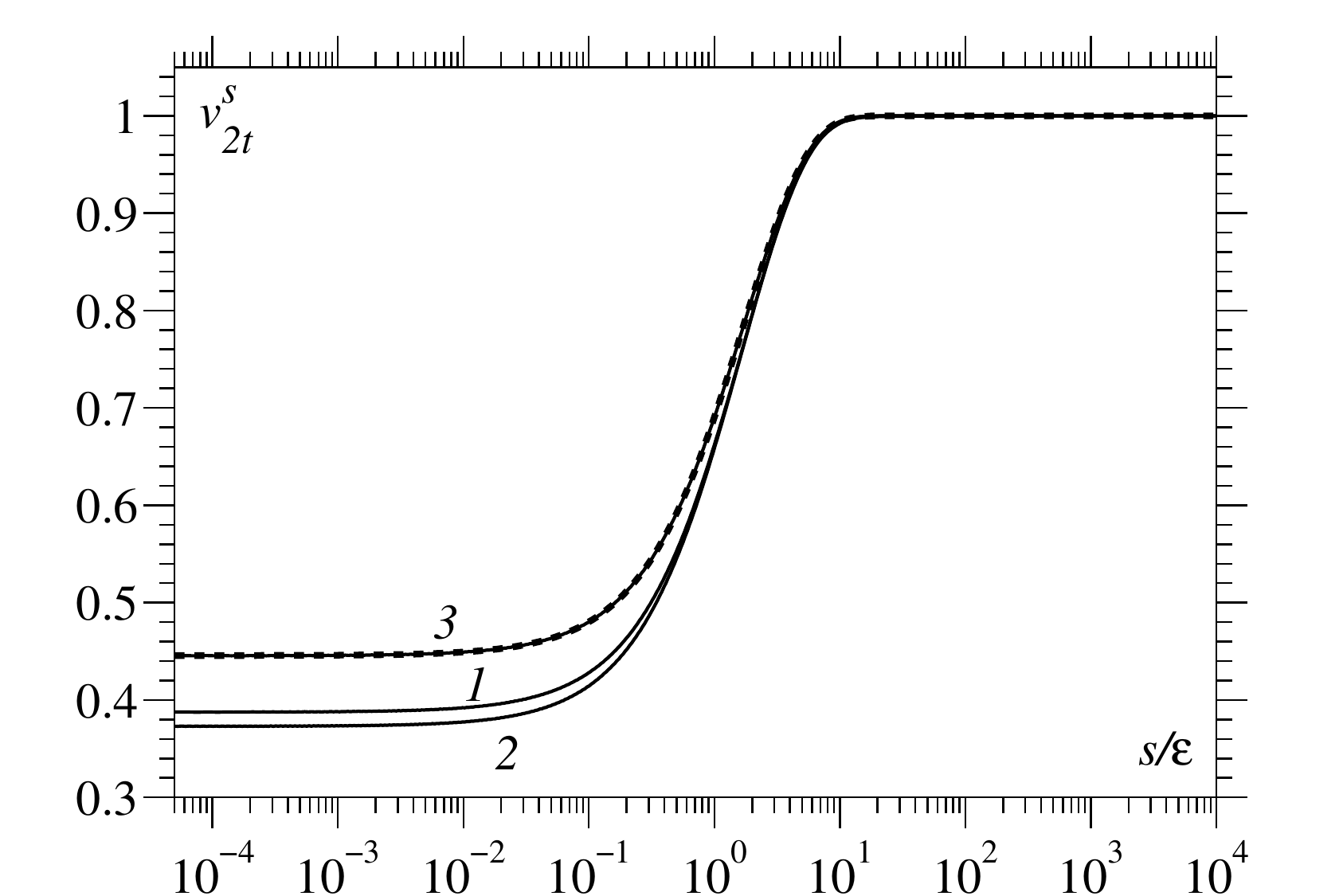}
\caption{Distribution of the surface variables along the free surface
 against the scaled distance from the contact line $s/\epsilon$ compared
 to the asymptotic distribution (dashed lines).   Curves are for
 $(Ca,\epsilon) = 1:(5\times10^{-2},5\times 10^{-4}),
 2:(10^{-3},2.5\times10^{-2}),
 3:(10^{-3},5\times10^{-4})$.  Left: surface density $\rho^s_2$.
 Right: surface velocity tangential to the solid $v^s_{2t}$.}
\label{F:benchss}
\end{figure}

The relaxation of the surface variables along the liquid-solid
interface to their equilibrium far-field values is shown in
Figure~\ref{F:benchss}. As one can see, the computed curves converge
to the asymptotic distributions of \S\ref{asymptotics} as $\epsilon,Ca\rightarrow 0$ with data set $3$ being practically
indistinguishable from the asymptotic curves (dashed lines).  In all
cases the surface density is below its equilibrium value at the
contact line, and it takes a finite distance for the interface to
fully form.  At the contact line, the surface velocity tangential to
the solid is lower than its equilibrium value and it increases to
reach the solid's velocity on the same characteristic length scale.

The convergence of our computational results to the asymptotic ones
in the limit considered has now been confirmed. Now, we can do the
reverse and use the code to ascertain the accuracy with which the
boundary-value problem for the ODEs (\ref{two_odes}) gives the dynamic contact angle and the
accuracy of the formula (\ref{val_cangle}) obtained under the
additional assumption $\lambda\gg1$ when both are applied in a range
of parameters where the assumptions of the asymptotics may not be
satisfied. In other words, it is worth checking how the asymptotics
works outside its formal limits of applicability.

In Table~\ref{table} a range of values of $\epsilon$ and $Ca$ are
considered with the parameters as above except for
$\bar{\beta}=\bar{\alpha}^{-1}=10^3$, which is now fixed, so that
$V^2$ varies.  In general, over the considered range the asymptotics
gives a remarkably good approximation to the dynamic contact angle.
The formula (\ref{val_cangle}) also gives a very reasonable
approximation which improves considerably for smaller values of
$\epsilon$. This good agreement is obtained despite the fact that
$V^2$ is smaller than the value of $Ca$ in certain cases, i.e.\ a
parameter considered finite in the asymptotics is smaller than the
one assumed to be asymptotically small, e.g.\ for $\epsilon=10^{-4}$
we have $V^2=8\times10^{-2}$, and yet for $Ca=10^{-1}$ the computed
angle is still only $1.6^\circ$ away from the asymptotic value.
\begin{table}
\begin{tabular}{|c|c|c|c|c|}
  \hline
                              & $Ca=1$      & $Ca=10^{-1}$   & $Ca=10^{-2}$  & Asymptotic angle (Approximate value from (\ref{val_cangle}))\\
  \hline
  $\epsilon=10^{-2}$          & 147.3$^\circ$        & 149.6$^\circ$          & 146.5$^\circ$          & 146.3$^\circ$~~~~~(136.9$^\circ$ )\\
  \hline
  $\epsilon=10^{-3}$          & 91.6$^\circ$          & 97.6$^\circ$           & 98.2$^\circ$           & 97.4$^\circ$~~~~~(93$^\circ$ ) \\
  \hline
  $\epsilon=10^{-4}$          & 55.6$^\circ$          & 59$^\circ$             & 60.5$^\circ$           & 60.6$^\circ$~~~~~(59.7$^\circ$  ) \\
  \hline
\end{tabular}\\
\caption{Comparison of computed solution with asymptotic value and formula (\ref{val_cangle}).}\label{table}
\end{table}

Thus, it has been shown that our computational scheme, which
incorporates the interface formation model, is mesh-independent once
all the length scales are resolved and converges to the asymptotic
results in the limiting case. The computational results reported so
far can be used as benchmarks for those who are interested in
putting the interface formation model into a numerical code.  It is
useful to know that the asymptotic result gives a good prediction of
the key feature of the wetting process, the value of the dynamic
contact angle, with the formula (\ref{val_cangle}) giving a
reasonable approximation, well outside their strict limits of
applicability. Now, we will use the developed numerical tool to make
novel predictions on the size-dependence of the dynamic wetting flow
through a capillary.

\section{Size dependence of flow characteristics in Problem A}\label{results2}

Now, consider how the flow characteristics of steady capillary rise
(Problem A) depend on the radius of the capillary. We will vary the
radii over three orders of magnitude, from the $R=100~\mu$m
considered previously down to $R=0.1~\mu$m. In all cases, the far
field is placed at a sufficient distance from the contact line,
ensuring that the liquid-solid interface is in equilibrium there and
that moving it further away does not alter the dynamics of the
interfacial relaxation process.

Conventional theories, where the dynamic contact angle is an input,
give that, because the material properties of the system remain
unchanged and the meniscus is travelling at the same speed, the
dynamic contact angle must be the same in all of these cases, i.e.\
independent of $R$. In contrast, in the interface formation model,
where the dynamic contact angle is an output determined by the
relaxation process along the interfaces, i.e.\ it is part of the
solution, changes to the flow field near the contact line due to
changes in $R$ can alter the angle.

\begin{figure}
\centering
\includegraphics[scale=0.27]{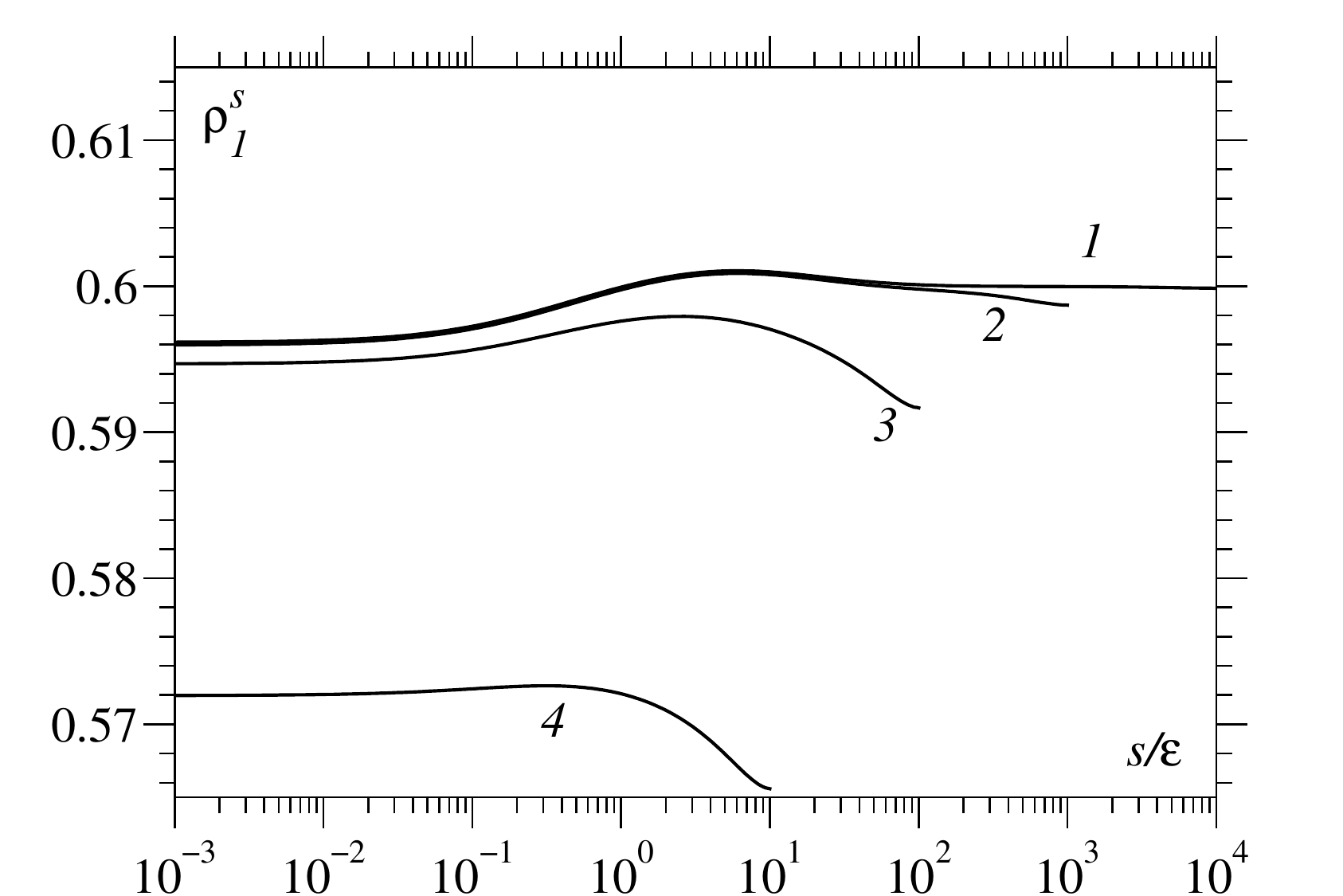}
\includegraphics[scale=0.27]{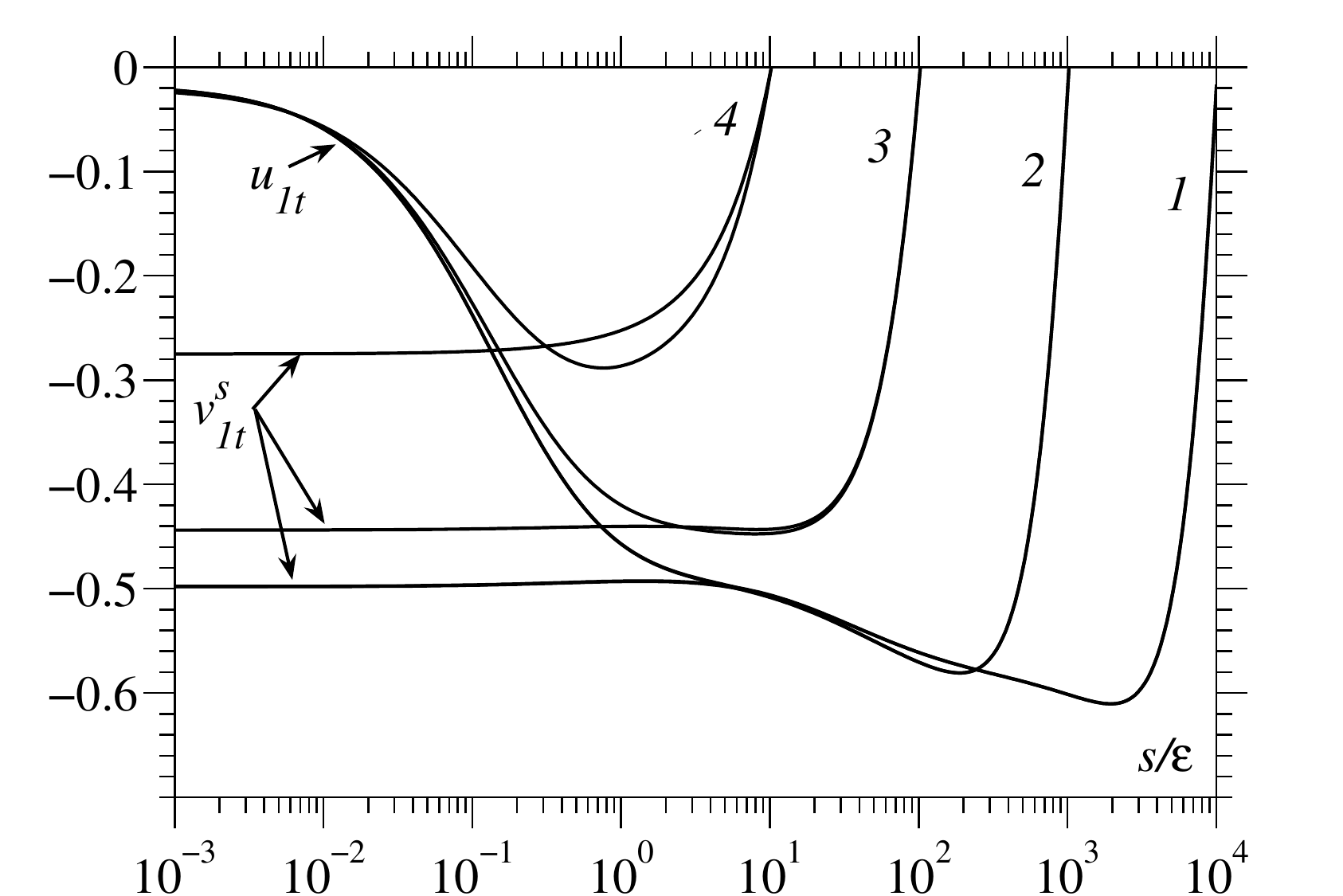}
\caption{Distribution of surface variables along the free surface in capillaries of sizes 1:$R=100~\mu m$, 2:$R=10~\mu m$, 3:$R=1~\mu m$, 4:$R=0.1~\mu m$, against the scaled distance from the contact line $s/\epsilon$. Left: surface density $\rho^s_1$.  Right: velocities tangential to the free surface from the bulk $u_{1t}$ and in the interface $v^s_{1t}$.}
\label{F:vs1t_sizes}
\end{figure}

In Figure~\ref{F:vs1t_sizes}, the bulk and surface velocities
tangential to the free surface are plotted against $s/\epsilon=10^2
(\mu \hbox{m})^{-1} R~s$. In the geometry of a capillary, the apex
is a stagnation point in the flow; it slows down the flow in its
vicinity so that, as one can see in the Figure, the smaller $R$
becomes, the lower the bulk velocity is along the free surface. As
the contact line is approached, the surface velocity follows the
bulk velocity until $s\approx10\epsilon$ and then the bulk velocity
decreases whilst the surface velocity remains approximately
constant.  Thus, when the capillary is relatively large
($R\ge10\mu$m), as for curves $1,2$, there is enough space for the
bulk velocity and surface velocity to increase from zero at the apex
to a `far field' value, which is not far from $u_f(\theta_d)=-0.57$
given by the asymptotic solution.  Then it is this value that the
surface velocity keeps up to the contact line. However, as the
radius of the capillary decreases and the apex gets closer to the
contact line, a `far field' no longer exists and the velocities do
not have enough room to accelerate to reach the same `far field'
value, and, as a result, the surface velocity at the contact line
becomes much reduced. As can be seen from Figure~\ref{F:vs1t_sizes},
the surface density on the free surface is less affected, remaining
close to its equilibrium value even for the smallest capillary and
consequently it is the surface velocity which determines the surface
mass flux into the contact line.  For smaller capillaries this flux
is significantly reduced.

\begin{figure}
\centering
\includegraphics[scale=0.27]{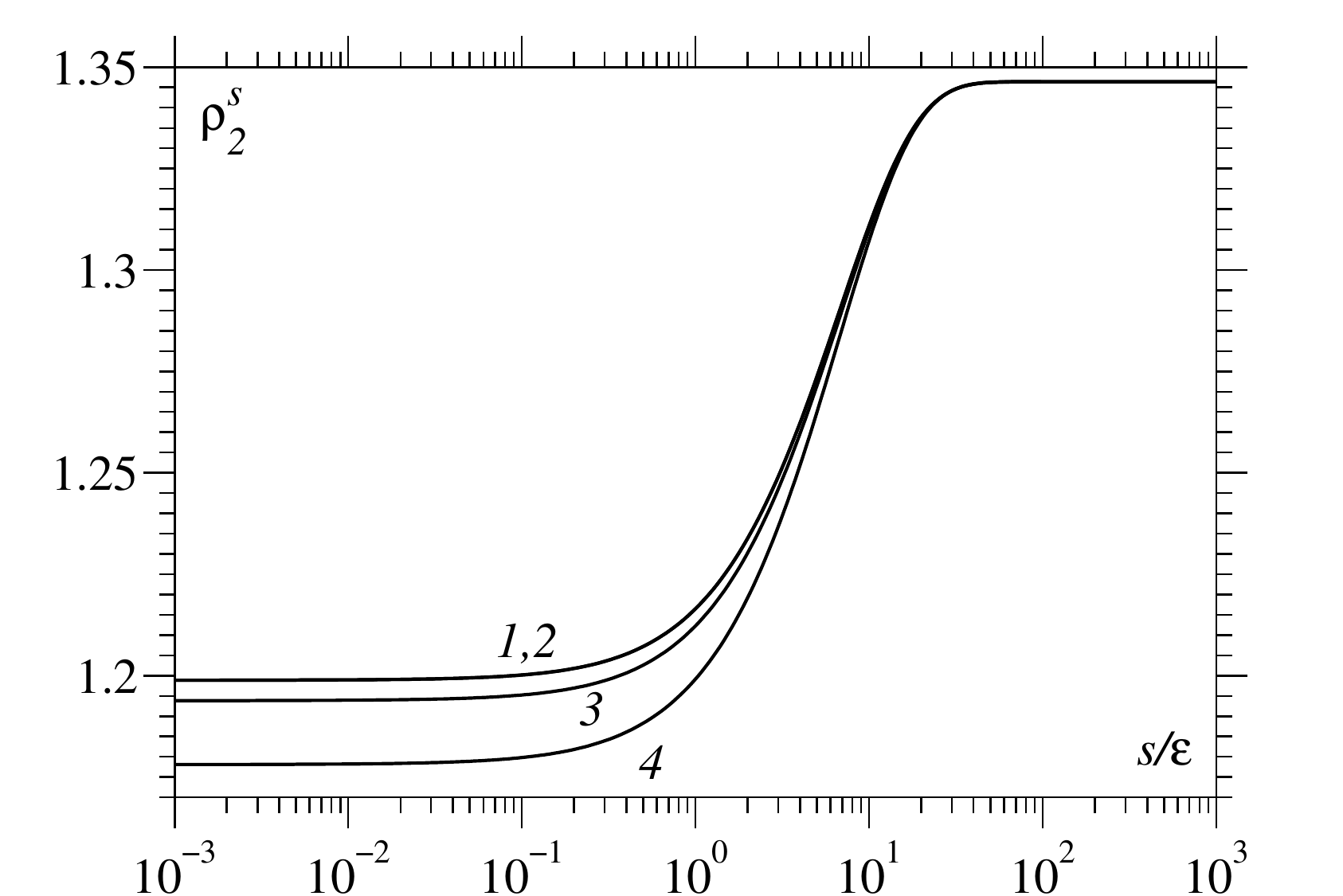}
\includegraphics[scale=0.27]{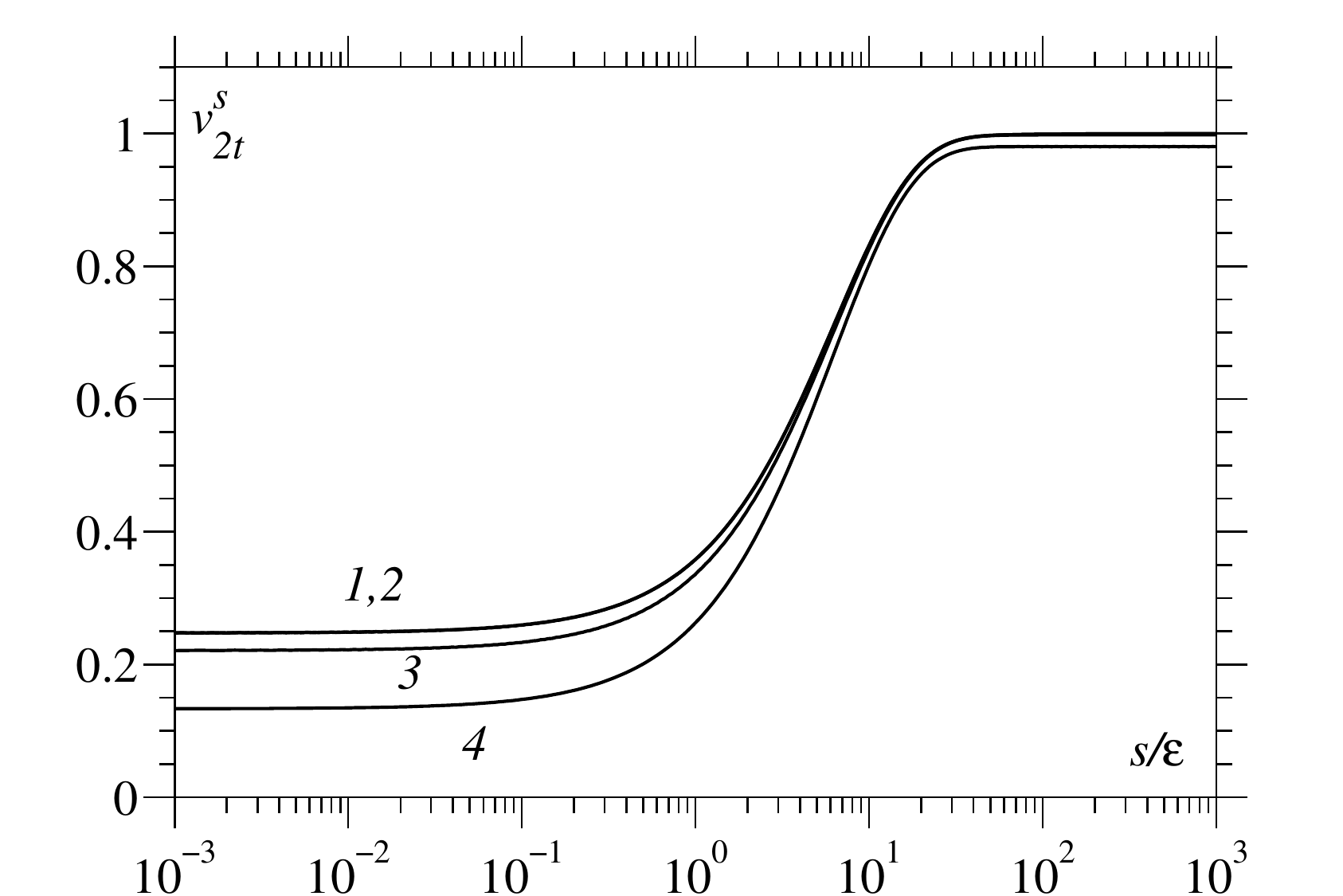}
\caption{Distribution of surface variables along the solid surface in capillaries of sizes 1:$R=100~\mu m$, 2:$R=10~\mu m$, 3:$R=1~\mu m$, 4:$R=0.1~\mu m$, against the scaled distance from the contact line $s/\epsilon$. Left: surface density $\rho^s_2$.  Right: velocities tangential to the solid surface from the bulk $u_{2t}$ and in the interface $v^s_{2t}$.}
\label{F:vs2t_sizes}
\end{figure}

Continuity of surface mass flux across the contact line means that,
as $R$ goes down, the flux into the liquid-solid interface is also
reduced, as can be seen in Figure~\ref{F:vs2t_sizes}, where, as is
the case for the free surface, changes in the surface velocity are
more pronounced than those in surface density. The general trend for
both of these surface variables along the liquid-solid interface is
the same exponential profile, with the value at the contact line
differing in each case. Notably, as the capillary size decreases,
the influence of slip on the velocity at the liquid-solid interface
becomes more pronounced and this accounts for curve~$4$ of the
surface velocity deviating from the other curves for large
$s/\epsilon$.

\begin{figure}
\centering
\includegraphics[scale=0.27]{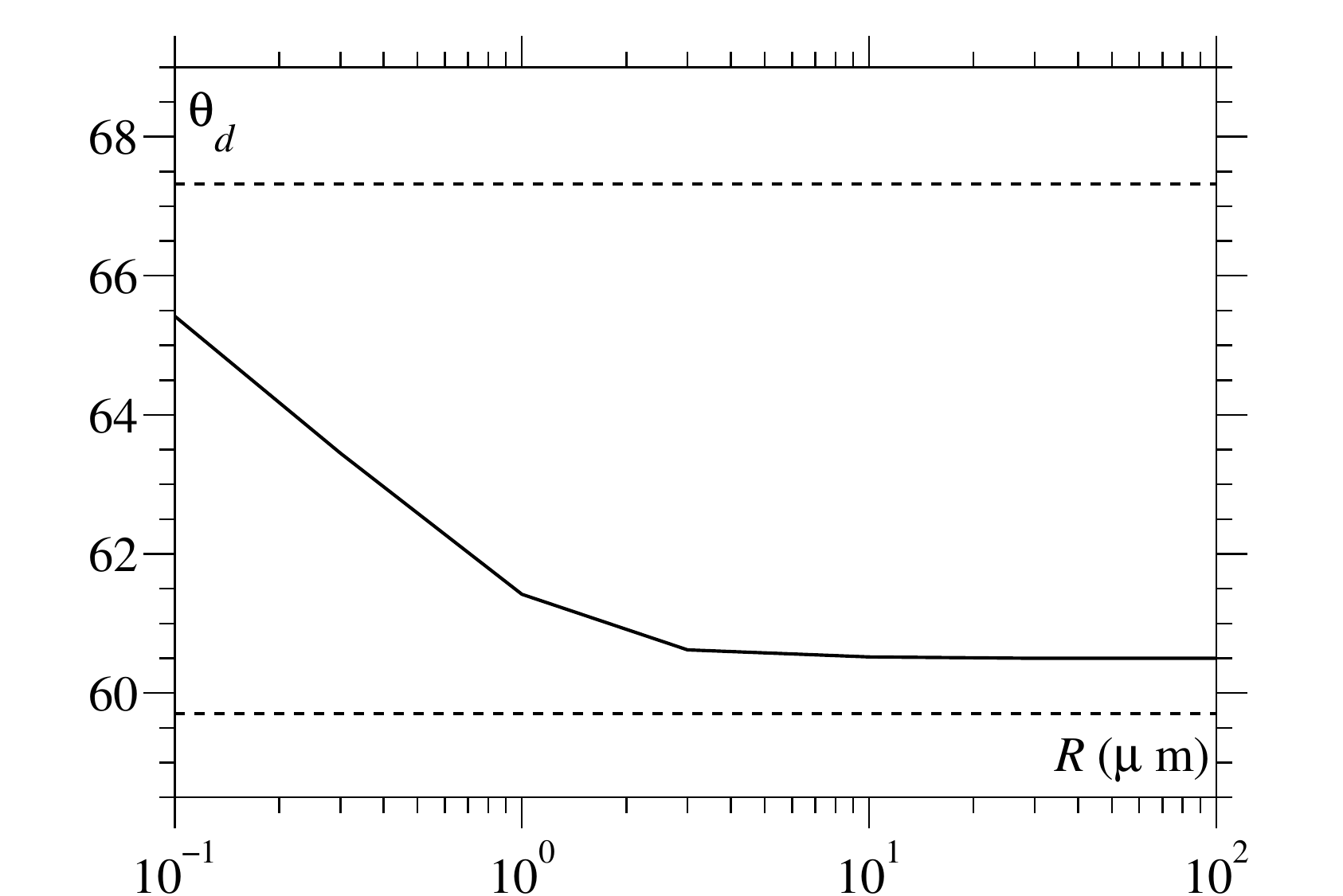}
\caption{Dynamic contact angle $\theta_d$ as the capillary radius $R$ is varied from $100$~nm up to $100~\mu$m.  The lower dashed line is the asymptotic value using $u_f(\theta_d)$ obtained from (\ref{val_moffatt_fs}) whilst the upper one is obtained using $u_f=0$.}
\label{F:theta_sizes}
\end{figure}

The value of the dynamic contact angle is `negotiated' (via Young's
equation) by the surface tensions at the contact line, and, although
the changes in the surface density, and hence surface tension, are
less pronounced than those in surface velocity, it can be seen in
Figure~\ref{F:theta_sizes} that the former still lead to appreciable
changes in the dynamic contact angle as the capillary radius is
reduced. There is a general trend upwards which is consistent with
previous studies \cite{shik07} showing that, if the forming
liquid-solid interface is `starved' of surface mass flux coming from
the free surface, this results in higher dynamic contact angles. In
Figure~\ref{F:theta_sizes}, the lower dashed-line corresponds to the
asymptotic value, whilst the upper dashed-line is the value for the
angle which would be obtained if the velocity in the far field is
taken as $u_f=0$ in the asymptotic formula (\ref{val_cangle}).  Our
curve is wedged between these two potentially limiting values, and
it will be the subject of further work to investigate whether this
simple estimate is actually a bound on the variation of the dynamic
contact angle in this geometry or whether additional factors come
into play in other parameter regimes.
%

\section{Problem B: Unsteady imbibition into a capillary}\label{results3}

Next, the time-dependent imbibition of a liquid into an initially
dry capillary is considered, i.e.\ the case in which a capillary
comes into contact with a reservoir of liquid leading to the fluid
being sucked into the capillary until the meniscus reaches Jurin's
height, at which gravitational forces balance the capillary ones
that drive the flow.  The capillary is aligned vertical to the
gravitational field so that the body force $\mathbf{F} =
-g\mathbf{e}_z$.  It is assumed that the base of the capillary is
maintained at a constant pressure, the atmospheric one, and that the
flow at the inlet is parallel to the walls of the capillary so that
\begin{equation}\nonumber u=0,\qquad p = p_g, \qquad (0<r<1,~z=0).\end{equation}
As an initial condition, we assume that a small amount of liquid is
in the capillary, of height one-tenth of the radius of the capillary
$h=0.1$, that this liquid is at rest, and that the meniscus has not
had time to adjust to its surroundings so that the interface is flat
with $z_1=h$ for $0<r_1<1$. The interface formation variables on the
free surface are taken to be the equilibrium values, so that
$\sigma_1=1$, and the liquid-solid interface is assumed not to be
formed yet so that $\rho^s_2 = \rho^s_{(0)}$.  This corresponds to
$\sigma_2 = 0$ and hence is consistent with our assumption of a flat
interface as the Young equation (\ref{P_473}) gives $\theta_d =
90^\circ$.  Thus, as initial conditions we have
\begin{equation}\nonumber\mathbf{u}=0,\qquad \rho^s_{1}=\rho^s_{1e},\qquad \rho^s_2 = \rho^s_{(0)},\qquad h=0.1.\end{equation}

As a benchmark test case, parameters are obtained from the
experiments in \cite{joos90}. In particular, we compare our
simulations to experimental data in Figures $6,~8$ of \cite{joos90},
where the imbibition of a silicon oil of density $\rho = 9.8\times
10^2$~kg~m${}^{-3}$, viscosity $\mu=12.25$~kg~m${}^{-1}$~s${}^{-1}$
and a surface tension with air of $\sigma =2.13\times
10^{-2}$~kg~s${}^{-2}$ into capillaries with perfectly wettable
walls, $\theta_e=0^\circ$, was studied.  The two capillary radii studied were
$R=0.036$~cm and $R=0.074$~cm.  Rather than fitting the interface
formation parameters for this liquid-solid combination, we use, as a
first approximation, estimates for the model's parameters considered
previously.  Using these parameters and taking as a characteristic
velocity $U=\sigma/\mu = 1.74\times10^{-3}$m~s$^{-1}$, so that
$Ca=1$, we have for $R=0.036$~cm that
\begin{align}\notag
Re& = 5\times10^{-5},\quad St = 5.8\times10^{-2}, \quad \epsilon = 4.1\times 10^{-4}, \quad \bar{\alpha} =5.6\times10^{-6},\quad \bar{\beta}=1.8\times 10^{5},\\ Q&=1.3\times 10^{-2},\quad \rho^s_{1e} = 0.6,\quad \rho^s_{2e}=1.4,\quad \lambda = 2.5,\quad \theta_e=0^\circ,
\end{align}
whilst for $R=0.074$~cm
\begin{equation}\notag
Re = 1\times10^{-4},\quad St = 2.5\times10^{-1},\quad \epsilon = 2\times 10^{-4}, \quad \bar{\alpha} =2.7\times10^{-6},\quad \bar{\beta}=3.7\times 10^{5},
\end{equation}
and the remaining parameters are the same.

A relevant reference case for comparison is the so-called
Lucas-Washburn approximation \cite{washburn21}, often used to
analyze experiments \cite{zhmud00} on the unsteady imbibition into
capillary tubes. In this approximation, the Navier-Stokes equations
are volume averaged to obtain the meniscus height $h$ as a function
of time under the assumptions that (a) Poiseuille flow occurs
throughout the entire capillary, (b) that the driving force for the
flow is the difference $p_g-p_c$ in pressure between the pressure at
the inlet $p_g$ and the pressure at the meniscus $p_c$, (c) the
meniscus is a spherical cap at all times so that
$p_g-p_c=2\sigma\cos\theta_d/R$ and, in the original form of the
theory, (d) the contact angle is assumed to be equal to its
equilibrium value the entire time, so that, in our case,
$\theta_d\equiv0^\circ$ and hence $p_g-p_c=2\sigma/R$. When there is
no gravitational field, the meniscus continues to propagate
indefinitely into the capillary, being slowed by the gradually
increasing viscous drag as the distance between the meniscus and the
inlet increases.  If gravity is taken into account, then the
meniscus will eventually reach its equilibrium Jurin height
$h_\infty$ calculated from balancing the driving force $p_g-p_c$
with the body force on the fluid $\rho g h_\infty$, i.e. $h_\infty =
2\sigma/(\rho R g)$. Inertial effects can also be easily included,
resulting in a nonlinear ODE to be solved for the meniscus height
which in dimensional form is given by
\begin{equation}\label{washburn}
\rho\left[hh'' + \left(h'\right)^2\right] = -\frac{8\mu}{R^2}hh' + \frac{2\sigma}{R} - \rho g h,
\end{equation}
where the prime denotes differentiation with respect to time and, to
compare with our calculations, the initial conditions at $t=0$ are
$h=0.1R,~h'=0$.  The terms on the left hand side of (\ref{washburn})
are inertial forces, the first term on the right hand side
represents viscous forces, the second one is the capillary driving
force and the last term is the body force of gravity.  There are
papers that have been concerned with improving the basic
Lucas-Washburn model \cite{joos90,popescu08,martic02}, but here we
just use the original model for comparison as a reference point.

As the Lucas-Washburn model predicts that the meniscus is always a spherical cap with, in the case of perfect wetting, $\theta_d=0^\circ$, we take the predicted averaged height $h$, and then fit the spherical cap to it ensuring mass conservation which results in the apex at $z_a = h-2R/3$ and the contact line at $z_c = h+R/3$.  This means that at the instant when the simulation starts, the interface jumps from its initially flat shape to a spherical cap which extends both above and below the initial interface, i.e.\ partly back into the reservoir.  This is a known unphysical feature of the model \cite{zhmud00}, and forces it to predict infinite speeds as $t\rightarrow0$.

\begin{figure}
\centering
\includegraphics[scale=0.45]{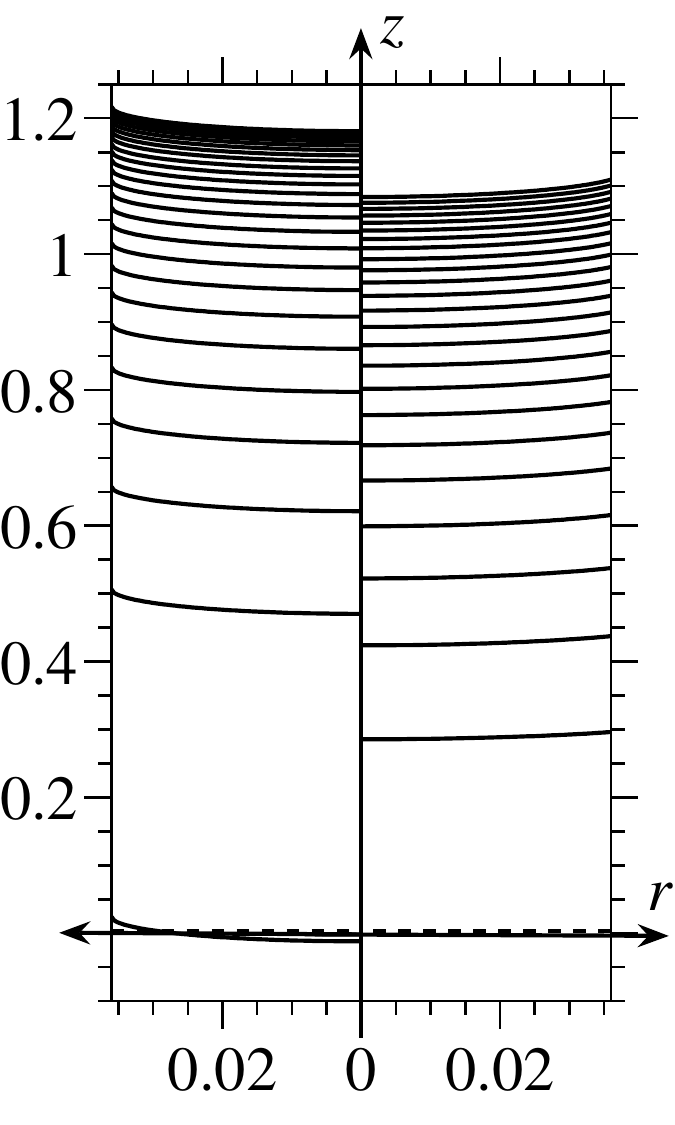}
\includegraphics[scale=0.45]{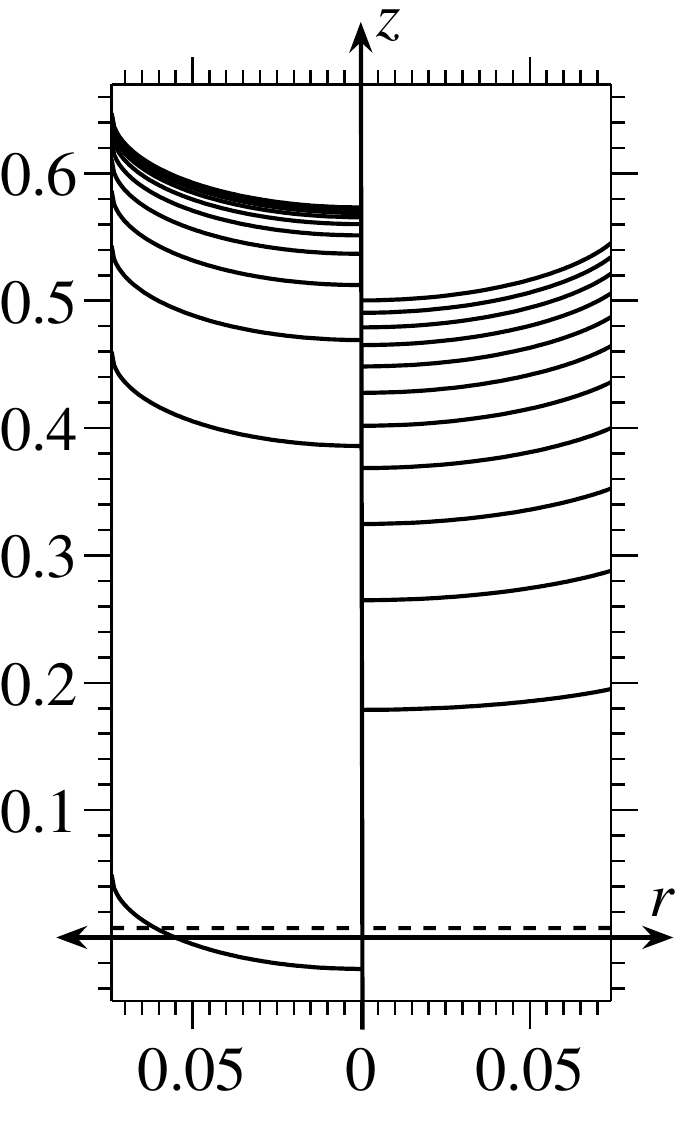}
\caption{Snapshots of the free-surface shape taken at constant time intervals as predicted by the Lucas-Washburn model (left of each figure) compared to our simulation (right of each figure), not to scale. Left:  Capillary of radius $R=0.036$cm with snapshots at intervals of 100 seconds for the first 2500 seconds. Right: Capillary of radius $R=0.074$cm with snapshots at intervals of 50 seconds for the first 600 seconds.}
\label{F:cap_evo}
\end{figure}

In Figure~\ref{F:cap_evo}, the predictions of the two models are compared for each capillary.  The Lucas-Washburn model always predicts a much faster uptake of the liquid, whilst in our simulation the meniscus is slowed as it takes time for the dynamic contact angle to decrease, and hence the capillary pressure to increase, from an initially flat interface, so that no unphysical jumps in variables are observed in the initial stages. Notably, at the end of the time periods considered, the Lucas-Washburn result predicts that the menisci are close to their Jurin height, $h_\infty=1.23$~cm and $h_\infty=0.6$~cm, respectively, whilst the results from our simulation are such a distance away from $h_\infty$ that a comparison of the results with experimental ones should be able to easily distinguish between the two models' predictions.

\begin{figure}
\centering
\includegraphics[scale=0.27]{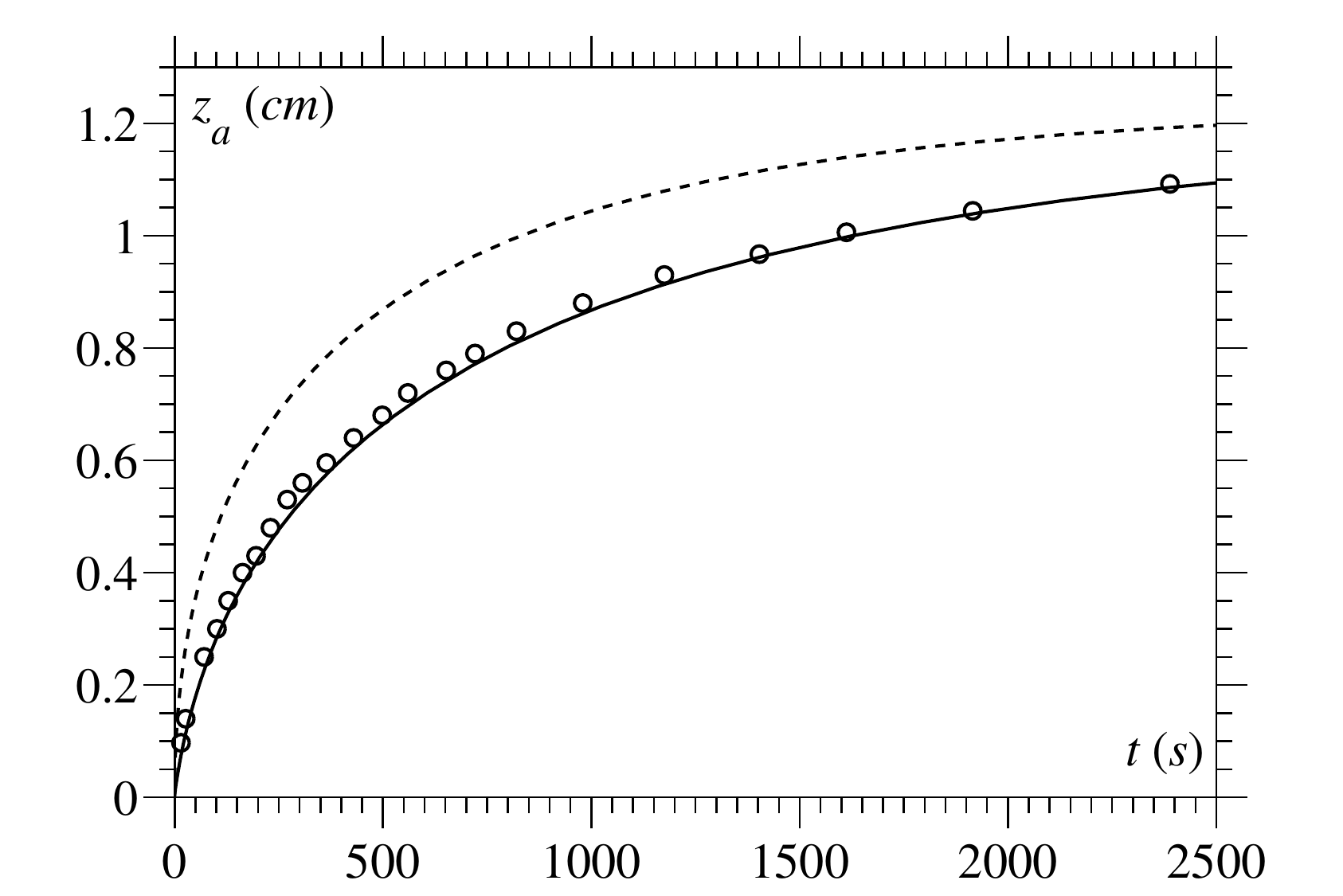}
\includegraphics[scale=0.27]{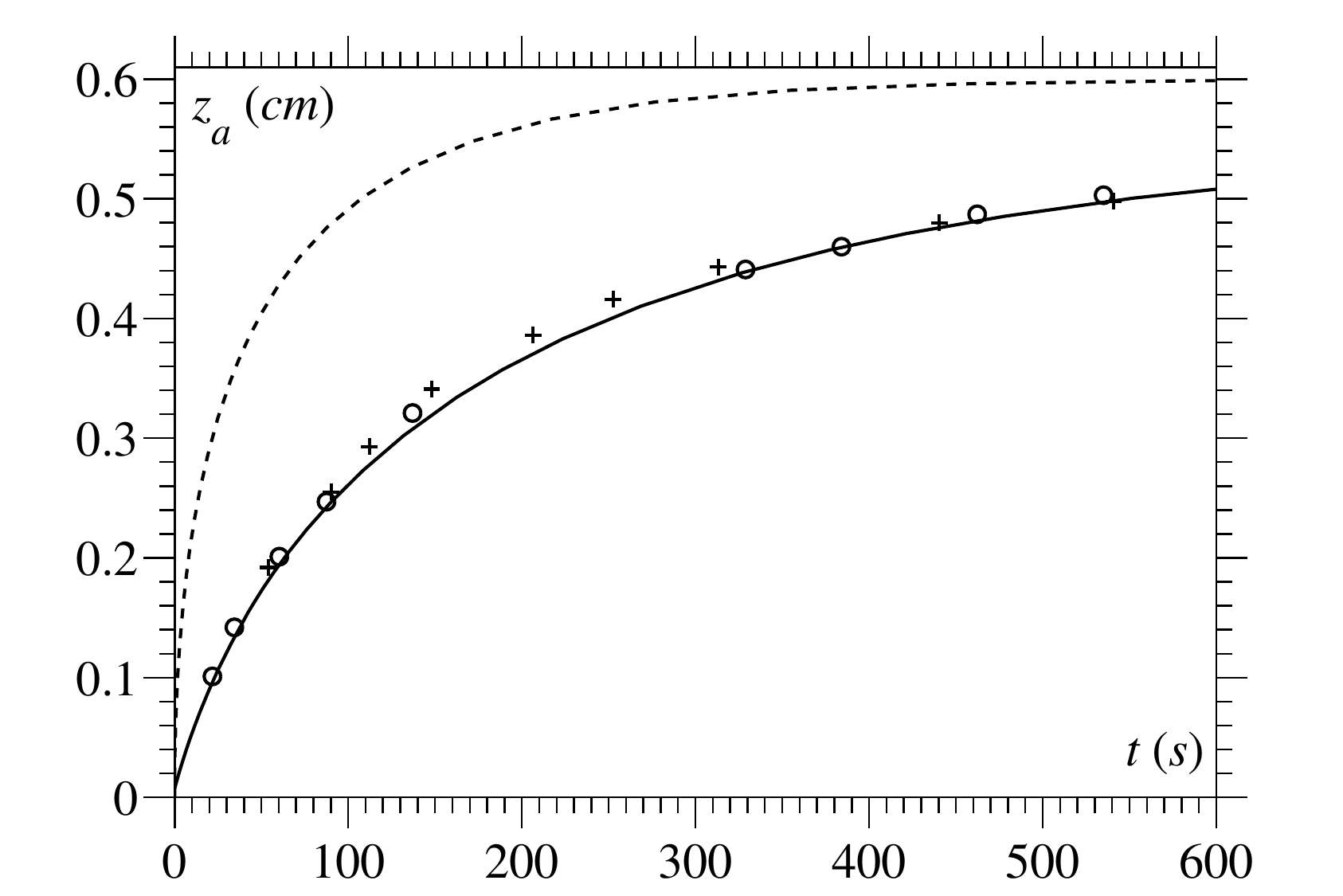}
\caption{Apex height of the meniscus (in cm) as a function of time (in seconds).  Experimental results \cite{joos90} are circles and crosses (which show a repeated experiment to demonstrate the reproducibility of the data), our simulations are the solid lines whilst the dashed line is the Lucas-Washburn result.}
\label{F:joos}
\end{figure}

The results from our simulations are compared to both the experimental results of \cite{joos90} as well as to the Lucas-Washburn curve in Figure~\ref{F:joos}, where two sets of data have been shown in the second graph to illustrate the repeatability of the experimental results.  Considering that no parameters have been fitted, our computations are seen to approximate the data exceptionally well and are vastly superior to the Lucas-Washburn curve which hugely overpredicts the speed at which the capillary uptakes the fluid. The inadequacy of the Lucas-Washburn equation has been recognized before and attributed to the need to account for a variable dynamic contact angle \cite{joos90}. In a forthcoming article, our code will be used to assess how all the models proposed for imbibition into a capillary perform, establish the individual effects contributing to the meniscus' behaviour and, since none of the models account for the effect of the capillary size on the contact angle, ascertain bounds of applicability of the models previously proposed.  Here, we are satisfied with the current result which demonstrates the code's ability to describe experimental data with ease.

\begin{figure}
\centering
\includegraphics[scale=0.27]{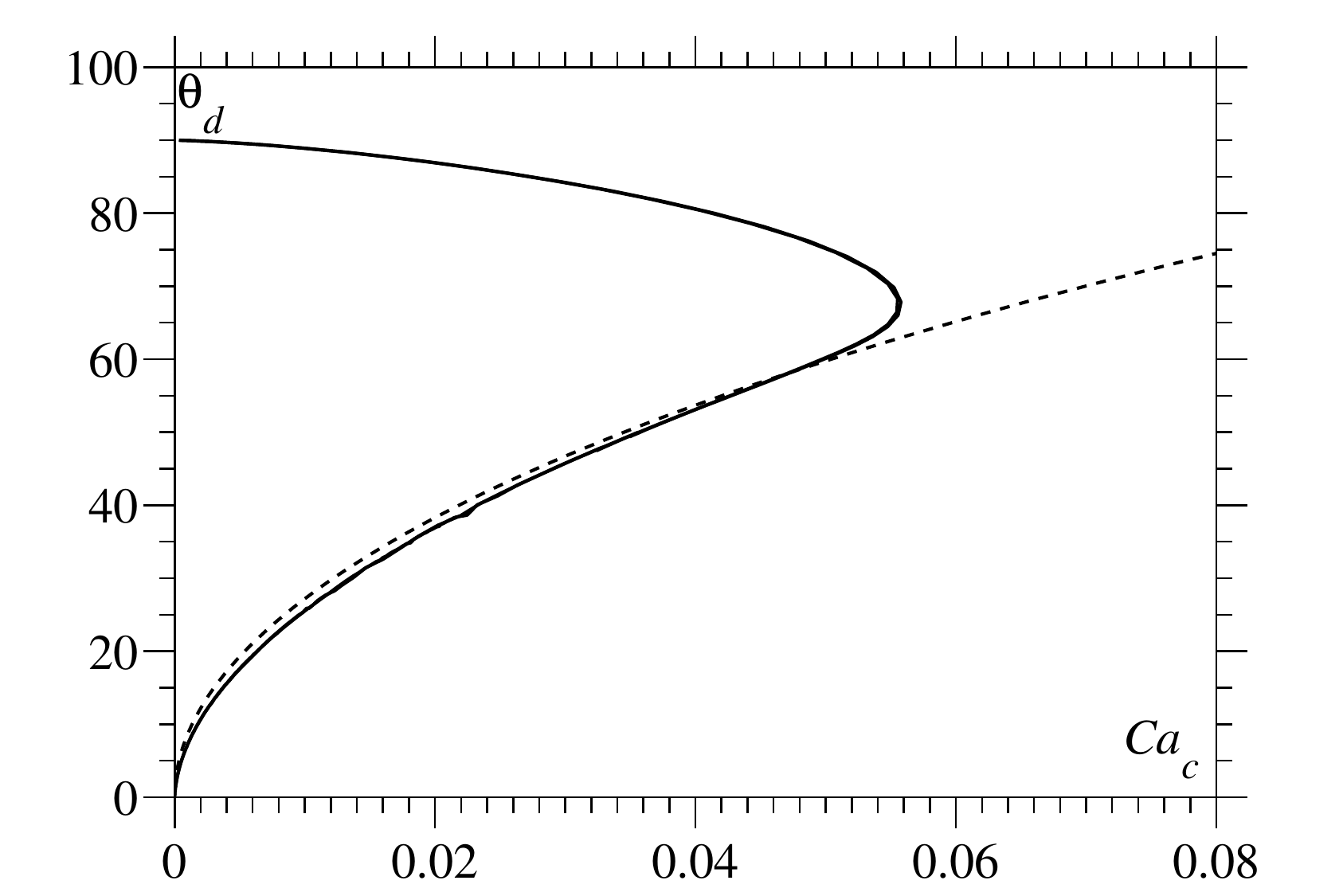}
\caption{The relationship between the capillary number based on the contact-line speed ($Ca_c$) and dynamic contact angle ($\theta_d$) for the two simulations considered, whose curves are graphically indistinguishable, compared to the asymptotic line (dashed)}
\label{F:joos_sa}
\end{figure}

In Figure~\ref{F:joos_sa}, the relationship between capillary number
based on the contact line speed and contact angle is presented.
Notably, as the two curves are for the same fluid, it is only the
variation of the contact-line speed that causes the variation of the
capillary number, so that the latter can be viewed just as the
dimensionless contact-line speed. Surprisingly, the curves for the
two different capillary sizes are graphically almost
indistinguishable. One may naively think that this is an argument in
favour of using a speed-angle formula; however, as one can see from
Figure~\ref{F:joos_sa}, the velocity dependence of the contact angle
is multivalued so that it is not a function.  For example, at
$Ca_c=0.03$ the contact angle $\theta_d=45.7^\circ$ or $84.2^\circ$.
Furthermore, the upper branch of the curve corresponds to a {\it
decrease\/} in the contact angle as the contact-line speed
increases. The upper branch of the curve is far away from the
asymptotic line (dashed line) computed from (\ref{val_cangle}) whilst the lower one is very close to
it (the asymptotic value at $Ca_c=0.03$ is $\theta_d=46.4^\circ$).

The reason for this double-valuedness of the dynamic contact angle
versus the contact-line speed plot can be explained very simply. The
liquid started moving from rest and at $t=0$ the contact angle had
the value of $\theta_d=90^\circ$. Therefore, if the model does not
produce unphysical singularities, with dynamic characteristics
experiencing instant jumps, then both the contact-line speed and the
contact angle have to evolve continuously from their initial values,
in our case, to $0$ and $90^\circ$, respectively. Hence there
appears a branch in the speed-angle plot stemming from the initial
state and leading towards the quasi-steady state. This manifestation in
the speed-angle relationship of the initial conditions rules out in
principle the possibility of treating the dynamic contact angle as a
prescribed input into the model. It is important to emphasize that
this is a general argument based on the necessity not to have
unphysical singularities in the mathematical model of a physical
phenomenon, and the plot produced by using the interface formation
model is merely an illustration that a model regarding the dynamic
contact angle as part of the solution satisfies the above
requirement in a natural way.

Following the initial period, where the dynamic contact angle
evolves from its initial value and the contact-line speed increases,
the flow `settles' into the `regular' quasi-steady pattern. There
the velocity-dependence of the contact angle falls onto the one
given by the asymptotics outlined in \S\ref{asymptotics}. It is
important to point out here that, although the radii of the
capillaries considered in this section are large enough for the
influence of $R$ on the flow to be negligible, the unsteadiness of
the process makes it necessary to use the interface formation model
in its entirely and not just the asymptotic result for the
speed-angle relationship that follows from the model in the limiting
case. The latter becomes applicable at the later stages of the
imbibition, whereas at the onset both (\ref{val_cangle}) and any of
the velocity-dependencies of the contact angle used in the
conventional models will make the contact-line speed jump from
zero straight up to a large finite value.

\section{Concluding remarks}\label{conclusions}

The necessity to develop a robust scheme of incorporating the model
of dynamic wetting as an interface formation process into a
numerical code is dictated by the need to describe the
experimentally observed effect of the dynamic contact angle being
dependent on the entire flow
field/geometry. This effect becomes more pronounced as the size of
the system goes down and has far-reaching implications for a host of
emerging technologies.

In this paper, such a scheme is described and the resulting
numerical code is thoroughly validated. The exposition is
purposefully laid out in a `digestible' manner, with specific
details given in the Appendix, allowing an interested reader to
reproduce the code and compare the results with the benchmark
calculations provided and, together with our previous publication
\cite{sprittles11c}, giving the first guide for simulating fluid
flows with forming interfaces. A particular emphasis is put on the
estimates for the required mesh resolution and the physics
associated with the smallest length scales for different regimes.
These estimates as well as the benchmark calculations remain valid
independently of the particular numerical scheme used.

The considered test problem of a steady imbibition into a capillary
highlights a unique feature of the interface formation model, namely
that the dynamic contact angle is an {\it output}, part of the
solution, as opposed to being a prescribed function of the
contact-line speed and other parameters. For this test problem, our
computations show that, as the radius $R$ of the capillary is
reduced, the velocity-dependence of the dynamic contact angle
becomes influenced by $R$, with the dynamic contact angle for the
same contact-line speed being higher for smaller radii. This is an
essentially new physical effect predicted using the interface
formation model. Given that, when the influence of the flow
field/geometry was first discovered experimentally \cite{blake94},
the term `hydrodynamic {\it assist\/} of dynamic wetting' was coined
to emphasize the possibility of {\it lowering\/} the dynamic contact
angle by manipulating the flow field, here we may term the newly
discovered phenomenon as `hydrodynamic {\it resist\/} to dynamic
wetting'. The focus of this paper is on numerical implementation of
the interface formation model and so we postpone the full
exploration of the effect of `hydrodynamic resist' to a future
publication in which our framework will be compared to experiments,
e.g.\ \cite{sobolev00}, where resist-like behaviour has been
observed but is yet to be theoretically described.

The problem of unsteady imbibition into a capillary demonstrates the
above unique feature of the interface formation model from a
different perspective. This problem shows that the
velocity-dependence of the dynamic contact angle bears the influence
of the initial conditions even in the situation where otherwise the
effect of the flow field/geometry would be negligible. In
particular, the dynamic contact angle can {\it decrease\/} with the
increasing contact-line speed. Qualitatively, this possibility
follows from a general argument about the continuity of variation of
the flow characteristics necessary to avoid unphysicality, and the
interface formation model provides a quantitative illustration of
how this unphysicality is removed.

The validation of our numerical code against experimental data
demonstrating excellent agreement without the need to fit any
parameters opens up two lines of inquiry. First, it would be
interesting to consider how scaling down of the system towards
submicron and nanoscales affects the flow and, by comparing the
results with experimental data, identify the scales where specific
`nanoeffects' become important. Secondly, the code can be used to
validate different simplified models for the capillary imbibition by
identifying the regions of parameters where different effects play
the dominant role.

What has been shown is only a small part of the capabilities of the
interface formation model fully incorporated into a numerical
framework. In \cite{sprittles_chem}, the dynamics of freely
oscillating liquid drops and of drops impacting and spreading on
chemically patterned surface is considered, demonstrating the
framework's capability to simulate flows in which changes in
free-surface shape are extreme. The ability of the interface
formation model to incorporate additional physical effects is shown
in \cite{sprittles07,sprittles09}, where a viscous flow over a solid surface
whose wettability varies has been described, with the variations in
wettability affecting the flow even in the absence of any meniscus.
Further influences on the wetting dynamics, such as those associated
with the solid substrate's properties, e.g.\ its porosity
\cite{shik_porous}, or thermal and electromagnetic effects, can be
added as required, safe in the knowledge that the underlying
framework's accuracy has been completely confirmed.

\section{Appendix}

In \cite{sprittles11c}, a user-friendly step-by-step guide to the finite element simulation of dynamic wetting flows using the conventional models is given.  Many of the aspects dealt with in that paper remain unchanged when we introduce the interface formation model into this framework, and these are not repeated here.  In particular, we do not repeat: the re-writing of tensorial expressions in terms of a particular coordinate system, the mesh design which utilizes the bipolar coordinate system, the mapping of integrals onto a master element, the calculation of the corresponding integrals using Gaussian quadrature or the solution procedure based on the Newton-Raphson method.  Here, the main change is that we are now solving many more equations along the boundaries of the domain, and, consequently, the element-level residuals change from those in \cite{sprittles11c} and are thus listed in full. Therefore, in parallel with \cite{sprittles11c}, this guide enables one to easily reproduce all the results presented in the paper. For a more detailed exposition about using the FEM to solve fluid flows the reader is referred to \cite{gresho1,gresho2}, whilst specific information on free surface flows, including alternative mesh designs, can be found in \cite{kistler84,tezduyar92,christodoulou92,ramanan96,cairncross00}. We have provided information of our own specific implementation of the framework described in the main body of the paper.  Many alternatives exist (e.g.\ different element types) which may produce equally accurate results but here we present what we have used and tested, which for a `practitioner' gives a ready-to-use algorithm.

Two-dimensional and three-dimensional axisymmetric flows are considered simultaneously by using a variable $n$ which takes the values $n=0$ in the former case and $n=1$ in the latter. Two-dimensional flow occurs in a Cartesian coordinate system $(r,z)$ whilst axisymmetric flow is in a cylindrical polar coordinate system $(r,z,\vartheta)$, where $r$ is now the radial coordinate and $\vartheta$ is the azimuthal angular coordinate around the $z$-axis about which symmetry is assumed (e.g.\ Figure~\ref{F:domain}). For both coordinate systems, the governing equations are solved in the $(r,z)$-plane.  In this coordinate system, the surface velocity is $\mathbf{v}^s = v^s_t\mathbf{t} + v^s_n\mathbf{n}$, and $\mathbf{v}^s_{||} = v^s_t\mathbf{t}$ as, in the $n=1$ case, $\mathbf{v}^s_{||}\cdot\mathbf{e}_\theta = 0$.  Then, notably, the divergence of the tangential component of a surface vector is given by
\begin{equation}
\nabla^s\cdot\mathbf{a}^s_{||} = \pdiff{a^s_t}{s} + \frac{n a^s_{t.r}}{r}, \qquad a^s_{t.r}=(\mathbf{a}^s\cdot\mathbf{t})(\mathbf{t}\cdot\mathbf{e}_r)
 \end{equation}
where we have used that $\mathbf{t}\cdot\pdiff{\mathbf{t}}{s}=0$.

\subsection{Finite element procedure}\label{procedure}

The finite element method presented here is not restricted to the capillary geometry and our only assumption about the free surface is that in the $(r,z)$-plane, i.e.\ the computational domain, it is a line parameterized by arclength $s$ with position defined by a function of one variable $h=h(s)$ which depends on the particular mesh design. In other words, to determine the free-surface shape we must determine the value of $h_\textrm{i}$ at every node on the free surface $\textrm{i}=1,\hbox{...},N_1$.

The domain is tesselated into $e=1,\hbox{...},N_e$ non-overlapping subdomains of area $E_e$, the \emph{elements}, which in our case will be two-dimensional curved-sided triangles whose positions are defined by the spines of the mesh and hence by the values of free surface unknowns $h_\textrm{i}$.  The boundary of the computational domain is composed of one-dimensional elements of arclength $s_e$ formed from the sides of the bulk elements which are adjacent to the boundary.  We refer to element-level quantities as \emph{local} and those defined over the entire domain, which were used in \S\ref{FEM}, as \emph{global}.   Each element contains a set number of local nodes, and it is the set of all local nodes which form the global nodes referred to in \S\ref{FEM}.  We have used Roman letters $(\textrm{i},\textrm{j})$ for global quantities and italicized letters $(i,j,k,l)$ for local ones. In the FEM, one must store a function $I$ which relates each local node $i$ in each element $e$ to its global node number $\textrm{i}$ so that ${\textrm{i}}=I(e,i)$.

The global functions and residuals are constructed in a piecewise manner from local quantities, by ensuring that, as required in the FEM, the
interpolating function associated with a given node is constructed to be zero outside the elements to which that node belongs.  This allows us to derive expressions for the residuals in an arbitrary element which, after summing up contributions from every element in the domain, will form a set of algebraic equations whose coefficients are integrals to be calculated by Gaussian integration as shown in \cite{sprittles11c}.

To avoid having to construct each local interpolating function in an element-specific manner, they are constructed on a master element with coordinates ($\xi,\eta$), so that $\phi_{i} = \phi_{i}(\xi,\eta),~\psi_{i} = \psi_{i}(\xi,\eta)$.  Mixed interpolation is achieved, to ensure the Ladyzhenskaya-Babu\u{s}ka-Brezzi \cite{babuska72} condition is satisfied, by using the V6P3 Taylor-Hood triangular element which approximates velocity by means of bi-quadratic local interpolating functions $\phi_{i}(\xi,\eta)~~(i=1,\hbox{...},6)$ and pressure using bi-linear ones $\psi_{i}(\xi,\eta)~~(i=1,\hbox{...},3)$, with explicit expressions given in \cite{sprittles11c}.
Over an element $e$ the following functional forms are used
\begin{equation}\nonumber
(u,w) = \sum^{6}_{j=1} (u_{j},w_{j})\phi_{j}(\xi,\eta),\qquad p
= \sum^{3}_{j=1} p_{j}\psi_{j}(\xi,\eta),\qquad (r,z)
= \sum_{j=1}^6 (r_j,z_j)\phi_{j}(\xi,\eta).
\end{equation}

By ensuring that elements whose sides form the free surface always have local nodes $2,6,3$ associated with them and that those on the solid surface are always $j=1,5,2$, we can define surface interpolating functions $\phi_{1,j}(\eta)=\phi_j(\xi,\eta=-1)$ and $\phi_{2,j}(\eta)=\phi_j(\xi=-1,\eta)$, respectively. Then, a surface variable $a^s$ and the free-surface shape $(r_1(h),z_1(h))$, which is determined from the free surface unknowns $h$, are approximated on the surfaces quadratically as
\begin{equation}\nonumber
a^s_1 = \sum_{j=2,6,3}a^s_{1,j}\phi_{1,j}(\xi),\qquad (r_1,z_1) = \sum_{j=2,6,3}(r_{1,j},z_{1,j})\phi_{1,j}(\xi),\qquad a^s_2 = \sum_{j=1,5,2} a^s_{2,j}\phi_{2,j}(\eta).
\end{equation}

The global residuals (the $R_{\textrm{i}}$'s) presented in \S\ref{FEM}, which involve integrals over the entire domain, are formed by summing up local residuals (the $R_{e,i}$'s), denoted with a subscript $e$ to indicate which element the local residual is calculated in, obtained by integrating over each of the $N_e$ elements in the global domain. Let $\bar{N}_1$ be the set of numbers of those elements that form part of the free surface and $\bar{N}_2$ be the numbers of elements forming a part of the liquid-solid interface.  Then, the continuity of mass, momentum equations\footnote{Note the typo in \cite{sprittles11c} where the limit for $i$ in $R^{C}_{{\textrm{i}}}$ and $R^{M,\alpha}_{{\textrm{i}}}$ should be $3$ and $6$, respectively}, kinematic equation on the free surface and impermeability equation at the liquid-solid interface are:
\begin{align}\label{summ_resid}
R^{C}_{{\textrm{i}}} = \mathop{\sum_{e=1}^{N_e} \sum_{i=1}^{3}}_{\textrm{i}=I(e,i)} R^{C}_{e,i},\quad R^{M,\alpha}_{{\textrm{i}}} = \mathop{\sum_{e=1}^{N_e} \sum_{i=1}^{6}}_{\textrm{i}=I(e,i)} R^{M,\alpha}_{e,i}, \quad  R^{K}_{{\textrm{i}}} = \mathop{\sum_{e=1}^{N_e} \sum_{i=2,6,3}}_{\textrm{i}=I(e,i),~e\in \bar{N}_1} R^{K}_{e,i}, \quad R^{I}_{\textrm{i}} = \mathop{\sum_{e=1}^{N_e} \sum_{i=1,5,2}}_{\textrm{i}=I(e,i),~e\in \bar{N}_2} R^{I}_{e,i},
\end{align}
where the constraint under the summation symbols ensures that, through $\textrm{i}=I(e,i)$, the local residuals are assigned to the correct global ones and, via $e\in \bar{N}$, that the surface equations are only evaluated when an element is adjacent to the boundary of the domain. In addition, we now have residuals from the interface formation equations given by
\begin{align}\label{summ_resid1}
R^{v^s_{1t}}_{{\textrm{i}}} = \mathop{\sum_{e=1}^{N_e} \sum_{i=2,6,3}}_{\textrm{i}=I(e,i),~e\in \bar{N}_1} R^{v^s_{1t}}_{e,i}, \quad
R^{v^s_{1n}}_{{\textrm{i}}} = \mathop{\sum_{e=1}^{N_e} \sum_{i=2,6,3}}_{\textrm{i}=I(e,i),~e\in \bar{N}_1} R^{v^s_{1n}}_{e,i}, \quad
R^{\rho^s_1}_{{\textrm{i}}} = \mathop{\sum_{e=1}^{N_e} \sum_{i=2,6,3}}_{\textrm{i}=I(e,i),~e\in \bar{N}_1} R^{\rho^s_1}_{e,i},\quad
R^{\sigma_1}_{{\textrm{i}}} = \mathop{\sum_{e=1}^{N_e} \sum_{i=2,6,3}}_{\textrm{i}=I(e,i),~e\in \bar{N}_1} R^{\sigma_1}_{e,i},\\
R^{v^s_{2t}}_{{\textrm{i}}} = \mathop{\sum_{e=1}^{N_e} \sum_{i=1,5,2}}_{\textrm{i}=I(e,i),~e\in \bar{N}_2} R^{v^s_{2t}}_{e,i}, \quad
R^{v^s_{2n}}_{{\textrm{i}}} = \mathop{\sum_{e=1}^{N_e} \sum_{i=1,5,2}}_{\textrm{i}=I(e,i),~e\in \bar{N}_2} R^{v^s_{2n}}_{e,i}, \quad
R^{\rho^s_2}_{{\textrm{i}}} = \mathop{\sum_{e=1}^{N_e} \sum_{i=1,5,2}}_{\textrm{i}=I(e,i),~e\in \bar{N}_2} R^{\rho^s_2}_{e,i},\quad
R^{\sigma_2}_{{\textrm{i}}} = \mathop{\sum_{e=1}^{N_e} \sum_{i=1,5,2}}_{\textrm{i}=I(e,i),~e\in \bar{N}_2} R^{\sigma_2}_{e,i},
\end{align}
and the residual from Young's equation $R^{Y}$ which is only applied at the contact line node.

Taking the equations of \S\ref{FEM} over an arbitrary element $e$, with area $E_e$ and boundary $s_e$, we now derive the local residuals required in (\ref{summ_resid}) and (\ref{summ_resid1}).

\subsection{Element-level residuals}\label{theequations}

For an arbitrary element $e$, with local nodes $i=1,...,6$, contributions from the bulk of the element $E_e$ to the momentum residuals are
\begin{align}\nonumber
&R^{M,r}_{e,i} =  Re~\left[M_{ij}\diff{u_j}{t} + A_{ij}\left(u_j - \diff{r_j}{t}\right)\right]+K^{11}_{ij} u_j + K^{12}_{ij}w_j + C^{1}_{ik}p_k,  \\ \label{sum_ns2}
&R^{M,z}_{e,i} =  Re~\left[M_{ij}\diff{w_j}{t} + A_{ij}\left(w_j - \diff{z_j}{t}\right)\right]+K^{21}_{ij} u_j + K^{22}_{ij}w_j + C^{2}_{ik}p_k + St~G_i, \\\notag
\qquad &j=1,...,6 \quad k=1,...,3,
\end{align}
where summation over repeated indices is henceforth assumed.  The $M$ terms are the mass matrices, the $A$ terms are from the nonlinear convective terms, $K$ terms are associated with viscous forces, the $C$ terms are with pressure forces and $G$ terms correspond to the body force acting on the liquid due to gravity.  Temporal derivatives will be considered in \S\ref{temporal}.

For $i=1,...,3$ we have the incompressibility residuals
\begin{equation}\nonumber
R^{C}_{e,i} = C^{1}_{ji} u_j + C^{2}_{ji} w_j \qquad j=1,...,6  .
\end{equation}

If an element $e$ forms a part of the free surface, $e\in \bar{N}_1$, then for $i=2,6,3$, i.e.\ the free surface nodes, there are additional contributions to the momentum equations from capillary stress terms, so that
\begin{align}\nonumber
R^{M,r}_{e,i} = &R^{M,r}_{e,i} + \hbox{$\frac{1}{Ca}$} F^{1}_{ij}\sigma_{1,j} - In^1_{ij} p_{g,j} \qquad j=2,6,3,  \\ \label{sum_fs2}
R^{M,z}_{e,i} = &R^{M,z}_{e,i} + \hbox{$\frac{1}{Ca}$} F^{2}_{ij}\sigma_{1,j} - In^2_{ij} p_{g,j} \qquad j=2,6,3,
\end{align}
where $p_{g,j}$ is the nodal value of the gas pressure, which in this work has been assumed to remain constant.  If an element $e$ forms the part of the free surface, $e\in\bar{N}_1$, adjacent to the contact line, so that local node $i=2$ is the contact-line node, then there is an additional contribution
\begin{align}\nonumber
R^{M,r}_{e,2} = &R^{M,r}_{e,2} +  \hbox{$\frac{1}{Ca}$}  T^{1}\sigma_{1,i=2}, \\ \label{sum_theta2}
R^{M,z}_{e,2} = &R^{M,z}_{e,2} +  \hbox{$\frac{1}{Ca}$}  T^{2}\sigma_{1,i=2},
\end{align}
which is where the contact angle is imposed into the weak formulation.  The contact angle is determined from the Young equation, now applied only at the contact-line node, whose residual is given by
\begin{equation}\nonumber
R^Y = \sigma_{1,i=2}\cos\theta_d+\sigma_{2,i=2}.
\end{equation}

Additionally, for $i=2,6,3$ one has the kinematic equation
\begin{equation}\nonumber
R^{K}_{e,i}  = In^{1}_{ij}\diff{r_{1,j}}{t} +  In^{2}_{ij}\diff{z_{1,j}}{t} - I_{ij}v^s_{1n,j} \qquad j=2,6,3.
\end{equation}

If an element $e$ forms a part of the liquid-solid interface, $e\in\bar{N}_2$, then for $i=1,5,2$ there are additional contributions to the momentum equations of
\begin{align}\nonumber
R^{M,r}_{e,i} = &R^{M,r}_{e,i} + \hbox{$\frac{\bar{\beta}}{Ca}$}\left(N^{11}_{ij} (u_j - U_j)   +  N^{12}_{ij}(w_j - W_j)\right) + In^{1}_{ij}\Lambda_j -\hbox{$\frac{1}{2Ca}$} S^{1}_{ij} \sigma_{2,j}\qquad j=1,5,2,  \\ \label{sum_ss2}
R^{M,z}_{e,i} = &R^{M,z}_{e,i} + \hbox{$\frac{\bar{\beta}}{Ca}$}\left(N^{21}_{ij}(u_j - U_j)  +  N^{22}_{ij} (w_j - W_j)\right) + In^{2}_{ij}\Lambda_j -\hbox{$\frac{1}{2Ca}$} S^{2}_{ij} \sigma_{2,j}\qquad j=1,5,2,
\end{align}
where the $N$ terms come from the slip terms in the generalized Navier condition (\ref{ns555}), whilst the $S$ terms are associated with gradients in surface tension in (\ref{ns555}) and the terms containing $\Lambda$ are from the normal stress on the interface. The substrate speed\footnote{Which should also have been present in \cite{sprittles11c}.} has been decomposed into scalar components as $\mathbf{U} = U\mathbf{e}_r + W\mathbf{e}_z$.   Additionally, for $i=1,5,2$ one has the impermeability equation (\ref{w_imp}) in the form
\begin{equation}\nonumber
R^{I}_{e,i}  = I_{ij}v^s_{2n,j} - \left(In^1_{ij}U_{j} + In^2_{ij}W_{j}\right) \qquad j=1,5,2.
\end{equation}

Moving on to the equations of the interface formation model, which do not appear in \cite{sprittles11c}, the Darcy-type equations (\ref{w_vt1}) and (\ref{w_vt2}) give
\begin{align}\nonumber
R^{v^s_{1t}}_{e,i} =  I_{ij} v^s_{1t,j} - \left(It^{1}_{ij}u_j + It^{2}_{ij}w_j\right) -\hbox{$\frac{1+4\bar{\alpha}\bar{\beta}}{4\bar{\beta}}$}D_{ij}\sigma_{1,j}, \qquad i,j=2,6,3,
\end{align}
\begin{align}\nonumber
R^{v^s_{2t}}_{e,i} =  I_{ij} v^s_{2t,j} -\hbox{$\frac{1}{2}$}\left[It^{1}_{ij}(u_j+U_j) + It^{2}_{ij}(w_j+W_j)\right] - \bar{\alpha}D_{ij}\sigma_{2,j}\qquad i,j=1,5,2.
\end{align}

The surface velocity normal to each interface $\gamma=1,2$ is determined from
\begin{align}\nonumber
R^{v_{\gamma n}}_{e,i} =  \left(In^{1}_{ij}u_j + In^{2}_{ij}w_j\right) - I_{ij}v^s_{\gamma n,j} - Q I_{ij}\left(\rho^s_{\gamma, j} - \rho^s_{\gamma e, j}\right) \qquad \gamma=1:~ i,j=2,6,3,\quad\gamma=2:~i,j=1,5,2
\end{align}
where we have allowed for the extension to problems in which $\rho^s_e$ varies spatially.

The terms appearing in the surface mass balance equation (\ref{w_rr1}) are more complicated due to the presence in this equation of temporal derivatives and the divergence of surface vectors.  First, recall that on each surface $\gamma=1,2$ the finite element surface mass balance equation is
\begin{align}\notag
R^{\rho^s_\gamma}_{\textrm{i}} &= \left(R^{\rho^s_\gamma}_{\textrm{i}}\right)_{A_\gamma} + \left(R^{\rho^s_\gamma}_{\textrm{i}}\right)_{C_\gamma} \\ \notag
\left(R^{\rho^s_{\gamma}}_{\textrm{i}}\right)_{A_\gamma} &= \int_{A_\gamma}\left[ - \epsilon\rho^{s}_{\gamma}\left(\mathbf{v}^{s}_{\gamma||}-\mathbf{c}^s_{\gamma||}\right)\cdot\nabla^s\phi_{\gamma,\textrm{i}}+\epsilon\phi_{\gamma,\textrm{i}}\left(\pdiff{\rho^{s}_{\gamma}}{t} + \rho^s_\gamma\nabla\cdot \mathbf{c}^s_{\gamma||} + \rho^s_\gamma(\mathbf{v}^s_\gamma\cdot\mathbf{n}_\gamma)\nabla\cdot\mathbf{n}_\gamma\right)+\phi_{\gamma,\textrm{i}}\left(\rho^{s}_{\gamma}-\rho^{s}_{\gamma e}\right)\right] ~dA_\gamma \\ \label{a_rr}
\left(R^{\rho^s_\gamma}_{\textrm{i}}\right)_{C_\gamma} &=-\int_{C_\gamma}\epsilon\phi_{\gamma,\textrm{i}}\rho^s_\gamma\left(\mathbf{v}^s_{\gamma||}-\mathbf{c}^s_{\gamma||}\right)\cdot\mathbf{m}_\gamma~dC_\gamma \qquad (\textrm{i}=1,...,N_\gamma).
\end{align}
Before deriving the individual terms, it proves convenient to introduce the component of the surface's mesh velocity tangential to that surface as a new variable which we label $c^s_t$.  This is discretized in the usual way on each surface $\gamma=1,2$
\begin{equation}\nonumber
c^s_{\gamma t} = \sum_{j} c^s_{\gamma t,j} \phi_{\gamma,j}\qquad \gamma=1:~ j=2,6,3,\quad\gamma=2:~j=1,5,2.
\end{equation}
This new variable is in the ($r,z$)-plane, so that $c^s_t = \mathbf{c}^s_{||}\cdot\mathbf{t}$ and is thus defined by the finite element equation
\begin{align}\nonumber
R^{c^s_{\gamma t}}_{e,i} = I_{ij}c^s_{\gamma t,j} - \left(It^1_{ij}\diff{r_j}{t} + It^2_{ij}\diff{z_j}{t}\right)\qquad \gamma=1:~ i,j=2,6,3,\quad\gamma=2:~i,j=1,5,2.
\end{align}

The first of the three nonlinear terms in (\ref{a_rr}) is
\begin{align}\nonumber
\int_{s_{\gamma e}}\pdiff{\phi_{\gamma,i}}{s} \rho^s_{\gamma} \left(v^s_{\gamma t}-c^s_{\gamma t}\right) \,r^n\,ds_{\gamma e} = \rho^s_{\gamma,j} \left(v^s_{\gamma t,k} - c^s_{\gamma t,k}\right) \int_{s_{\gamma e}}\pdiff{\phi_{\gamma,i}}{s}\phi_{\gamma,j}\phi_{\gamma, k} \,r^n\,ds_{\gamma e}  =\\ p_{ijk}\rho^s_{\gamma,j} \left(v^s_{\gamma t,k} - c^s_{\gamma t,k}\right) = P_{ij}\left(v^s_{\gamma t,k},c^s_{\gamma t,k}\right)\rho^s_{\gamma,j}.
\end{align}
The second one is
\begin{align}\nonumber
\int_{s_{\gamma e}}\phi_{\gamma,i} \rho^s_{\gamma}\nabla^s\cdot\mathbf{c}^s_{\gamma ||} \,r^n\,ds_{\gamma e} = \rho^s_{\gamma,j} c^s_{\gamma t,k} \int_{s_{\gamma e}}\phi_{\gamma,i}\phi_{\gamma,j}\pdiff{\phi_{\gamma,k}}{s} \,r^n\,ds_{\gamma e} + n\rho^s_{\gamma,j}c^s_{t,k}\int_{s_{\gamma e}}\phi_{\gamma,i}\phi_{\gamma,j}\phi_{\gamma,k}t_r ds_{\gamma,e} = \\ w_{ijk}\rho^s_{\gamma,j}c^s_{\gamma t,k} + n w^r_{ijk}\rho^s_{\gamma,j}c^s_{\gamma t,k} = W_{ij}(c^s_{\gamma t,k})\rho^s_{\gamma,j}.
\end{align}
The third term gives
\begin{align}\nonumber
\int_{s_{\gamma e}}\phi_{\gamma,i} \rho^s_{\gamma}v^s_{\gamma n}\nabla^s\cdot\mathbf{n}_{\gamma} \,r^n\,ds_{\gamma e} = \rho^s_{\gamma,j} v^s_{\gamma n,k}\int_{s_{\gamma e}}\phi_{\gamma,i}\phi_{\gamma,j}\phi_{\gamma, k}\kappa \,r^n\,ds_{\gamma e}  = y_{ijk}\rho^s_{\gamma,j} v^s_{\gamma n,k} = Y_{ij}(v^s_{\gamma n,k})\rho^s_{\gamma,j},
\end{align}
where $\kappa = t_{\gamma r}\pdiff{n_{\gamma r}}{s}+t_{\gamma z}\pdiff{n_{\gamma z}}{s}+n(n_{\gamma r}/r)$ is the curvature.

Finally, this gives the surface mass balance equation as
\begin{align}\nonumber
\left(R^{\rho^s_\gamma}_{e,i}\right)_{A_\gamma} = \epsilon\left[I_{ij}\diff{\rho^s_{\gamma, j}}{t} - P_{ij}\rho^s_{\gamma,j} + W_{ij}\rho^s_{\gamma,j} + Y_{ij}\rho^s_{\gamma,j}\right] + I_{ij}\left(\rho^s_{\gamma,j} -  \rho^s_{\gamma e,j}\right)
\end{align}

Whenever the interface reaches the contact line, where now $\mathbf{m}_1=\mathbf{t}_1$ and $r=r_c$, we have an additional contribution on both the free surface and solid surface sides as:
\begin{align}\nonumber
\left(R^{\rho^s_1}_{\textrm{i}}\right)_{C_{cl}}  = -\epsilon\left[\rho^s_1\left(v^s_{1t,i=2}-c^s_{1t,i=2}\right)r_c^n\right]_{cl};
\qquad \left(R^{\rho^s_2}_{\textrm{i}}\right)_{C_{cl}}  =  \epsilon\left[\rho^s_1\left(v^s_{1t,i=2}-c^s_{1t,i=2}\right)r_c^n\right]_{cl}
\end{align}

Lastly, the equations of state (\ref{w_imp}) are given by
\begin{equation}\nonumber
R^{s_\gamma}_{e,i} = I_{ij}\sigma_{\gamma,j} - \lambda\left(I_{i}- I_{ij}\rho^s_{\gamma, j}\right)
\end{equation}
where $I_i = \sum_{j} I_{ij}$.

\subsubsection{Element-level matrices}\label{thecomponents}

It should be pointed out that, in contrast to \cite{sprittles11c}, in this paper non-dimensional parameters are not contained in the element-level matrices below as they have already been factored out.

The mass matrix is
\begin{equation}\nonumber
M_{ij} = \int_{E_e}\phi_{i}\phi_{j}\,r^n\,dE_e.
\end{equation}
and temporal derivatives are handled in \S\ref{temporal}.

The viscous terms are
\begin{equation}\nonumber
K^{11}_{ij} = 2k_{ij11} + k_{ij22} + 2 n~k^r_{ij},\quad K^{12}_{ij} = k_{ij21},\quad K^{21}_{ij} = k_{ij12},\quad K^{22}_{ij} = k_{ij11}  + 2k_{ij22},
\end{equation}
where
\begin{eqnarray}\nonumber
k_{ijkl} = \int_{E_e} \pdiff{\phi_{i}}{y_k} \pdiff{\phi_{j}}{y_l} \,r^n\,dE_e, \qquad k^r_{ij} =  \int_{E_e} \frac{\phi_{i}\phi_{j}}{r^2} \,r^n\,dE_e,
\end{eqnarray}
with  $y_1 = r,~y_2=z$.   Nonlinear convective terms are
\begin{equation}\nonumber
A_{ij}(u,w) = a_{ijk1}u_{k}+a_{ijk2}w_{k}\qquad k=1,...,6,
\end{equation}
where
\begin{equation}\nonumber
a_{ijkl} = \int_{E_e}\phi_{i} \phi_{k} \pdiff{\phi_{j}}{y_l}  \,r^n\,dE_e.
\end{equation}
The $C$ matrices represent pressure terms whilst their transpose is required in the continuity of mass equation:
\begin{equation}\nonumber
C^{1}_{ij}= -\int_{E_e}\psi_{j}\left(\pdiff{\phi_{i}}{r} + \frac{n\phi_{i}}{r}\right)\,r^n\,dE_e,\qquad
C^{2}_{ij}=-\int_{E_e}\psi_{j}\pdiff{\phi_{i}}{z} \,r^n\,dE_e.
\end{equation}
The gravitational force contributes through the $G$ term and is given by
\begin{equation}\nonumber
G_{i} = \int_{E_e}\phi_{i}\,r^n\,dE_e.
\end{equation}

On a free surface (one-dimensional) element $s_{1e}$, the capillary stress terms are
\begin{equation}\nonumber
F^{1}_{ij}=\int_{s_{1e}} t_{1r}\left(\pdiff{\phi_{1,i}}{s}+\frac{n\phi_{1,i}}{r}\right)\phi_{1,j}
\,r^n\,ds_{1e},\qquad
F^{2}_{ij}=\int_{s_{1e}} t_{1z}\pdiff{\phi_{1,i}}{s}\phi_{1,j} \,r^n\,ds_{1e}.
\end{equation}
The components of the vectors tangential $\mathbf{t}= (t_{\gamma r},t_{\gamma z})$ and normal $\mathbf{n} = (n_{\gamma r},n_{\gamma z})$ to surface $\gamma=1,2$ are found in the usual way, see \cite{sprittles11c}.

At the contact line ($r_c,z_c$) contributions will occur at local node $i=2$ of the first free surface element so that
\begin{equation}\nonumber
T^{1} = \left[\left( t_{2r}\cos\theta_d + n_{2r} \sin\theta_d \right)
r_c^n\right],\qquad T^{2} = \left[\left(t_{2z}\cos\theta_d  + n_{2z}\sin\theta_d
\right)r_c^n\right].
\end{equation}

On a (one-dimensional) element $s_{2e}$ on the liquid-solid interface, the tangential stress terms from the Navier-slip boundary condition for $k,l=1,2$ are
\begin{equation}\nonumber
N^{kl}_{ij}= \int_{s_{2e}} \phi_{2,i}\phi_{2,j} t_{2k}t_{2l}   \,r^n\,ds_{2e},
\end{equation}
with $t_{21}=t_{2r}$ and $t_{22}=t_{2z}$. The terms associated with gradients in surface tension are given by
\begin{equation}\nonumber 
S^{1}_{ij}=\int_{s_{2e}}\phi_{2,i}\pdiff{\phi_{2,j}}{s} t_{2r} \,r^n\,ds_{2e},\qquad
S^{2}_{ij}=\int_{s_{2e}}\phi_{2,i}\pdiff{\phi_{2,j}}{s} t_{2z} \,r^n\,ds_{2e},
\end{equation}

On a surface $\gamma=1,2$ the following other expressions have been used
\begin{equation}\nonumber
I_{ij} = \int_{s_{\gamma e}} \phi_{\gamma,i}\phi_{\gamma,j} \,r^n\,ds_{\gamma e}.
\end{equation}
\begin{equation}\nonumber 
In^{1}_{ij}=\int_{s_{\gamma e}}\phi_{\gamma,i}\phi_{\gamma,j} n_{\gamma r} \,r^n\,ds_{\gamma e},\qquad
In^{2}_{ij}=\int_{s_{\gamma e}}\phi_{\gamma,i}\phi_{\gamma,j} n_{\gamma z} \,r^n\,ds_{\gamma e}.
\end{equation}
\begin{equation}\nonumber 
It^{1}_{ij}=\int_{s_{\gamma e}}\phi_{\gamma,i}\phi_{\gamma,j} t_{\gamma r} \,r^n\,ds_{\gamma e},\qquad
It^{2}_{ij}=\int_{s_{\gamma e}}\phi_{\gamma,i}\phi_{\gamma,j} t_{\gamma z} \,r^n\,ds_{\gamma e}.
\end{equation}
\begin{equation}\nonumber 
D_{ij}=\int_{s_{\gamma e}}\phi_{\gamma,i}\pdiff{\phi_{\gamma,j}}{s}  \,r^n\,ds_{\gamma e}, \qquad
P_{ij}=p_{ijk}\left(v^s_{\gamma t,k} - c^s_{\gamma t,k}\right),
\end{equation}
where
\begin{equation}\nonumber
\qquad p_{ijk} = \int_{s_{\gamma e}}\pdiff{\phi_{\gamma,i}}{s} \phi_{\gamma,j}\phi_{\gamma,k}  \,r^n\,ds_{\gamma e}.
\end{equation}
\begin{equation}\nonumber 
S^{1}_{ijk}=\int_{s_{\gamma e}}\pdiff{\phi_{\gamma,i}}{s}\phi_{\gamma,j}\phi_{\gamma,k} t_{\gamma r} \,r^n\,ds_{\gamma e},\qquad
S^{2}_{ijk}=\int_{s_{\gamma e}}\pdiff{\phi_{\gamma,i}}{s}\phi_{\gamma,j}\phi_{\gamma,k} t_{\gamma z} \,r^n\,ds_{\gamma e}
\end{equation}
\begin{equation}\nonumber
W_{ij} = w_{ijk}c^s_{\gamma t,k} + n  w^r_{ijk}c^s_{\gamma t,k},
\end{equation}
where
\begin{equation}\nonumber
w_{ijk}   = p_{kji},\quad
w^r_{ijk}  = \int_{s_{\gamma e}}\phi_{\gamma,i}\phi_{\gamma,j}\phi_{\gamma,k}t_{\gamma r} \,ds_{\gamma e}.
\end{equation}
\begin{equation}\nonumber
Y_{ij} = y_{ijk}v^s_{\gamma n,k},
\end{equation}
where
\begin{equation}\nonumber
\qquad y_{ijk} = \int_{s_{\gamma e}}\phi_{\gamma,i}\phi_{\gamma,j}\phi_{\gamma, k}\kappa\,r^n\,ds_{\gamma e}.
\end{equation}

\subsection{Temporal discretization}\label{temporal}

The result of our spatial discretization is a system of non-linear  Differential Algebraic Equations (DAEs) of index 2 \citep{gresho2}. However, it is well known that the  same methods that apply to ODEs can be used for DAEs \cite{lotstedt86}, and for a review of the methods available we refer the reader to \cite{gresho2}.

The second-order Backward Differentiation Formula (BDF2), which has been applied successfully to similar problems \cite{heil04}, is implemented into our scheme. Below, the method is applied to a scalar equation $\dot{y}=f(y,t)$, with the extension to derivatives in the Navier-Stokes and interface formation equations being a straightforward task. For a time $(n+1)$ with step $\triangle t$, the method applied to the scalar equation gives
\begin{equation}\nonumber \frac{y_{n+1}-y_{n}}{\triangle
t}=\frac{1}{3}\frac{y_{n}-y_{n-1}}{\triangle
t}+\frac{2}{3}\dot{y}_{n+1},
\end{equation}
where the subscript indicates the time step at which a variable is evaluated and
$dy/dt=\dot{y}$. Alternatively, it may be written as
\begin{equation}\nonumber \frac{3y_{n+1}-4y_{n}+y_{n-1}}{2\triangle
t}=\dot{y}_{n+1},
\end{equation}
which is the second-order accurate one-sided Taylor series expansion
of $y$ at $t_{n+1}$.

Often, during a physical process there will be different stages that are characterized by different time scales. It is important that our temporal discretization takes these different scales into account so that the largest possible time step, that enforces a certain accuracy, is chosen
automatically. This is achieved by choosing a `step' so that the local truncation error
$d_{n}=y_{n+1}-y(t_{n+1})$, where $y(t_{n+1})$ is the exact solution, is maintained below a certain
tolerance.  By using an explicit second-order Adams-Bashforth method (AB2) \cite{gresho1} to
predict the solution $y^{p}_{n+1}$, and then comparing the difference between the actual solution
and the predicted one, after solving the non-linear equations we are able to deduce this error. For
the equation, $\dot{y}=f(y,t)$, having obtained the solution $y_{n+1}$ we follow the analysis of
\cite{gresho1} to obtain a new time step. The BDF2 for variable step size gives
\begin{equation}\label{FEM_bdf2_var} \frac{y_{n+1}-y_{n}}{\triangle
t_{n}}=\frac{\triangle t_{n}}{2\triangle t_{n}+\triangle
t_{n-1}}\frac{y_{n}-y_{n-1}}{\triangle t_{n-1}}+\frac{\triangle
t_{n}+\triangle t_{n-1}}{2\triangle t_{n}+\triangle
t_{n-1}}\dot{y}_{n+1},
\end{equation}
while the predictor gave
\begin{equation}\nonumber
y^{p}_{n+1}=y_{n}+\left(1+\frac{\triangle t_{n}}{\triangle t_{n-1}}\right)\triangle t_{n}
\dot{y}_{n}-\left(y_{n}-y_{n-1}\right)\left(\frac{\triangle t_{n}}{\triangle t_{n-1}}\right)^{2}.
\end{equation}
The local truncation error of (\ref{FEM_bdf2_var}) is given by
\begin{equation}\label{FEM_lte1}
d_{n}=\frac{\left(\triangle t_{n}+\triangle
t_{n-1}\right)^{2}}{\triangle t_{n}(2\triangle t_{n}+\triangle
t_{n-1})}\frac{\triangle t_{n}^{3}
\ddot{y}_{n}}{6}+\textrm{O}(\triangle t_{n}^{4})
\end{equation}
whilst the predictor's error is
\begin{equation}\label{FEM_lte2}
y^{p}_{n+1}-y(t_{n+1})=-\left(1+\frac{\triangle t_{n}}{\triangle
t_{n-1}}\right) \frac{\ddot{y}_{n}}{6}+\textrm{O}(\triangle
t_{n}^{4}).
\end{equation}
The exact solution's contribution $y(t_{n+1})$, which, of course, in general will
not be known, is eliminated from (\ref{FEM_lte1}) and (\ref{FEM_lte2}) to give
\begin{equation}\nonumber
d_{n}=\frac{(1+\triangle t_{n-1}/\triangle t_{n})^{2}}{1+3(\triangle t_{n-1}/\triangle
t_{n})+4(\triangle t_{n-1}/\triangle t_{n})^{2}+2(\triangle t_{n-1}/\triangle
t_{n})^{3}}(y_{n+1}-y_{n+1}^{p}),
\end{equation}
so that the error is linearly proportional to the difference between the predicted and actual
solution. This error estimate is then used to compute the next step size
\begin{equation}\nonumber
\triangle t_{n+1}=\triangle t_{n}(\epsilon/\| d_{n}\|)^{1/3},\qquad \| d_{n}\|^{2}=d_{n}^{T}
d_{n}/(N y_{max}^{2}).
\end{equation}
Here, $N$ is the total number of nodes, $y_{max}$ is an estimate of the maximum value of $y$ in the
domain and $\epsilon$ is the relative error tolerance parameter. Then the error of the approximate
solution is bounded by $\| d_{n+1}\| \le \epsilon y_{max}$. This allows us to choose the largest
possible time step whilst ensuring that the error of the temporal integration remains below a
chosen tolerance.

The variable step method outlined above can be extended for use with the Navier-Stokes and interface formation equations and the reader is referred to \cite[][p.~797]{gresho2}, for details. In this book a number of `rules of thumb' are suggested based on the ratio of the new and previous
step sizes, the so-called Delta T Scale Factor, $DTSF=\frac{\triangle t_{n+1}}{\triangle t_{n}}$, and these can be adapted to the problem of interest as required.

\subsection{Far-field velocity profile for the steady propagation of a meniscus through a capillary}\label{ff_derivation}

To ensure mass is conserved in our domain for this steady problem in which an adsorption-desorption process will occur along the interfaces, we must derive appropriate conditions at the base of the capillary, i.e. in the truncated far field. Assuming the problem is steady, so that $\mathbf{c}^s_{||}=\pdiff{\rho^s}{t}=0$, noting that there is no surface mass flux through the axis of symmetry and that there is no sink or source of mass at the contact line, we have
\begin{align}\nonumber
0&=& \int_V \nabla\cdot\mathbf{u} \,dV = \int_A \mathbf{u}\cdot\mathbf{n} \,dA = \int_{A_{ff}} \mathbf{u}\cdot\mathbf{n} \,dA_{ff} + Q\int_{A_{1}} \left(\rho^s_1 - \rho^s_{1e}\right) \,dA_1+ Q\int_{A_{2}} \left(\rho^s_2 - \rho^s_{2e}\right) \,dA_2 \\ \nonumber
 &=& \int_{s_{ff}} w \, dA_{ff} - \epsilon Q \left[ \int_{s_1} \nabla\cdot\left(\rho^s_{1}\mathbf{v}^s_{1||}\right) \,dA_1 + \int_{s_2} \nabla\cdot\left(\rho^s_{2}\mathbf{v}^s_{2||}\right) dA_2\right] \\ \nonumber
 &=& (2\pi)^n\left(\int_{s_{ff}} w  \,r^n \,dr - \epsilon Q \left( - \left.\rho^s_{1}\mathbf{v}^s_{1||}\cdot\mathbf{m}_1 r^n\right|_{apex} - \left.\rho^s_{1}\mathbf{v}^s_{1||}\cdot\mathbf{m}_1 r^n\right|_{cl} - \left.\rho^s_{2}\mathbf{v}^s_{2||}\cdot\mathbf{m}_2 r^n\right|_{cl} - \left.\rho^s_{2}\mathbf{v}^s_{2||}\cdot\mathbf{m}_2 r^n\right|_{ff}  \right)\right) \\ \nonumber
 &=& (2\pi )^n\left(\int_{s_{ff}} w  r^n dr + \epsilon Q \left. \rho^s_{2}v^s_{2t}\right|_{ff} \right)
\end{align}
with subscripts $ff$ referring to far field variables.

Given that $w=w(r)$ in the far-field, from the equations of motion we see that
\begin{equation}\nonumber \diff{p}{z} = \frac{1}{r^n}\diff{}{r}\left(r^{n}\diff{w}{r}\right),
\end{equation}
and, after integrating once, with $\diff{p}{z} = G$, this gives
\begin{equation}\nonumber \frac{G r}{n+1} + Ar^{-n} = \diff{w}{r}.
\end{equation}
Noting that $\diff{w}{r}=0$ at $r=0$, so that $A=0$, we integrate again to find
\begin{equation}\nonumber w = \frac{Gr^{2}}{2(n+1)} + B.
\end{equation}
Applying the Navier condition $-\diff{w}{r} = (\bar{\beta}/Ca) \left(w-W\right)$ at $r=1$, assuming that in the far field there are no gradients in surface tension, we have that
\begin{equation}\nonumber - \frac{G}{n+1} = \frac{\bar{\beta}}{Ca} \left(\frac{G}{2(n+1)} + B - W\right),
\end{equation}
so that
\begin{equation}\nonumber
B = W - \frac{G}{2(n+1)}\left[1+\frac{2}{\bar{\beta}/Ca} \right],
\end{equation}
and, finally
\begin{equation}\nonumber w = W + \frac{G}{2(n+1)}\left[r^2 - 1 - \frac{2}{(\bar{\beta}/Ca)}\right].
\end{equation}

Now, we calculate the pressure gradient, G, required to maintain a steady flux in and out of the domain for a given $W$ by calculating
\begin{align}\nonumber  \int^{1}_{r=0} w r^n dr &=& \left[ W\frac{r^{n+1}}{n+1} + \frac{G}{2(n+1)}\left(\frac{r^{n+3}}{n+3} - \frac{r^{n+1}}{n+1} - \frac{2r^{n+1}}{(\bar{\beta}/Ca)(n+1)}\right)\right]^{1}_{r=0} \\
&=&W\frac{1}{n+1} + \frac{G}{2(n+3)(n+1)^2}\left[n+1 - n - 3 - \frac{2(n+3)}{(\bar{\beta}/Ca)}\right]
\end{align}
Noting that the surface velocity in the far field is $v^s_{2f} = (1/2)\left(w+W\right)$, where $q = \epsilon Q \rho^s_{2e}$, we have that
\begin{equation}\nonumber 0 = W\frac{1}{n+1} - \frac{G}{(n+3)(n+1)^2} \left[1+ \frac{n+3}{(\bar{\beta}/Ca)}\right] + q W - \frac{Gq}{2(n+1)(\beta/Ca)}
\end{equation}
so that
\begin{equation}\nonumber G = \frac{2W(n+1)(1+\bar{q})}{2/(n+3)+ (2+\bar{q})/(\bar{\beta}/Ca)}
\end{equation}
with $\bar{q} = q(n+1)$ and therefore finally
\begin{equation}\nonumber w = W + \frac{W(1+\bar{q})}{2/(n+3)+ (2+\bar{q})/(\bar{\beta}/Ca)}\left(r^2 - 1 - 2/(\bar{\beta}/Ca)\right),
\end{equation}
where, in \S\ref{results}, we had $n=1$ and $W=-1$.

\section*{Acknowledgements}
The authors would like to thank Dr Mark Wilson, Dr Paul Suckling and Dr Alex Lukyanov for many stimulating discussions about the FEM implementation of dynamic wetting phenomena and Jonathan Simmons for carefully proof reading the manuscript. JES kindly acknowledges the financial support of EPSRC via a Postdoctoral Fellowship in Mathematics.

\bibliographystyle{model1-num-names}
\bibliography{Bibliography}

\end{document}